%% file: ESWA_2020.tex
\documentclass[12p,numbers,sort&compress]{elsarticle}
\renewcommand{\footnotesize}{\scriptsize}
\usepackage{textcomp}
\usepackage{listings} 
\usepackage{graphicx}
\usepackage{caption}
\usepackage{subcaption}

\usepackage{csquotes}

\usepackage{tikz}
\usetikzlibrary{tikzmark}
\usetikzlibrary{fit} 
\usetikzlibrary{positioning}
\usetikzlibrary{arrows}
\usetikzlibrary{shapes.multipart}

\usepackage[hidelinks,bookmarks=false]{hyperref}
\usepackage[numbered]{bookmark}
\usepackage{color,soul}
\usepackage{booktabs}
\usepackage{multirow}
\usepackage{url}
\usepackage{tcolorbox}
\usepackage{amsmath,amssymb,amsfonts}
\usepackage{algorithmic}
\usepackage{xcolor}
\usepackage{url}
\usepackage{tabularx} 
\usepackage{array}
\usepackage{lineno,hyperref}
\usepackage{longtable}
\modulolinenumbers[5]
\journal{Journal of \LaTeX\ Templates}

\usepackage{pgfplots}
\pgfplotsset{compat=1.14}

\usepackage{adjustbox}
\usepackage{float}
\usepackage{rotating}
\usepackage{tablefootnote}
\usepackage{nth}
\DeclareCaptionType{TextBox}
\newcolumntype{L}{>{\centering\arraybackslash}m{3cm}}

\tcbset{tab1/.style={fonttitle=\bfseries\large,fontupper=\normalsize\sffamily,
colback=yellow!6!white,colframe=red!75!black,colbacktitle=black!75!black,
coltitle=white,center title,freelance,frame code={
\foreach \n in {north east,north west,south east,south west}
{\path [fill=red!75!black] (interior.\n) circle (3mm); };},}}

\tcbset{tab2/.style={enhanced,fonttitle=\bfseries,fontupper=\tiny\sffamily,
colback=yellow!6!white,colframe=red!50!black,colbacktitle=black!75!black,
coltitle=white,center title}}
\usepackage[T1]{fontenc}
\usepackage{array, booktabs, makecell}
\usepackage[utf8]{inputenc}
\usepackage{dirtytalk}
\usepackage{tcolorbox}

\usepackage[british]{babel}
\usepackage{enumitem}

\newlist{SubItemList}{itemize}{1}
\setlist[SubItemList]{label={$-$}}

\let\OldItem\item
\newcommand{\SubItemStart}[1]{%
    \let\item\SubItemEnd
    \begin{SubItemList}[resume]%
        \OldItem #1%
}
\newcommand{\SubItemMiddle}[1]{%
    \OldItem #1%
}
\newcommand{\SubItemEnd}[1]{%
    \end{SubItemList}%
    \let\item\OldItem
    \item #1%
}
\newcommand*{\SubItem}[1]{%
    \let\SubItem\SubItemMiddle%
    \SubItemStart{#1}%
}

\usepackage{framed}
\usepackage{mdframed}
\usepackage{pstricks}
\usepackage{xspace}
\usepackage{pgf-pie}
\usepackage{comment}
\usepackage{colortbl}
\tcbuselibrary{skins}
\tcbuselibrary{listings}
\usepackage[normalem]{ulem}
\usetikzlibrary{patterns,}
\usepgfplotslibrary{colorbrewer}
\usepackage{verbatim}
\usepackage{tikz}
\usetikzlibrary{tikzmark}
\usetikzlibrary{fit} 
\usetikzlibrary{positioning}
\usetikzlibrary{arrows}
\usetikzlibrary{shapes.multipart}
\usepackage{lscape}
\usepackage{placeins}
\usepackage{afterpage}
\newlength\mylength
\usepackage{fontawesome}
\usepackage{xspace}

\newcommand{\etal}{\textit{et al. \xspace}}
\newcolumntype{L}{>{\arraybackslash}m{18cm}}
\usepackage[flushleft]{threeparttable}
\usepackage{mdframed}

\newtcolorbox{boxK}{
    sharpish corners, 
    boxrule = 0pt,
    toprule = 4.5pt, 
    enhanced,
    fuzzy shadow = {0pt}{-2pt}{-0.5pt}{0.5pt}{black!35} 
}
\begin{document}

\begin{frontmatter}

\title{\textcolor{black}{Software Refactoring Research with Large Language Models: A Systematic Literature Review}}

\author[STEVENS]{Sofia Martinez\fnref{contrib}}
\fntext[contrib]{Authors contributed equally}
\ead{smartine1@stevens.edu}

\author[STEVENS]{Luo Xu\fnref{contrib}}
\ead{lxu41@stevens.edu}

\author[STEVENS]{Mariam Elnaggar}
\ead{melnagga@stevens.edu}

\author[STEVENS]{Eman Abdullah AlOmar\corref{mycorrespondingauthor}}
\cortext[mycorrespondingauthor]{Corresponding author}
\ead{ealomar@stevens.edu}


\address[STEVENS]{Stevens Institute of Technology, Hoboken, NJ, USA}

\begin{abstract}
Background: Code refactoring is the improvement of code internally without changing the external functionalities of the program. Due to its exhaustive nature, developers often avoid manually refactoring code. Researchers have since looked into utilizing Large Language Models (LLMs) to automate the task of refactoring. 

Aim and Method: Despite the promising results, there is a lack of clear understanding of LLMs' effectiveness in automated refactoring. In order to address this issue, we conducted a Systematic Literature Review (SLR) of \textcolor{black}{50} primary studies. We categorized the studies into different refactoring methods studied, prompt engineering and techniques conducted, LLM tools used, languages used, and datasets used. \color{black}We touched upon the benchmarks each studies had used, how accurate LLM-generated refactorings are, and the challenges that this field faces currently.\color{black}

Result: From our literature review we found that: (i) \color{black}there are various tools that different studies use to enhance and study LLM-driving refactoring, with tools that were used to detect code smells, generate code bases, and compare refactoring outcomes. \color{black}
(ii) Various datasets were collected from multiple open-source projects in multiple programming languages for analysis. These platforms included GitHub, Apache, and F-Droid, with the most popular language collected and analyzed being Java. (iii) One-Shot, Few-Shot, Context-Specific, and Chain-of-Thought prompting methods have been shown to be the most effective depending on the language used. In some instances, being capable of reducing code smell by up to 89\%. (iv) \textcolor{black}{The definition of ``Accuracy'' varies significantly across the literature surveyed, as this depends on the context of the study}, thus calling for a need to have a standardized measurement for Accuracy. \color{black}(v) The most often mentioned code smells were Large Class and Long Method, with a lot of studies also not specifying, while the most often applied refactoring type is Extract Method, showing promising results in using LLM to perform this refactoring type. \color{black}(vi) When working with LLMs, they often generate erroneous code, struggle with more complex refactoring, and often misunderstand the developers' requests and miss refactoring requests. 

Conclusion: Our study serves to be a collection of knowledge on the topic of LLM refactoring found by various other studies, and to highlight any issues that are often missed by the researchers. We hope to empower and guide the future development of LLM-driven refactoring with our findings. 

\end{abstract}
\begin{keyword}
refactoring, quality, literature review.
\end{keyword}

\end{frontmatter}


\section{Introduction}
\label{Section:Introduction} 
\color{black}As Large Language Models (LLMs) become more complex and capable, developers now seek to incorporate these tools into their developmental processes. One of the most important tasks in development that is often overlooked is refactoring. Refactoring refers to the activity of taking preexisting code and changing its internal structure to improve it without changing its functionalities \cite{fowler2018refactoring,alomar2021preserving}. Code refactoring is imperative to the readability, maintainability, and efficiency of the program, especially for legacy code bases that have grown over time. This code held a large amount of technical debt that many human developers would avoid fixing, and rather, focus on code development over maintainability or efficiency \cite{yanakiev2025applying}. Due to this habit, when developers do have time to go back to fix the existing code base, it often ends up taking them a substantial amount of time.

Therefore, previous studies have started exploring how to automate code refactoring to build better code and to cut down on manual labor \cite{baqais2020automatic}. Many studied the effects of external frameworks and tools used alongside LLMs \cite{pomian2024assist,pomian2024next,dilhara2024unprecedented,cordeiro2024empirical,gao2024context,wang2025testing, palit2025reinforcement, guenoune2025llm}, the kinds of refactoring done by LLMs \cite{gehring2023deterministic,white2023prompt,shirafuji2023refactoring,depalma2024exploring,alomar2024refactor,chavan2024analyzing}, and the different ways to prompt LLMs \cite{gehring2023deterministic,white2023prompt,shirafuji2023refactoring,depalma2024exploring,alomar2024refactor}. 
However,  there is a lack of Systematic Literature Review  (SLR) on evaluating the overarching impacts of LLMs on automated code refactoring. Oftentimes, literature exists to address only one part of the problem in LLM-developer interactions (such as analyzing Few-Shot prompting \cite{shirafuji2023refactoring} or analyzing effect of LLMs with Java refactoring \cite{gao2024context}).

The specifications of the studies also needed to be addressed; this study isolates the evaluation of refactoring by LLMs as a separate case compared to using LLMs in other ways as code refactoring holds special usages for models. In using LLMs for refactoring, one must consider the added layer of how the code will exist within a predefined environment while also replicating what already previously existed. In other words, how can developers interact with LLMs to get the perfect response, that is the perfect blend of style, utility, and maintainability? This approach in refactoring addresses several aspects of LLMs' capabilities that LLMs have shown to be capable of while also showing errors in; thus we must address what are these errors and how to best maximize the usage of LLMs in this case.

This SLR can then help address questions across the board in different programming languages and the best techniques for interacting with LLMs as a developer for refactoring instead of being limited to one topic, tool, method or language. 
To address this issue, we reviewed the most relevant studies done with LLMs-driven refactoring. We extracted key information from \textcolor{black}{50} studies and categorized the papers, identifying the most often used techniques by researchers in their evaluations--such as prompting strategies, evaluation methods, types of refactorings performed--and identified the major challenges faced when automating refactoring using LLMs. We found that prompt structure, context of the project, and providing examples can significantly improve the quality of LLM-generated refactorings, mitigating critical concerns such as hallucinations, context loss, and scalability issues. We hope that this SLR can help bridge the gaps between studies and further the study on the topic of LLM refactoring. \color{black}

When defining our research questions, we framed them around the quantitative and qualitative aspects of the papers as follows:
\begin{itemize}
    \item \textbf{RQ$_1$: \textcolor{black}{What are the tools used in LLM-driven refactoring?}} \textcolor{black}{We posed this question to categorize and study the effects of the existing approaches and tools for LLM-driven refactoring. When implementing different tools, frameworks, and approaches alongside LLMs, we see a considerably large amount of changes in output.}
    
    \item \textbf{RQ$_2$: What datasets and benchmarks have been used to evaluate LLMs for refactoring tasks?} We posed this question to examine the various datasets and benchmarks used to evaluate LLMs for refactoring and further understand the design of the experiments in each study.
    \item \textbf{RQ$_3$: What prompt techniques have been used to perform refactoring tasks?} We posed this question to examine what prompting methods researchers have been exploring. By employing the best prompting strategies, we can help reduce hallucinations and maximize refactoring efficiency.
    \item \textbf{RQ$_4$: How accurate are LLM-generated refactoring suggestions?} This RQ investigates the Accuracy of refactoring suggestions generated by LLMs. To this end, we perform quantitative, qualitative, comparative, and correctness data analysis of the LLM-based tools and approaches for each study.
    \item \textcolor{black}{\textbf{RQ$_5$: What refactoring characteristics are addressed in LLM-driven refactoring?} This RQ investigates the main attributes of refactoring in LLM-assisted refactoring through the lens of 4 identified data collection: qualitative, quantitative, comparative, and correctness.}
    \item \textbf{RQ$_6$: What are the key challenges in automating refactoring using LLMs?} We pose this question to identify and understand the difficulties that arise with LLM-based refactoring and analyze possible methods to address them.
    
\end{itemize}
The contributions of this paper are summarized as follows:
\begin{itemize}
    \color{black}
    \item A comprehensive analysis of the existing literature on LLM-based refactoring across tool supports, datasets used, prompt engineering techniques, and refactoring types.
    \color{black}
    \item A breakdown of LLM-driven refactoring approaches and an outline of strengths and applications of the major refactoring methods.
    \item Identifying open issues in existing research and recommendations for future research directions, including concerns, prompt engineering strategies, tools, and refactoring methods implementations.
\end{itemize}

The remainder of the paper is structured as follows: Section \ref{Section:RelatedWork}, a review of existing studies related to LLM-driven refactoring. Section \ref{Section:Methodology}, an outline of our study setup in terms of search strategy and study selections. Section \ref{Section:Result}, the findings of our research and analysis. Section \ref{Section:Discussion}, concerns and call-to-action for possible future research. Section \ref{Section:Threats}, the threats to the validity of our work, and concluded with Section \ref{Section:Conclusion}.

\section{Related Work}
\label{Section:RelatedWork}

\textcolor{black}{Baqais and Alshayeb \cite{baqais2020automatic} presented a systematic literature review on automated software refactoring. The authors applied multiple filtering steps to identify 41 relevant studies. Their analysis highlights growing interest in automated and search-based code refactoring, but model refactoring remains underexplored. Singh and Kaur \cite{singh2018systematic} conducted a systematic literature survey on 238 publications related to refactoring with respect to code smells. The authors generally covered refactoring, but were more focused on code smells and antipatterns. The study revealed various data sets and tools that are associated with performing software refactoring. AlDallal and Abdin \cite{al2017empirical} reviewed 76 selected primary studies on the impact of refactoring on different software quality attributes. The results show that different refactoring methods can have opposite impacts on different quality attributes, indicating that refactoring does not necessarily always increase all attributes of the quality of the software. The authors also used a vote-counting approach to determine the level of consistency among the results of different studies on the effect of refactoring on software quality, which sheds light on how specific refactoring methods affect some internal quality attributes. However, due to insufficient findings, the study did not identify the impacts of refactoring methods on external and other internal attributes.}

\textcolor{black}{Zhang \etal \cite{zhang2011code} investigated the effectiveness of using code bad smells to direct code refactoring. To explore this, the authors performed a systematic review of 319 articles published between 2000 and 2009, of which 39 extremely relevant studies were analyzed. The study revealed that Duplicated Code received the most research attention, while other types of code bad smells, such as Message Chains, were not studied as much, indicating insufficient knowledge of certain code bad smells. Additionally, most studies focus on finding code bad smells instead of evaluating their impacts, which suggests that there is little evidence to support the use of Code Bad Smells to help guide software refactoring. AlOmar \etal \cite{alomar2024behind} presented a systematic literature review on Extract Method refactoring, a popular technique for improving code quality. To classify and evaluate Extract Method research, the authors compiled 1,367 papers and filtered them down to 83 primary studies. Their findings indicate that several methods have been proposed, with 38.6\% of publications focusing on code clones, some of the tools requiring developer involvement, and since existing benchmarks are not standardized, making comparisons between tools is difficult.} 

\textcolor{black}{More recently, Jiang \etal \cite{10.1145/3747588} surveyed the role of Large Language Models in code generation, giving a systematic overview of how LLMs can be applied to software development tasks such as transformation, synthesis, and completion. While the focus of the paper was not explicitly on refactoring, many of the identified challenges, such as ensuring semantic preservation, handling complex code structures, and evaluating code quality, are similar to the issues faced in automated refactoring. This connection suggests that LLMs face common underlying challenges across a wide range of applications within software engineering. He \etal \cite{he2025llm} presented a systematic literature review on the usage of LLM agents in software development with 71 published papers and provided a breakdown of how LLM-powered Multi-Level Agents (LMAs) worked and what are some existing usages for LMAs. The authors found that there are different ways of setting up LMAs, such as having agents working in agile or waterfall models, having agents being capable to cross-examine different answers and check on its own reasoning, making LMAs less prone to mistakes and hallucinations. The authors has also found that LMAs are useful across different software engineering fields such as requirements engineering, code generation, software QA, and maintenance; but for future studies, it should be examined how can agents be best trained for different specialized roles and how show tasks be delegated between LMAs and human developers. Hou \etal \cite{hou2024large} reviewed 395 papers from 2017 - 2024, the paper reported seeing over 70 different LLMs being used for software engineering tasks, and these LLMs can be separated into 3 different kinds of architectures: Decoder-only, Encoder-only, and Decoder-Encoder. Out of these 3, Decoder-only models are seen to be the most flexible and are used for both code generation and completion. In the datasets examined, the author reported seeing open-source datasets being the most prevalent, at 62.83\% of the datasets collected, calling into question the need to see more industry datasets. The authors have stated what tasks in software engineering LLMs are most often tasked to complete, which are categorized into: Requirements engineering, Software design, Software development, Software quality assurance, Software maintenance, and Software management, with Software development having the most references. The authors have found there to be a lack of consistency and standard across the industry, making it hard to come to conclusive evaluations across studies.}

\textcolor{black}{Zapkus and Slotkienė \cite{zapkus2024unit} conducted a systematic literature review on the usage of LLMs on creating unit tests, the authors reported that the LLMs Codex, GPT-3, StarCoder, and GPT-4 were used to generate unit tests. The authors reported that the topic of LLM code generation is on the rise, increasing by 56.82\% per year. Wang and Chen \cite{wang2023review} conducted a systematic literature review with 20 existing papers and found that the research that goes into LLM-generated code fails to keep up with the fast-moving pace of the applications, and that there is a lack of standardization of quality assessment along with lack of assessment beyond security and safety, calling these two qualities the baseline. The authors also pointed out that there is a lack of consideration for the human developer when conducting the evaluation. Husein \etal \cite{husein2025large} conducted a systematic literature review with 23 papers and found that LLMs can perform code generation/completion on different levels, where token level is the most widely studied. They have found that with bigger datasets and transformer-based models, Accuracy is increased. They also report that dynamically typed languages, such as python, are more prone to error than static languages such as Java when generated by LLMs.} 

\section{Study Design}
\label{Section:Methodology}
\subsection{Survey Planning}
\subsubsection{Identifying the Need for a Systematic Literature Review} Since the use of LLMs for refactoring is still a developing field, a comprehensive review of the application of LLMs for refactoring is needed. Although previous studies have explored various aspects of refactoring, including quality impact and code smell detection, few have specifically focused on the use of LLMs for refactoring. Therefore, the primary objectives of this study are to:
\begin{itemize}
\item Gather existing research on the application of LLMs for refactoring. 
\item Analyze research findings and conclusions to identify benefits and limitations of LLM-based refactoring. 
\item Highlight gaps and challenges to guide future research in this evolving field. 
\end{itemize}
\subsubsection{Specifying the Research Questions} During the process
of conducting an SLR, it is of paramount importance to pinpoint
pertinent research questions that have the potential to provide
clear answers. We identified \textcolor{black}{six} such research questions:
\begin{itemize}
    \item RQ$_1$. \textcolor{black}{What are the tools used in LLM-driven refactoring?}
    \item RQ$_2$. What datasets and benchmarks have been used to
evaluate LLMs for refactoring tasks?
    \item RQ$_3$. What prompt techniques have been used to perform refactoring tasks?
    \item RQ$_4$. How accurate are LLM-generated refactoring suggestions?
    \item \textcolor{black}{RQ$_5$. What refactoring characteristics are addressed in LLM-driven refactoring?}
    \item RQ$_6$. What are the key challenges in automating refactoring using LLMs?
\end{itemize}

\subsection{Primary Studies Selection}
\subsubsection{Search Strategy}
\begin{figure}[t]
 	\centering
 	\includegraphics[width=0.7\columnwidth]{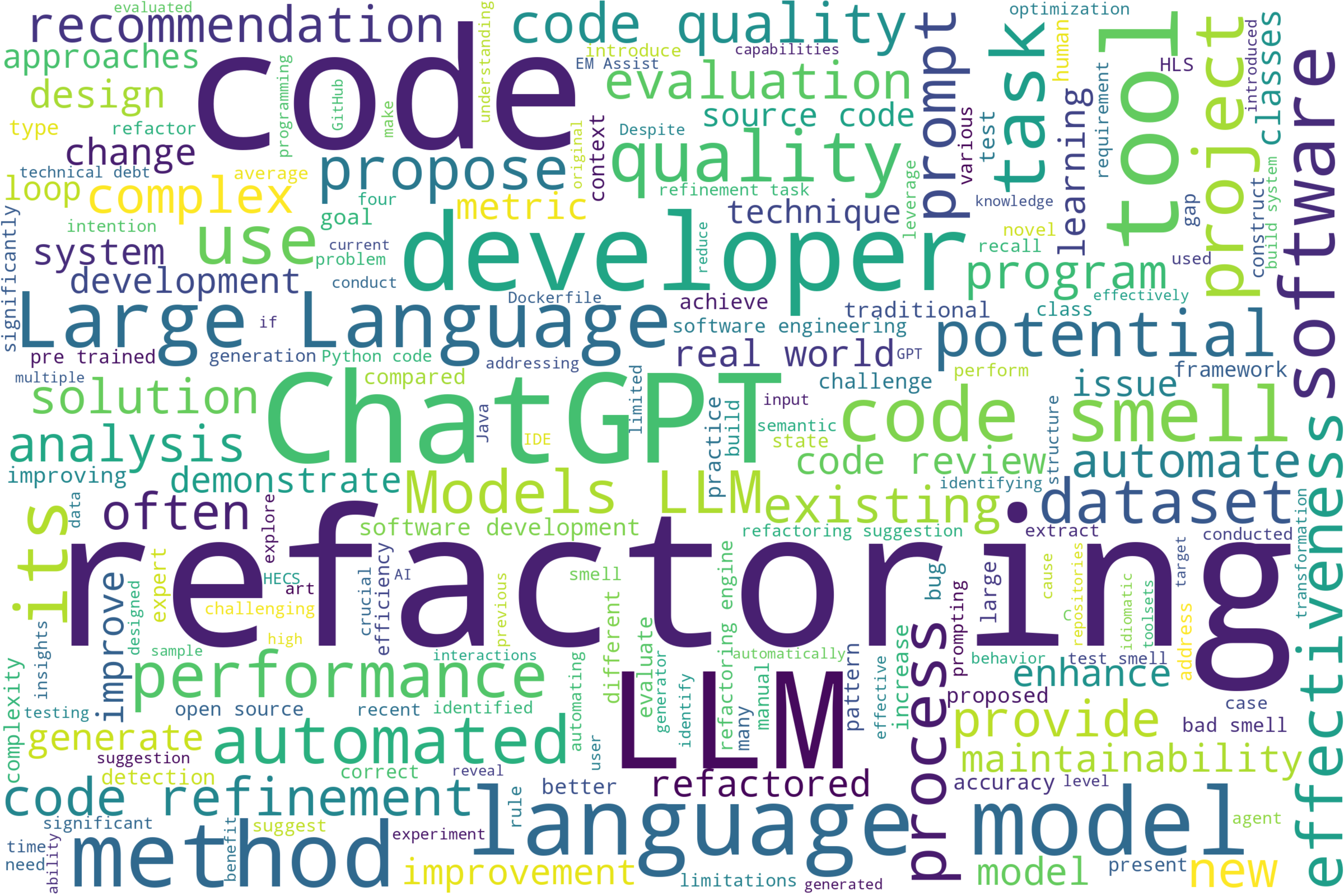}
 	\caption{\textcolor{black}{Word cloud of paper abstracts of primary studies.}}
 	\label{fig:wordcloud}
\end{figure}

To gather relevant studies, we searched eight digital libraries (i.e.,  ScienceDirect, Scopus,
Springer Link, Web of Science, ACM Digital Library, IEEE Xplore, Wiley, and Google Scholar) that store published computing research papers. Given the rapid development of LLMs, we limited our search to studies published between \textcolor{black}{Nov 2022 and Sept 2025}. To ensure that we included research that combines both refactoring and LLMs, we constructed a search string that combines terms related to both fields \textcolor{black}{and applied it to the abstract and keyword fields}. \textcolor{black}{TextBox \ref{txt:search_string} shows our search string in these search engines.} \textcolor{black}{To get a high-level picture of
the covered topics, we generated a word cloud\footnote{The figure is generated using Manus.} of paper abstracts, as depicted in Figure \ref{fig:wordcloud}.}  
\begin{TextBox}[h]
\centering
\fbox{\begin{minipage}{26em}
\textcolor{black}{``refactor*'' OR ``code refinement'' OR ``code restructure'' OR ``transforming code'' OR ``code modification'' combined with ``LLM*'' OR ``language model'' OR ``language models'' OR ``GenAI'' OR ``Gen AI'' OR ``Generative AI'' OR ``GenerativeAI'' OR ``ChatGPT'' OR ``GPT'' OR ``pre-trained'' OR ``BERT'' OR ``Copilot''}
\end{minipage}}
\captionof{TextBox}{Search string.\label{txt:search_string}} 
\end{TextBox}

However, some changes were made due to variations in search functionality across databases:  
\begin{itemize}
    \item In Google Scholar, filtering by abstract \textcolor{black}{and keyword} was not possible, so we conducted a title-based search instead. \textcolor{black}{To further restrict the results and improve the relevance of retrieved studies, we limited our search terms to "refactor*", "LLM*","language models", and "language agents". Since Google Scholar does not support the wildcard (*) operator, we explicitly included each variation of the relevant terms.}  
    \item In ScienceDirect, filtering by abstract \textcolor{black}{and keyword} also includes the title \textcolor{black}{field. Because this search mode does not support wildcard (*) operators and permits only up to eight Boolean operators, we explicitly included each variation of the relevant terms and divided the full search string into three separate queries to capture all relevant combinations of refactoring and LLM related terms.}
    \item In Springer, \textcolor{black}{searching the abstract was not possible, so we performed only a keyword search.}
    \item In Wiley, we performed a full-text search due to the limited number of results when searching \textcolor{black}{only the abstract and keywords fields.}
\end{itemize}
\subsubsection{Study Selection}
We applied a set of inclusion and exclusion criteria, so we could choose papers that would be useful for our search. We included papers that were peer-reviewed and published in journals or conference proceedings within the field of computing. The studies also need to explicitly discuss the use of LLMs in software refactoring, not just refactoring or LLMs, to maintain our focus. We excluded non-peer-reviewed papers, such as preprints \cite{xu2025mantra,chen2025together,cordeiro2024empirical,gheyi2025evaluating}, papers written in languages other than English\textcolor{black}{, and position papers.} 

\textcolor{black}{Notably, we also exclude papers where the primary contribution is not the application of a generative LLM for refactoring. This includes studies that utilize encoder-only models like BERT for classification tasks. For instance, we exclude works like Palit \etal \cite{palit2023automatic}, which uses GraphCodeBERT, a BERT-based model not designed for sequence to sequence generation, to identify refactoring candidates, and Alazba \etal \cite{alazba2025smellybot}, which focuses solely on code smell detection, not implementing any actual refactoring. This deliberate scoping decision is grounded in a fundamental methodological distinction. While studies like Liu \etal \cite{liu2023refbert} demonstrate the effective application of BERT-based models to classification-specific tasks, like predicting a new identifier name, these models operate under a different approach than the generative LLMs that are the focus of this review. Encoder models like BERT are typically applied to refactoring by presenting it as a classification or simple prediction task (e.g., predicting a single token for a new method name), which is naturally evaluated with metrics like Accuracy and F1-score. For example, the SmellyBot tool \cite{alazba2025smellybot} uses CodeBERT, a BERT-based model not designed for sequence-to-sequence generation, to achieve high performance (e.g., 95.93\% Accuracy, 96.82\% Precision) on the specific, classification-based task of code smell detection. The tool operates by analyzing code to classify it into predefined categories of smells, functioning as a classifier rather than a generator, and does not implement any actual refactoring. In contrast, the generative LLMs covered in our primary studies, such as ChatGPT and LLaMA, are designed for open-ended sequence generation, which is essential for producing the diverse and complex code transformations inherent in most refactoring operations. The subset of the primary studies that reported metrics such as Precision and Recall to evaluate generative LLMs, applied them to the evaluation of their generated output as a whole (e.g., comparing an entire generated code snippet to a ground truth version), rather than assessing a single classification decision. Therefore, to maintain a consistent analytical framework for synthesizing results, particularly for  RQ$_4$ (accuracy of suggestions) and RQ$_5$ (refactoring characteristics), we limit our primary studies to those investigating this distinct class of generative models. It is important to note, however, that several of the included primary studies do use static analysis tools or BERT-based models as baselines for comparison against their proposed generative LLM approaches. Where these comparative results are reported within our selected primary studies, we synthesize and discuss them to provide insight into the relative performance of generative LLMs against these alternative methodologies.}

\textbf{Stage 1: Identification of potentially relevant articles.} We started the selection process by conducting an initial search across our databases, applying our search string to titles, abstracts, and keywords based on each library's search capabilities. This search yielded a total of \textcolor{black}{867} research publications. The text-string search's results were unrestricted, ensuring an extensive collection of potentially relevant studies. 

\textbf{Stage 2: Removal of duplicates.} After merging all the search results, we identified and removed duplicate publications, which reduced the dataset to \textcolor{black}{558} unique research papers and eliminated \textcolor{black}{309} articles.

\textbf{Stage 3: Exclusion of articles based on title and abstract.} To narrow down our selection, we applied our inclusion and exclusion criteria to the titles and abstracts of the remaining studies. This process narrowed down our dataset to \textcolor{black}{79} papers, and when we could not determine relevance from just the title and abstract, the study was kept for further analysis. 

\textbf{Stage 4: Exclusion of articles based on full text.} The remaining publications were fully analyzed to assess their relevance, and research that did not align with our inclusion and exclusion criteria was filtered out. As a result, \textcolor{black}{50} papers were selected for our study. 


\subsection{Study Quality Assessment} To assess the quality of PSs, we followed the guidelines
proposed in \cite{kitchenham2007guidelines,li2021understanding,dybaa2008empirical}. We selected three quality assessment questions that could apply to all PSs. These questions serve to evaluate the study quality of each PS in the following three dimensions: objective, method, and coverage of the studies. The questions are as follows: Q1) \textit{Does the study’s primary objective explicitly focus on the LLM-based refactoring?}; Q2) \textit{Does the study include automatic or semi-automatic LLM-based refactoring approaches?}; Q3) \textit{Does the study sufficiently describe the LLM-based technique, underlying algorithms, and evaluation methods?}. These questions are implicitly used in the above filtering process. If a PS passes these quality criteria, we believe that it has valuable information for SLR. The answer to each of these questions is either “Yes”, “Partially”, or “No” with numerical values of 1, 0.5, or 0, respectively. If the questions did not apply to the context of a PS, they were not evaluated. The overall quality of each PS is calculated by summing up the scores of the applicable questions. Overall, all the published articles in the accepted literature scored well on the quality assessment questions.

\subsection{Data Extraction, Categorization, and Analysis} 

To seek the gaps of current studies in LLM refactorings, we screened a total of \textcolor{black}{50} different papers on such topic. 

After collecting relevant papers we created a spreadsheet categorizing important key attributes we wanted to point out in the papers. Within our spreadsheet we had attributes such as title, link to the paper, year it was published, authors, venue, data source, research questions addressed, topics investigated, tools used, type of analysis, LLMs used, refactoring types used, programming language examined, prompting techniques, final prompt, results, and extra notes for miscellaneous items that we wanted to point out. We had taken turns in reviewing the papers. There are two authors involved in reviewing and cross-checking the manual analysis of the research papers. In our process, after one author reviewed a paper, the other would circle back and review it as well, then check the other's notes to ensure correctness. \\ 
After reviewing the key points of the articles, we analyzed each article again. For our second viewing of the papers, we set forth to write a summary of each of the papers. The summaries are broken down into three specific parts: goal, approach, and results. This helped us get a clear reminder of each paper when we needed a refresher while creating this paper, especially when writing RQ$_1$, RQ$_3$, and RQ$_4$. After the initial summary, we took note of the methodology of the research paper, specifically, how the data was collected, and what the authors did to collect the data. This can be anything ranged from how many datasets were collected, to how the data was preprocessed and processed, what the refactoring process (RQ$_1$\textcolor{black}{, RQ$_5$)} was, the prompting process (RQ$_3$), filtering processes (RQ$_2$), any specific methods used (RQ$_1$), the kinds of data collected (RQ$_2$), any tools that were used (RQ$_1$), and models used (RQ$_2$). These small attributes that made up the various dataset points gave us an insight into some of the most popular and effective usages of LLMs for refactoring, along with giving us snippets of possibilities of what may be missing or lacking in terms of research and understanding. Then, after collecting the methodologies of the papers, we also collected the evaluations from the papers. Evaluations include the qualitative analysis and quantitative analysis, along with extra notes of interest (RQ$_2$). We took note of quantitative analysis by taking a look at the numbers provided, what was calculated (RQ$_2$), important metrics and data, such as the number of tests (RQ$_2$), the percentage of success (RQ$_4$), and comparisons with previous data and tools (RQ$_2$). Qualitative analysis depended more on the overarching non-numeric information from the papers, such as effectiveness (RQ$_4$), baseline comparisons (RQ$_2$), surveys (RQ$_2$), manual comparisons (RQ$_2$) and categorizations (RQ$_3$, RQ$_1$). To ensure complete and correct collections of data, we also asked ChatGPT-4 to check over our written evaluations and methodologies to see if there are important data points missed and added to our notes. This process included sending ChatGPT-4 all of the written notes of summaries with the original papers, asking if there are any more notes or topics that should be included in the summary. This usually resulted in improved grammar, added with some sentences for extra details. The authors also reviewed each other's notes at the end of each iteration and met to communicate and come to a consensus whenever there were disagreements.\\
Once all the data and information had been collected and studied, we found that there are specific methods of approaches to LLM refactoring, these included Prompt Engineering, External Frameworks and Tools, and Refactoring Types. These are different approaches that we see resurfacing across multiple papers repeatedly. These categories have been further analyzed in RQ$_1$ and RQ$_3$ of this study. \\
In RQ$_3$ of the study, we studied the specific prompting techniques that were used in the papers, several of which were repeatedly mentioned and analyzed across multiple papers. 
Throughout our time collecting data and writing summaries of the primary studies, we developed a classification of the key challenges (RQ$_6$) encountered when automating refactoring using LLMs. We began the categorization process by comprehensively reviewing the spreadsheet and study summaries, listing the difficulties that were reported when refactoring with LLMs. Then, we grouped studies with similar limitations (e.g., code hallucinations, intent misinterpretation). Finally, we analyzed these groups to identify and write clear descriptions of each main challenge, as originally documented by the authors of the primary studies.

\subsection{Final Primary Studies Selection} The selection process outlined in the previous section resulted in \textcolor{black}{50} relevant publications. The primary sources of these studies are summarized in Table \ref{Table:Publication_Sources}, spanning a total of \textcolor{black}{6} different databases published at various journals, conferences, and workshops. The earliest relevant study was published in 2023, while the most recent appeared in 2025, highlighting an increasing research interest in the use of LLMs for refactoring and emphasizing the need for further research in this field. 

\input{Tables/Venue}

\section{Experimental Results}
\label{Section:Result}

\subsection{RQ$_1$: \textcolor{black}{What are the tools used in LLM-driven refactoring?}}


\textbf{Motivation.} 
Due to the nature of progressing code bases and developments, it has become increasingly hard for developers to create and update code that maintains readability, maintainability, and efficiency \cite{yanakiev2025applying}. \textcolor{black}{One way to aid in this challenge is by deploying the power of LLMs, but studies vary in their use of tools to compare or aid LLMs in refactoring, so this section is primarily discussing the variety of tools used, their application in the study, and the role they play in the LLM-based system.}


\textbf{Approach.} 
To address these issues, multiple studies have explored LLM-driven refactoring approaches that involve more effective tools with the LLMs. We reported some of these studies' characteristics in Table \ref{Table:Related_Work_in_llms_Refatoring}, \color{black}giving a quick view of all of the papers regarding what tools, LLMs, and programming languages they used. \textcolor{black}{The table differentiates between pretrained models based on their architecture and functional role. Models that include decoder components like CodeT5, PLBART, CodeGPT-adapt, CodeGen, and CodeGPT can be considered as either tools or LLMs based on their purpose. When used within a larger LLM system, they function as supporting tools to an LLM, but when they independently generate or refactor code and are used without a supporting larger LLM, they are regarded as LLMs. By contrast, encoder-only models like BERT and its variants lack generative capability and are therefore classified as tools, as discussed earlier.}
\color{black}

\textbf{Results.} 
\textcolor{black}{We found that the use of tools provided valuable insight for improving LLM refactoring systems and acted as benchmarks to measure LLMs' refactoring ability. Here is the breakdown of the different tools and how they were used.}
\textcolor{black}{External frameworks and tools} are defined as extra helpers used alongside LLMs to enhance their accuracy and efficiency \cite{pomian2024assist, pomian2024next, dilhara2024unprecedented, cordeiro2024empirical, gao2024context, wang2025testing, ksontini2025refactoring, gao2024preference, zhang2024refactoring, alturayeif2025refactoring, wu2024ismell, dong2025chatgpt}. Models and tools used in code smell detection have been found to be highly effective in working together with LLMs to implement higher accuracy in detection and automated refactoring. One such tool that was studied and examined was the iSMELL model \textcolor{black}{which used CodeBERT to encode deep semantic information from code snippets into a vector capturing its multifaceted features.} When using the iSMELL model combined with LLMs, it is found to achieve a 28.7\% improvement over standalone LLMs \cite{wu2024ismell}. 
\textcolor{black}{Additionally, in multiple papers, pretrained code models were implemented to encode code into vectors used as node features in hypergraph neural networks to be provided to an LLM to validate the generated refactoring suggestions to implement Extract Method and Move Method refactoring \cite{cui2024three, cui2024one, wang2024hecs}. A hybrid approach combining structured LLM guidance with rule-based techniques for Python refactoring has achieved over 90\% Accuracy, demonstrating that structured LLM-driven methodologies can significantly enhance the reliability of code refactoring \cite{zhang2024refactoring}. 
One paper used IDE-based static analysis to filter out unrelated classes and VoyageAI to measure semantic similarity between source class and candidate target class to improve Move Method and effectively eliminate all hallucinations that previously affected 80\% of generated results \cite{batole2025leveraging}. Palit and Sharma \cite{palit2025reinforcement} integrated reinforcement learning into their approach to improve LLM refactoring output from 41 successful tests for the CodeT5 model to 66 successful tests when implementing reinforcement learning. The RefactoringMiner tool was used to identify instances of refactoring to generate datasets and provide the LLMs with examples in multiple papers \cite{liu2024empirical, zhang2024refactoring, xu2025mantra, batole2025leveraging, palit2025reinforcement}}. 

Plugins are also shown to be effective in improving current performance in refactoring by LLMs. One such plugin that was studied was EM-Assist \cite{pomian2024assist,pomian2024next}, a tool that integrates LLMs with IntelliJ IDEA, and aids in using ranking mechanisms and validation techniques to refine generated refactoring suggestions. This approach of improving generation was compared with traditional static analysis-based tools such as JExtract, and has been found to create an improvement in Recall from 39.4\% to 53.4\% \cite{pomian2024assist,pomian2024next}. LLMs were also effective in testing existing refactoring engines in IntelliJ IDEA, Eclipse, NetBeans, and VSCode-Java, improving refactoring bug detection by identifying 86 bugs compared to the baseline approach of 24 bugs \cite{dong2025chatgpt}. Across all fields of metrics, the impact of these tools\textcolor{black}{, frameworks, plugins, and approaches} has been largely positive. 

\textcolor{black}{Multiple studies utilized refactoring tools as benchmarks to compare LLM refactoring approaches. CodeReviewer was a popular benchmark when discussing code reviewing tasks with 4 papers mentioning it \cite{wang2025exploration, guo2024exploring, caumartin2025exploring, guo2025intention}, and refactoring tools like JDeodorant and JMove, both Java refactoring tool, were compared to multiple LLM refactoring approaches. \cite{cui2024three, cui2024one, wu2024ismell, wang2024hecs, batole2025leveraging, zhang2025move}. The results of these comparisons are further explored in RQ$_4$ and RQ$_5$.}

\input{Tables/Approach}

\begin{boxK}
\textit{\textbf{Summary for RQ$_1$.} Tools are often used in the studies alongside LLMs to help bettering the performances of LLM refactoring and to collect data of the performances. There are external frameworks and tools used to aid LLMs for identifying code smells, generating data sets, ease of plugin, and measuring effectiveness. We have collected and categorized the tools used along with the LLMs, the languages used, and the focus of the specific study, giving an overview of the tools used in LLM-driven refactorings.} 
\end{boxK}

\subsection{RQ$_2$: What datasets and benchmarks have been used to evaluate LLMs for refactoring tasks?} 

\textbf{Motivation.} Harnessing the capabilities of LLMs has the potential to automate the costly and time-consuming task of refactoring in software engineering. Automating this process with LLMs could significantly reduce developer effort, minimize human error, and enhance software quality at a large scale. Numerous new LLM-based tools and approaches have been proposed in the wake of the widespread accessibility to these models in recent years. \textcolor{black}{However, with each study relying on different datasets and benchmarks to evaluate these models, there is variation in how effectiveness is measured. In this section, our analysis focuses on examining the datasets and benchmarks referenced across the papers, looking at their source, composition, and language. This analysis allows us to better identify patterns in the datasets and sources of the papers, as well as further understand how these factors could bias the evaluations and ultimate conclusions reached about the refactoring capabilities of these LLMs.} Therefore, it is important to analyze the data they use in their approaches and compare them to understand the nature of these evaluations.

\textbf{Approach.} To further understand the quality, workflow, and design of the experiments in each of the primary studies, we report in Table \ref{Table:Dataset_in_llms_Refatoring} the following three characteristics:
\begin{itemize}
    \item \textit{Source:} Indicates where the dataset or benchmark originated from.
    \item \textit{Dataset:} Indicates the type and quantity of the data within each dataset or benchmark.
    \item \textit{Language:} Indicates the programming language(s) represented in the dataset or benchmark.
\end{itemize}
We report the source of any dataset or benchmark from previously published research as ‘Previous work’ along with the link to that work. Additionally, the source of any datasets or benchmarks extracted from unnamed projects is labeled `Various open-source’. 

\textbf{Results.} In the following section, we discuss the sources, datasets, and languages used to evaluate LLMs for refactoring tasks.

\noindent\textbf{Sources.} There were a variety of sources used to evaluate LLMs for refactoring. Eleven out of the 50 primary studies, \textcolor{black}{22\%}, relied on previously published work and referenced established datasets to ensure the credibility of their evaluations. However, we identified \textcolor{black}{35} primary studies \textcolor{black}{(70\%)} that collected data from open-source projects, with code repositories hosted on platforms such as GitHub, Apache, and F-Droid. This trend toward data extraction from open-source projects for conducting these studies is due to the accessibility of this source code, as researchers can analyze real-world refactoring instances without restriction. Additionally, two studies \textcolor{black}{(4\%)} sourced their datasets from educational environments, using student program submissions to test their refactoring approach. Finally, two studies \textcolor{black}{(4\%)} generated new datasets by systematically prompting LLMs and integrating dataset creation into the evaluation process. The diversity in dataset sources highlights the absence of a standardized approach to collecting data for LLM-based refactoring research. 

\noindent\textbf{Datasets.}
There is significant variation in the composition of the datasets and benchmarks used to evaluate LLMs for refactoring tasks, reflecting the variety of sources and overall diversity in approaches to evaluating the refactoring capabilities of LLMs. If the number of projects used is not explicitly mentioned in a study, we report the other quantities of data that are mentioned. Data types range from commits to full methods, and even include types of prompt patterns. The largest datasets contained up to 30,000 code samples and 41,311 refactoring cases, while the smallest contained only 116 methods. Moreover, some datasets \cite{ishizue2024improved,shirafuji2023refactoring} contained full program submissions for multiple programming problems. \textcolor{black}{This range highlights that while some evaluations focus more on scale, others prioritize depth and context. Consequently, datasets built from commits or real student submissions may provide more realistic scenarios, corresponding with the more common use cases of these LLMs, leading to different evaluation outcomes than those based on isolated methods or artificially constructed examples, which may test narrower and more basic capabilities of LLMs.} This variation in dataset composition and sources again reinforces the lack of a standardized benchmark for evaluating LLM-based refactoring, meaning there is inconsistency in how model effectiveness is measured across primary studies. 

\noindent\textbf{Languages.} 
The most common programming language contained in the datasets is Java, which is present in \textcolor{black}{33} \textcolor{black}{(66\%)} of the primary studies \cite{pomian2024assist, pomian2024next, gehring2023deterministic, depalma2024exploring, alomar2024refactor, 
liu2024empirical,
gao2024context, choi2024iterative,
kim2025comparative, wang2025towards, wang2025exploration, midolo2025refactoring, batole2025leveraging, tkachuk2025determining, ghammam2025build, wu2024ismell, baumgartner2024ai,
gao2024preference, cui2024three, zhang2025move, ben2024assessing, wang2024hecs, cui2024one, wang2025testing, menolli2024teaching, guo2024exploring, guo2025intention, palit2025reinforcement, 
caumartin2025exploring, pandini2025exploratory, guenoune2025llm, dong2025chatgpt}. This is followed by Python \cite{white2024chatgpt, gehring2023deterministic, shirafuji2023refactoring, alomar2024refactor, chavan2024analyzing, dilhara2024unprecedented, zhang2024refactoring, wang2025exploration, ksontini2025refactoring, gao2024preference, ishizue2024improved, gautam2024refactorbench, zhang2024copilot, midolo2025automated, alturayeif2025refactoring, guo2024exploring, guo2025intention, alomar2025chatgpt, caumartin2025exploring, guenoune2025llm, pucho2025refactoring, icmi2025evaluating} which makes up 22 \textcolor{black}{(44\%)} studies. \textcolor{black}{The next two most prevalent languages were JavaScript \cite{gehring2023deterministic, alomar2024refactor, chavan2024analyzing, wang2025exploration, icmi2025evaluating, guo2024exploring, guo2025intention, alomar2025chatgpt, caumartin2025exploring, amaral2025improving}  and C++ \cite{alomar2024refactor, wang2025exploration, xu2025hlsrewriter, guo2024exploring, guo2025intention, caumartin2025exploring, guenoune2025llm}, which appeared in ten (20\%) and seven (14\%) of the studies respectively. There are multiple instances of other languages including C\#, Kotlin, SQL. However, they appear much less frequently than the aforementioned languages.} For example, there are instances of Java and Kotlin \cite{pomian2024next,pomian2024assist}, Java and Python \cite{gao2024preference}, Python and Javascript \cite{chavan2024analyzing}, Python, Bash and SQL \cite{ksontini2025refactoring}, as well as Python, Java, Javascript, C++, C\#, Typescript, Go and Ruby together. \textcolor{black}{Additionally, Ghammam \etal used mainly Java, Kotlin, and PHP, but incorporated 27 other languages \cite{ghammam2025build,alomar2024refactor}}. \textcolor{black}{There was a study \cite{de2025effectiveness} that exclusively used languages other than Java, Python, JavaScript, and C++, making up 2\% of the total of 50 primary articles.} The dominant trend of Java and Python data is due to the prevalence of the languages’ within existing datasets and benchmarks in software engineering research, refactoring tools, university curricula, and personal and industry projects. The variety of programming languages indicates that software engineering research is expanding to explore refactoring of queries and scripts, although it is still mainly skewed towards traditional codebase refactoring.

\textcolor{black}{\noindent\textbf{Data Leakage.} 
When it comes to preventing data leakage, there are a few common methods for preventing data leakage when evaluation of language models. The ones considered in Table 3 are training/testing splits, unseen projects or repositories, held-out datasets, and cross-validation. A training/testing split means dividing the datasets into separate sections for training and for evaluation, so that the model is not tested on data it has already seen. Unseen projects or repositories go a step further by evaluating models on entirely different sources than those used when training a model. A held-out dataset is a collection of data reserved only for testing that was never used during training. Finally, cross-validation splits the data into multiple folds (smaller subsets) and then cycles through multiple rounds where the model is tested on one fold and trained on the remaining folds, rotating the test fold each round.}

\textcolor{black}{A constant focus when it comes to an experiment is eliminating as much bias as possible. The critical need for these precautions is underscored by research on LLM threats within software engineering. As highlighted by Sallou \etal \cite{sallou2024breaking}, a primary concern is implicit data leakage due to pre-training. LLMs are trained on vast public datasets like GitHub, which may include the very benchmarks (e.g., Defects4J) used for evaluating their capabilities. This can lead to models memorizing existing solutions rather than generating new ones, severely threatening the validity of research findings.}

\textcolor{black}{In the context of LLM-based refactoring, data leakage refers to language models being tested on the same data that was used to train them. For the most exact, unbiased study of LLMs, precautions such as held-out datasets, training/testing splits, cross-validation, etc., should be taken to prevent data leakage. In our analysis of the various datasets and benchmarks used to evaluate LLMs on refactoring tasks, shown in Table 3, we found a widespread lack or omission of data leakage precautions by the authors. Overall, a large majority—36 out of 50 primary papers or 72\%—made no mention of deliberately implementing data leakage precautions when training and evaluating their LLMs. Of the 14 papers that did specify taking such precautions, six (12\%) used training/testing splits \cite{cui2024three, zhang2025move, ksontini2025refactoring, ghammam2025build, palit2025reinforcement, caumartin2025exploring}, with the rest using held-out datasets \cite{cui2024one, alturayeif2025refactoring, guo2025intention} at 6\%, unseen projects or repositories \cite{zhang2024refactoring, gautam2024refactorbench, liu2024empirical, guo2024exploring} at 8\%, or cross-validation \cite{wu2024ismell} at 2\%. For the remaining five papers, the application of data leakage prevention measures was unclear, and they were thus reported as ‘Not clear’ in the table. While the mention of preventative measures in almost a third, 28\%, of the reviewed primary papers is encouraging, this highlights a significant gap in current evaluation practices or reporting of data leakage precautions, which fails to adequately address the fundamental threats of data contamination inherent in LLM-based research.}

\textcolor{black}{Notably, this review does not confirm the presence or absence of data leakage in the analyzed studies. Rather, we report only whether authors explicitly mentioned implementing any preventive measures and, when specified, what those measures were. The inclusion of such precautions in a paper does not guarantee complete mitigation of data leakage, as verifying that falls beyond the scope of this review.}

\textcolor{black}{Figure \ref{fig:language} shows the distribution of programming languages used in the primary studies to evaluate refactoring tools. Java, Python, JavaScript, and C++ emerged as the most prevalent languages. The reported percentages are non-exclusive; a single study utilizing multiple languages contributes to the count for each language used. As the data indicate, Java is the most popular language in LLM-based refactoring research, featured in 66\% (33) of the primary studies. It is followed by Python, employed in 44\% (22) of the studies. JavaScript, in 20\% (10) of the studies, and C++, in 14\% (7) of the studies, were also common. Additionally, 24\% included other less frequently used languages besides the four mentioned above. This overview provides a detailed picture of the current programming language focus in the field, offering useful information that could guide future researchers within this domain concerning the diversity in their language selections.}

\input{Tables/Dataset}

\begin{boxK}
\textit{\textbf{Summary for RQ$_2$.} \textcolor{black}{The datasets and benchmarks used to evaluate LLMs for refactoring tasks vary greatly in both their composition and scale, making their standardization for more consistent comparison difficult. A majority of studies, 70\%, sourced data from open-source repositories, with Java (66\%) and Python (44\%) dominating the evaluation landscape, reflecting their prevalence in software engineering research. Additionally, 70\% of studies fail to report measures against data leakage, underscoring a substantial validity concern within this emerging field.}} 
\end{boxK}

\subsection{RQ$_3$: What prompt techniques have been used to perform refactoring tasks?} 

\textbf{Motivation.} The need for studying different prompting techniques in LLM-driven code refactoring comes from the need to improve the effectiveness, reliability, and adaptability of AI-assisted software development. While LLMs have shown promise in automating refactoring tasks, their performance depends heavily on how prompts are structured and syntactically formed. Poorly designed prompts can lead to inaccurate refactoring, making it crucial to identify and refine effective prompt engineering strategies. It is seen through studies that different refactoring tasks require varying levels of context and reasoning. Simple refactoring methods, such as renaming or adjusting indentation, can often be handled with basic prompts. However, more complex refactoring methods, such as Extract Method, demand more structured, iterative, and context-based prompts to ensure correctness and maintainability. 

\textbf{Approach.} To address these challenges, researchers have explored different prompting approaches to determine which methods yield the most accurate, efficient, and maintainable code transformations. \color{black} We reported the prompting technique, final prompts used, along with the best methods used \color{black} by each study in Tables \ref{Table:prompt_in_llms_Refatoring} and \ref{Table:prompt_in_llms_Refatoring-continue}.

\textbf{Results.} \textcolor{black} {Prompt engineering is defined as the process of creating and formulating inquires to LLMs to guide it in order to retain the most efficient and accurate results, \textcolor{black} {thus falling under the category of LLM based approach} \cite{gehring2023deterministic, white2023prompt, shirafuji2023refactoring, depalma2024exploring, alomar2024refactor, chavan2024analyzing, pomian2024assist, pomian2024next, dilhara2024unprecedented, gao2024context, zhang2024refactoring, choi2024iterative, wu2024ismell, ishizue2024improved, gautam2024refactorbench, baumgartner2024ai, cui2024three, ben2024assessing, wang2024hecs, gao2024preference, cui2024one, zhang2024copilot, wang2025testing, midolo2025automated, ksontini2025refactoring, menolli2024teaching, alturayeif2025refactoring, liu2024empirical, icmi2025evaluating, guo2024exploring, batole2025leveraging, kim2025comparative, wang2025towards, wang2025exploration, alomar2025chatgpt, midolo2025refactoring, pucho2025refactoring, caumartin2025exploring, pandini2025exploratory, guo2025intention, xu2025hlsrewriter, tkachuk2025determining, guenoune2025llm, amaral2025improving, de2025effectiveness}. We have found that several studies utilized prompt engineering in their research to guide LLMs in producing structured and semantically correct refactoring. There are multiple forms of prompt engineering implemented across various studies, and improvements and changes in prompting have been shown to be more efficient in creating refactored code.} Below are the seven categories \textcolor{black}{of prompt techniques} we've encountered, listed, explained\textcolor{black}{, and illustrated in Figure \ref{prompt-techniques}}.
\begin{itemize}
    \item Zero-Shot \cite{gehring2023deterministic, shirafuji2023refactoring, depalma2024exploring, alomar2024refactor, gautam2024refactorbench, baumgartner2024ai, zhang2024refactoring, ben2024assessing, midolo2025automated, ksontini2025refactoring, menolli2024teaching, icmi2025evaluating, ghammam2025build, guo2024exploring, kim2025comparative, alomar2025chatgpt, pucho2025refactoring, caumartin2025exploring, guo2025intention, tkachuk2025determining, amaral2025improving}: \textcolor{black}{Implemented in 19 studies (16\%)}, and provides minimal context and asks the model to refactor code directly. 
    \item One-Shot \cite{white2023prompt, shirafuji2023refactoring, alomar2024refactor, chavan2024analyzing, choi2024iterative, wu2024ismell, ishizue2024improved, baumgartner2024ai, ksontini2025refactoring, ghammam2025build, alomar2025chatgpt, midolo2025refactoring}: \textcolor{black}{Implemented in 12 studies (10\%), and }provides a single example to enhance the model's ability to generate and generalize across multiple different refactoring tasks.
    \item Few-shot \cite{white2023prompt, shirafuji2023refactoring, alomar2024refactor, pomian2024assist, pomian2024next, dilhara2024unprecedented, zhang2024refactoring, choi2024iterative, gautam2024refactorbench, baumgartner2024ai, cui2024three, wang2024hecs, gao2024preference, cui2024one, zhang2024copilot, wang2025testing, midolo2025automated, ksontini2025refactoring, alturayeif2025refactoring, liu2024empirical, icmi2025evaluating, wang2025towards, wang2025exploration, alomar2025chatgpt, guo2025intention, xu2025hlsrewriter, guenoune2025llm}: \textcolor{black}{Implemented in 25 studies (21\%), and} provides multiple examples to enhance the model's ability to generalize across different refactoring tasks.
    \item Chain-of-Thought \cite{gehring2023deterministic, white2023prompt, dilhara2024unprecedented, gao2024context, choi2024iterative, zhang2024copilot, wang2025testing, alturayeif2025refactoring, guo2024exploring, batole2025leveraging, alomar2025chatgpt, caumartin2025exploring, pandini2025exploratory, guo2025intention, xu2025hlsrewriter, guenoune2025llm, de2025effectiveness}: \textcolor{black}{Implemented in 18 studies (15\%), and} provides a step-by-step process of reasoning to improve the LLM's decision making in refactoring.
    \item Context-Specific \cite{white2023prompt, chavan2024analyzing, pomian2024next, zhang2024refactoring, choi2024iterative, wu2024ismell, ishizue2024improved, gautam2024refactorbench, cui2024three, zhang2024refactoring, ben2024assessing, wang2024hecs, gao2024preference, cui2024one, wang2025testing, midolo2025automated, menolli2024teaching, alturayeif2025refactoring, liu2024empirical, dong2025chatgpt, guo2024exploring, alomar2025chatgpt, caumartin2025exploring, pandini2025exploratory, guo2025intention, xu2025hlsrewriter, guenoune2025llm}: \textcolor{black}{Most commonly implemented with 27 studies (22\%) testing it, and} provides additional information on the codebase, such as usage and location of the code that should be fixed, along with quality attributes such as readability or maintainability refactoring goals to achieve
    \item Output Constraints \cite{white2023prompt, chavan2024analyzing, ishizue2024improved, gautam2024refactorbench, cui2024three, zhang2024refactoring, ben2024assessing, gao2024preference, wang2025testing, midolo2025automated, alturayeif2025refactoring, liu2024empirical, alomar2025chatgpt, xu2025hlsrewriter, de2025effectiveness}: \textcolor{black}{Implemented in 15 studies (12\%), and} gives specifications of what not to do or what to do in terms of formatting or correctness.
    \item Ranking \cite{pomian2024assist, pomian2024next}: \textcolor{black}{Implemented in 2 papers (2\%), and} asks the LLM to return back multiple versions of the refactored code, and to also rank the versions from best to worst.
\end{itemize}
By applying these prompting techniques, studies have assessed the effectiveness of improving refactoring outcomes in various code refactorings. The evaluations focused on factors such as accuracy, maintainability, and retention of the original functionalities, comparing different prompting strategies to determine which yields the most reliable results. Out of all of these techniques, it is evident that One-Shot prompting is the most effective for achieving correctness in Java code refactoring, improving test pass rates compared to Zero-Shot by 6.15\% and reducing code smells at a 3.52\% higher rate. One-Shot prompting is also shown to have yielded the highest unit test pass rate of 34.51\% and code smell reduction rate of 42.97\% outperforming compared to both Zero-Shot and Chain-of-Thought prompting \cite{cordeiro2024empirical}, and One-Shot prompting preforms better than Zero-Shot in built system refactoring receiving a Precision, Recall and F1-measure of  0.79, 0.78, and 0.76 compared to Zero-Shot's 0.67,
0.72, and 0.66 \cite{ghammam2025build}. But in certain cases, such as Python-based refactoring tasks, it is shown that Few-Shot prompting performed the best, improving refactoring quality by reducing cyclomatic complexity (CC) by 17.35\% and decreasing lines of code by 25.85\% while maintaining correctness \cite{shirafuji2023refactoring}. \textcolor{black}{Additionally, when tested in multiple programming languages, Few-Shot prompts outperformed both Chain-of-Thought and Zero-Shot prompts \cite{guo2025intention}.} Chain-of-Thought prompting is shown to improve diversity in refactoring, but in terms of correctness, it did not outperform One-Shot \cite{cordeiro2024empirical}. One study shows that context-enhanced prompting significantly reduces test smells, improving test code quality in automated test refactoring through the use of the UTRefactor framework. It was capable of reducing test smells from 2,375 to 265, achieving 89\% smell elimination, compared to 55\% for standard LLM-based approaches \cite{gao2024context}.
\input{Tables/Prompt}

\begin{boxK}
\textit{\textbf{Summary for RQ$_3$.} There are various ways of prompt engineering when interacting with LLMs for refactoring, these types are: Zero-Shot, One-Shot, Few-Shot, Chain-of-Thought, Context-Specific, Output Constraints, and Ranking. We found that for different use cases in different studies, certain prompts are better than others. Which prompting technique to employ for prompt engineering thus become a highly situational dependent decision.} 
\end{boxK}

\subsection{RQ$_4$: How accurate are LLM-generated refactoring suggestions?} %
\input{Tables/Analysis}

\textbf{Motivation.} LLMs offer the potential to automate refactoring, a time-consuming yet crucial task for maintaining software quality. Using LLMs to automate the refactoring process could significantly reduce development time and effort while enhancing the efficiency of countless software systems. Still, the accuracy of these suggestions is critical if they are to be integrated into existing systems, especially those managing sensitive information or impacting people's lives. The primary studies use different datasets, languages, and evaluation methods to measure the accuracy of their refactoring tools. Therefore, it is imperative to examine not just the accuracy of LLM-generated refactoring suggestions but also the processes used to assess this accuracy in order to make sufficiently informed decisions about their reliability, usability, and ultimate adoption within the software engineering industry, research, and beyond.

\textbf{Approach.} To further understand the accuracy of LLM-generated refactoring suggestions in the primary studies, we report in Table \ref{Table:Analysis_in_llms_Refatoring} the following types of data analysis:
\begin{itemize}
    \item \textit{Quantitative:} Indicates the dataset used to assess the accuracy of LLM-generated refactoring suggestions.
    \item \textit{Qualitative:}  Indicates the number and type of surveyed participants.
    \item \textit{Comparative:} Indicates the tools used for comparison.
    \item \textit{Correctness:} Indicates the results for measuring the accuracy of LLM-generated refactoring suggestions.
\end{itemize}
We report the Correctness metrics of each study as ‘Precision’, ‘Accuracy’, ‘Recall’, or ‘F1-measure’ when explicitly mentioned in the primary studies, and as ‘Unknown’ if no performance metrics are provided. We mark the Qualitative column as ‘No’ if no qualitative analysis was done. For the Comparative column, we document the names of tools used for comparison, marking ‘No’ if no comparative analysis was conducted, or ‘Unknown’ if the tools used for comparison were not named.

\textbf{Results.} This section discusses the quantitative, qualitative, comparative, and
correctness data analysis of LLM-based refactoring tools.\\
\textbf{Quantitative.} The datasets used for evaluating the accuracy of LLM-based refactoring vary greatly in composition and size between the primary studies. As mentioned earlier in RQ$_2$, if the number of projects used is not explicitly reported in a study, we report the other quantities of data
that are mentioned. \textcolor{black}{For example, DePalma \etal \cite{depalma2024exploring} specified 40 files, each containing 15-30 lines of code.} Besides full projects and files, data types include methods, commits, and even different prompt patterns, among others.  The sizes of these datasets go from being as small as 1 program \cite{gehring2023deterministic} or 116 methods \cite{ben2024assessing}, to as much as 11,336 projects \cite{cui2024one,wang2024hecs}. Notably, earlier studies tended to employ smaller datasets for evaluation. This is likely due to the increased capabilities of LLMs as of late, with models such as ChatGPT-4o being released on May 14, 2024. Continuous updates to these models include further token limit increases, allowing for greater dataset size. \textcolor{black}{More than a third (46\%) of the studies tested their tools on varying numbers of 'projects', implying the use of larger programs composed of multiple files, components, and classes, mainly from open-source repositories.} However, there is still vast variability in the datasets used for evaluating LLM-based refactoring approaches. Beyond the data types mentioned above, differences in programming languages and project sources can also affect the observed accuracy of refactoring suggestions. This lack of standardization makes comparisons between studies more difficult, emphasizing the need for unified datasets in future work.

\textbf{Qualitative.} The reviewed primary studies also exhibited variation in their qualitative methodology. There were \textcolor{black}{13} studies \textcolor{black}{(26\%)} that reported the use of human feedback in their evaluation \cite{depalma2024exploring, pomian2024next, pomian2024assist, cui2024three, cui2024one, wang2024hecs, menolli2024teaching, alturayeif2025refactoring, liu2024empirical, dong2025chatgpt, ghammam2025build, batole2025leveraging, pandini2025exploratory}, with the remaining 37 \textcolor{black}{(74\%)} not mentioning any participant involvement \cite{gehring2023deterministic, white2024chatgpt, shirafuji2023refactoring, alomar2024refactor, chavan2024analyzing, dilhara2024unprecedented, gao2024context, zhang2024refactoring, choi2024iterative, wu2024ismell, ishizue2024improved, gautam2024refactorbench, baumgartner2024ai, zhang2025move, ben2024assessing, gao2024preference, zhang2024copilot, wang2025testing, midolo2025automated, ksontini2025refactoring, icmi2025evaluating, guo2024exploring, guo2025intention, palit2025reinforcement, rajendran2025multi, kim2025comparative, wang2025exploration, wang2025towards, alomar2025chatgpt, midolo2025refactoring, pucho2025refactoring, caumartin2025exploring, xu2025hlsrewriter, tkachuk2025determining, guenoune2025llm, amaral2025improving, de2025effectiveness}. Those that did employ a survey to evaluate their refactoring tool or approach had different types of participants and sample sizes. Professional developers and engineers were surveyed in \textcolor{black}{eight} of the studies \textcolor{black}{(16\%)} \cite{pomian2024assist,pomian2024next,cui2024three,cui2024one,wang2024hecs,liu2024empirical, dong2025chatgpt, ghammam2025build}, while four \textcolor{black}{(8\%)} studies used only student participants \cite{depalma2024exploring,menolli2024teaching, batole2025leveraging, pandini2025exploratory}, and one used both professors and students \cite{alturayeif2025refactoring}. The number of participants throughout the studies ranged from 3-\textcolor{black}{60}, with the student sample sizes residing in the bottom half of the range-10, 11, 15, 23, and 30 students respectively. This variation suggests that the sample sizes may be due to the assumptions of the research teams, as well as the resources that were available to them. While the professional developers and engineers can offer knowledgeable and insightful responses due to their experience with real-world refactoring challenges, the use of student participants is also very valuable and necessary when it comes to evaluating tools or approaches designed to teach refactoring concepts. Ultimately, it is important that authors justify their choice of participants and sample size to better understand these nuanced answers and their context within the study’s goals and target applications. \textcolor{black}{While most studies did not employ user surveys, those that did gathered valuable insights into user experience beyond the authors' own observations, such as Cui \etal \cite{cui2024three}, where 68\% of surveyed software engineers preferred the authors' tool HMove, over other existing alternatives. This suggests this is a beneficial dimension for future work.} \\
\textbf{Comparative.} Out of 50 studies, 37 \textcolor{black}{(74\%)} studies directly compared LLM-generated refactorings against existing refactoring tools and baselines or manual refactoring \cite{gehring2023deterministic, pomian2024assist, pomian2024next, dilhara2024unprecedented, gao2024context, zhang2024refactoring, wu2024ismell, ishizue2024improved, gautam2024refactorbench, cui2024three, zhang2025move, ben2024assessing, wang2024hecs, gao2024preference, cui2024one, midolo2025automated, ksontini2025refactoring, alturayeif2025refactoring, liu2024empirical, dong2025chatgpt, ghammam2025build}. Specifically, Ghammam \etal \cite{ghammam2025build} compared the performance of their tool using different prompting techniques. Notably, the inclusion of comparison, as well as the range and quantity of the refactoring tools used for comparison, in the primary studies can be influenced by the financial and time resources available to the researchers. The remaining \textcolor{black}{13} studies \textcolor{black}{(26\%)} did not make comparisons \cite{white2024chatgpt, shirafuji2023refactoring, depalma2024exploring, alomar2024refactor, chavan2024analyzing, choi2024iterative, baumgartner2024ai, zhang2024copilot, menolli2024teaching, icmi2025evaluating, alomar2025chatgpt,tkachuk2025determining}, or did not specify the name of the baseline \cite{wang2025testing}. \textcolor{black}{A majority of the studies did perform some comparative analysis, yet the tools and baselines used varied significantly based on individual methodologies. This lack of a standardized benchmark makes it more difficult to get a full picture of LLMs’ relative performance in refactoring tasks.} For instance, some used existing refactoring tools and models \cite{pomian2024assist, pomian2024next, dilhara2024unprecedented, zhang2024refactoring, wu2024ismell, cui2024three, zhang2025move, wang2024hecs, cui2024one, midolo2025automated}, just other LLMs \cite{gehring2023deterministic,ishizue2024improved,gao2024preference,ben2024assessing}, or both \cite{gao2024context}, and \textcolor{black}{13.46\%} compare performance against manual refactoring \cite{gautam2024refactorbench,alturayeif2025refactoring,liu2024empirical, dong2025chatgpt, pucho2025refactoring, guenoune2025llm, ksontini2025refactoring}. One such study did a comparative analysis against an existing refactoring tool and manual refactoring \cite{ksontini2025refactoring}. \textcolor{black}{Wang \etal \cite{wang2025exploration} differentiated between pretrained models (T5-Review, CodeT5, CodeReviewer, and CodeBERT) and LLMs (CodeLLaMA, StableCode3B, LLaMA3, and ChatGPT 4o) in code review tasks revealing that pretrained models preformed better than LLMs due to LLMs preforming more code changes.} There were some tools that were used for comparison in multiple studies, such as JDeodorant, JMove, and PathMove, which could be a good start for establishing a standardized set of refactoring tools and baseline methods for comparison, allowing for a more consistent evaluation of LLM performance across future studies that the field would benefit from greatly.\\
\textbf{Correctness.} Each of the primary studies reported evaluation metrics for their respective tools, however, there was variation in the types of metrics that were reported. Similar to the comparative and qualitative analysis, there was a significant difference in the number of studies that reported performance metrics (e.g., Accuracy, Precision, Recall, F1-measure), \textcolor{black}{22} to be exact \cite{gehring2023deterministic, depalma2024exploring, dilhara2024unprecedented, zhang2024refactoring, wu2024ismell, baumgartner2024ai, cui2024three, zhang2025move, wang2024hecs, cui2024one, alturayeif2025refactoring, dong2025chatgpt, ghammam2025build, batole2025leveraging, kim2025comparative, caumartin2025exploring, guo2024exploring, xu2025hlsrewriter, tkachuk2025determining, guenoune2025llm, amaral2025improving,de2025effectiveness}, and those that reported quality metrics (e.g.,  cyclomatic complexity, lines of code, etc.) with 19 \cite{white2024chatgpt, shirafuji2023refactoring, alomar2024refactor, chavan2024analyzing, pomian2024assist, pomian2024next, gao2024context, choi2024iterative, ishizue2024improved, gautam2024refactorbench, ben2024assessing, gao2024preference, zhang2024copilot, wang2025testing, midolo2025automated, ksontini2025refactoring, menolli2024teaching, icmi2025evaluating, liu2024empirical}. \textcolor{black}{The Accuracy and F1-score metrics reported the highest percentages out of the performance metrics and almost always increased when compared to existing refactoring tools. While one study reported an Accuracy of 58.58\% \cite{amaral2025improving}, the vast majority of results were substantially higher with none under 78.61\% and 74\% respectively.} Reported Recall percentages ranged from 48\%-100\%, with a majority being 84.3\% or higher, indicating that the tools were decent at identifying valid refactoring opportunities. When compared to other automated approaches or manual refactoring, the Precision metric saw both increases \cite{zhang2024refactoring,cui2024three,zhang2025move,wang2024hecs,cui2024one} and decreases \cite{zhang2024refactoring,wu2024ismell}, suggesting that the quality of generated refactoring suggestions varied depending on the context of each study. \textcolor{black}{One interesting metric, BLEU-T, was reported by Caumartin \etal \cite{caumartin2025exploring} at 77\%. This measures the similarity of token sequences between the LLM generated code and ground truth code. This metric could be a useful metric to include in further studies, as it is suited to evaluating LLM output by the basic units, tokens, by which they interpret text.} Although the available performance metrics display promising results, it would be beneficial for future studies to include the more common Accuracy, Precision, Recall, and F1-measure in their evaluations so that the effectiveness of the latest refactoring tools can be compared using consistent and established benchmarks. \\
\textcolor{black}{Figure \ref{fig:metrics} displays the count of papers that report each of the four metrics (Quantitative, Qualitative, Comparative, Correctness) from Table \ref{Table:Analysis_in_llms_Refatoring} respectively. This view more clearly expresses the distributions of methodological focus, emphasizing which types of analysis are more prominent in studies concerning LLM-based refactoring approaches. As shown in the chart, quantitative analyses can be done with all 50 papers (100\%), followed by comparative evaluations at 74\%, correctness-focused assessments at 44\%, and qualitative analyses at 26\%. This overview provides a clear snapshot of current research practices and can help guide future studies toward underrepresented evaluation approaches for more balanced and comprehensive assessments of LLM-based refactoring tools.}

\begin{boxK}
\textit{\textbf{Summary for RQ$_4$.} \textcolor{black}{Reported performance is generally high, with Accuracy and F1-scores often exceeding 78.6\% and 74\%, respectively, showing improvement over existing tools. While the precision of suggestions is context-dependent, the use of non-standardized benchmarks and metrics, along with a gap in user-centered evaluation (74\% of studies did not integrate human feedback), makes it difficult to establish a generalizable conclusion about LLM accuracy. This indicates a need for a more unified evaluation framework that incorporates standardized benchmarks, consistent performance metrics, and diverse evaluation methods.}} 
\end{boxK}

\color{black}
\subsection{RQ$_5$: What refactoring characteristics are addressed in LLM-driven refactoring?}
\textbf{Motivation.} As LLMs become more popular in the field of refactoring, it is crucial to examine which refactoring methods LLMs excel in and which refactoring methods LLMs need to be studied more to gather data in terms of quantitative, qualitative, comparative, and correctness. Understanding this will help us best to recognize the gaps in the studies that pre-exists in the field of refactoring with LLMs. This research question is connected with RQ$_4$, as we dive deeper into the specific characteristics of refactoring we can examine them each within the quantitative, qualitative, comparative, and correctness frame.

\textbf{Approach.} We looked closely at each paper to examine the quantitative, qualitative, comparative, and correctness of each research study. Here is a breakdown of how we categorized each of the data points from the papers: \\
\textbf{Quantitative.} \textcolor{black}{This refers to primary studies that use numerical data to assess the performance of refactoring with LLMs.} For quantitative data points, if a paper has many numerical characteristic breakdowns of their studies aspects, such as the dataset size, type of data points, metrics such as Accuracy, Precision, Recall, F1, or performance time, we qualify the study as having quantitative data points. \\
\textbf{Qualitative.} \textcolor{black}{This refers to primary studies that explore developers' perceptions and motivations regarding refactoring and LLM-assisted tools, typically using non-numerical methods.} For qualitative data points, if a paper mentions or focuses on human-centric evaluations, or descriptive evaluations, including developer/student feedback, user studies, expert reviews/analysis, or assessments of maintainability, readability or usability, we qualify the study as having qualitative data points. \\
\textbf{Comparative.} \textcolor{black}{This refers to primary studies that systematically compare outcomes across two or more groups, typically examining LLM-assisted refactoring against human-only refactoring, traditional tools, or other LLM methods.} For comparative data, if a paper involves benchmarking or performance comparisons between pre-existing baseline, such as other LLMs, human developers, or traditional refactoring tools, we qualify the study as having comparative data. \\
\textcolor{black}{\textbf{Behavior Preservation.} This refers to primary studies that specifically measure the preservation of behavior and functional soundness of the generated refactoring output, ensuring that the transformation does not introduce bugs or alter the program's intended behavior. For functional validity, if a paper involves how accurately the refactoring preserves functionality, we qualify the study as having correctness data.} \\
In collecting these points we also looked into how each study performed code smell detection, what level of refactoring they performed on, the refactoring types they performed and how did they check for behavior preservation after performing in Table \ref{Table:refactoring_in_llms_Refatoring} along with Figure \ref{fig:mapping-refactoring}\footnote{The figure is generated using Manus.}. To further explain, “Refactoring Identification” refers to the code smells that were addressed by the authors. Within “Abstraction Layer,” we addressed whether the authors were talking about code level refactoring, design level refactoring or architecture level refactoring. For code level refactoring authors completed refactoring within a few lines of code inside a single file, for design level refactoring authors addressed code smells across different files, classes, and modules, and for architecture level refactoring, authors addressed refactoring on a higher level regarding folders, files, and the overall organization of the software. For "Behavior Preservation", we talked about how the authors checked for whether or not after refactoring, code stayed the same functionally, by following the categories listed in AlOmar \etal study \cite{alomar2021preserving}. Figure \ref{fig:mapping-refactoring}, shows a clear mapping of code smell, layer of refactoring, and what was the exact refactoring type applied with the studies; eliminating the lowest frequency data points such as Architectural Layer, Refused Bequest, Data Clumps, Long Parameter List, Lazy Class, Message Chain, Lingusitic smells, Test code Smells, Docker file code smells, Architectural code smells, Hardware code smells, and SQL code smell for a more clear and organized view.
\\
In addition, we also pointed out the exact methods of detection and refactoring to better understand what and how exactly each paper approaches refactoring in Table \ref{Table:Methods_of_Correction_and_Detection}.
Here are the code smells that we found, where we labeled labeled what each study found and what their definitions are in our own words, referencing Refactoring Guru\footnote{https://refactoring.guru/}, and also addressed special cases that do not exist on Refactoring Guru but rather defined by the papers themselves in Table \ref{Table:Code_Smells_Defintion}.
\input{Tables/Code_Smells}

\textbf{Results.} This section discusses the quantitative, qualitative, comparative, and correctness data analysis of LLM-based refactoring tools, grouped by refactoring type: Rename Method, Extracting, Simplifying Method, Code Formatting, Split File, Inline Method, JUnit Migration, Loop Optimization, and Move Method.\\
\textbf{Rename Method.} Rename Method refactoring involves changing the names of variables, functions, or classes. The studies examining this task \cite{gehring2023deterministic, shirafuji2023refactoring, depalma2024exploring, alomar2024refactor, ishizue2024improved, ben2024assessing, gautam2024refactorbench, liu2024empirical, dong2025chatgpt, ksontini2025refactoring, guo2024exploring, wang2025towards, pucho2025refactoring, caumartin2025exploring, tkachuk2025determining} showed variation in quantitative dataset sizes, ranging from very small datasets, such as a single Java program in Gehring \cite{gehring2023deterministic}, to larger collections of up to 880 Python programs in Shirafuji \etal \cite{shirafuji2023refactoring}, or 600 Dockerfiles in Ksontini \etal \cite{ksontini2025refactoring}. While only a few studies reported correctness metrics, those that did showed generally high performance, with 100\% Accuracy in Gehring \cite{gehring2023deterministic}, 96.8\% in Depalma \etal \cite{depalma2024exploring}, and 92.9\% Precision in Dong \etal \cite{dong2025chatgpt}, suggesting that LLMs can reliably perform Rename Method refactoring when prompted appropriately. Five studies performed qualitative evaluations involving human developers \cite{dong2025chatgpt, liu2024empirical, depalma2024exploring, guo2024exploring, wang2025towards, tkachuk2025determining, pucho2025refactoring, caumartin2025exploring}, giving feedback on generated refactoring outputs. Comparative analyses were conducted in multiple papers, either against existing refactoring tools, like ChatGPT, JDeodorant, PARFUM, or manual refactoring performed by developers \cite{pomian2024assist, gehring2023deterministic, zhang2024refactoring, ishizue2024improved, gautam2024refactorbench, wang2025towards, tkachuk2025determining, pucho2025refactoring}, often showing that LLM-based solutions can match or slightly outperform baseline approaches in identifying suitable renaming targets. Five studies \cite{liu2024empirical, ksontini2025refactoring, dong2025chatgpt, gautam2024refactorbench, pucho2025refactoring} specifically compared LLM-based tools with manual, human refactoring, which is relevant given that human comprehension and agreement are crucial to successful renaming. LLM-based solutions sometimes matched or outperformed human experts, but human judgment was ultimately used to validate output. Overall, the breadth of studies on Rename Method refactoring demonstrates that LLM-assisted approaches are effective for renaming tasks, and their output is generally consistent with what developers would consider correct and appropriate, under guided conditions. The absence of some metrics across the papers indicates the need for structured, systematic evaluations so that stronger patterns can be identified from consolidating the studies’ results. \\
\textbf{Extracting.} Extracting refactoring involves two related types: Extract Method, which moves code snippets to a new method and replaces the original code with a method call, and Extract Class, which creates a new class and places the fields and methods responsible for the relevant functionality in it \cite{gehring2023deterministic, shirafuji2023refactoring, alomar2024refactor, pomian2024assist, pomian2024next, zhang2024refactoring, choi2024iterative, wu2024ismell, ishizue2024improved, gautam2024refactorbench, gao2024preference, wang2025testing, menolli2024teaching, icmi2025evaluating, liu2024empirical, ksontini2025refactoring, dong2025chatgpt, guo2024exploring, wang2025towards, alomar2025chatgpt, pucho2025refactoring, caumartin2025exploring, guo2025intention, tkachuk2025determining, guenoune2025llm}. This task was the most widely studied among the refactoring types, with evaluations spanning a variety of datasets and approaches. Quantitative dataset sizes varied widely, from small sets of a single Java program in Gehring \cite{gehring2023deterministic} to large-scale analyses of tens of thousands of code samples \cite{gao2024preference}. Correctness metrics were reported in a subset of studies, with generally high values such as 100\% Accuracy in Gehring \cite{gehring2023deterministic}, 92.9\% Precision in Dong \etal \cite{dong2025chatgpt}, and median Recall of 48\% in Baumgartner \etal \cite{baumgartner2024ai}, suggesting that LLMs can reliably perform Extract Method refactoring when prompted appropriately. Several studies incorporated qualitative evaluations involving human developers \cite{liu2024empirical, gautam2024refactorbench, pomian2024assist, pomian2024next, wang2025towards, guenoune2025llm, guo2025intention, caumartin2025exploring, caumartin2025exploring, alomar2025chatgpt}, which provided insight into the practical usability, maintainability, and adherence to developer expectations of LLM-generated refactorings, although most studies did not include such feedback. Comparative analyses were conducted in multiple papers, either against existing refactoring tools such as ChatGPT, JDeodorant, RIdiom, SSECS, LLMRefactor, or manual refactoring performed by developers \cite{pomian2024assist, gehring2023deterministic, zhang2024refactoring, cui2024one, ishizue2024improved, wang2024hecs, gautam2024refactorbench, guo2024exploring, palit2025reinforcement, wang2025towards, guenoune2025llm, guo2025intention, tkachuk2025determining, pucho2025refactoring, caumartin2025exploring}, often showing that LLM-based solutions can match or outperform baseline approaches in identifying suitable code segments for extraction. Notably, the Extract Class studies highlight the added complexity of determining which fields and methods belong together in a new class, requiring models to reason about functional cohesion and class-level organization, a task that appears more challenging than method-level extraction but still benefits from LLM guidance. The inclusion of both method-level and class-level extraction studies indicates that LLMs are capable not only of identifying extractable code segments but also of reasoning about higher-level structural organization, suggesting potential for broader software design support. Overall, the breadth of studies on Extract Method and Extract Class refactoring demonstrates the great potential of LLM-assisted refactoring in real-world contexts, though the variability in reported metrics is likely a reflection of differences in dataset sizes, composition, and evaluation methods when it comes to applying LLMs to this task. \\
\textbf{Simplifying Method.} Simplifying Method refactoring involves making method calls simpler and easier to understand, such as modifying conditionals or adjusting parameters. Evaluations of this task \cite{wang2025testing, shirafuji2023refactoring, depalma2024exploring, alomar2024refactor, chavan2024analyzing, ben2024assessing, alomar2025chatgpt, pucho2025refactoring, xu2025hlsrewriter, tkachuk2025determining} were less extensive than for the Rename or Extract Method types. Quantitative dataset sizes varied widely, such as small sets of 40 files with student feedback in Depalma \etal \cite{depalma2024exploring}, 24 tasks in Xu \etal \cite{xu2025hlsrewriter} or large-scale analyses of hundreds of programs or conversations in Shirafuji \etal \cite{shirafuji2023refactoring}, Wang \etal \cite{wang2025testing}, and Midolo \etal \cite{midolo2025automated}. Notably, qualitative evaluation was more prominent in this set than in other refactoring types, with Depalma \etal \cite{depalma2024exploring} including feedback from 15 students and some studies \cite{chavan2024analyzing, ben2024assessing, tkachuk2025determining, pucho2025refactoring, alomar2025chatgpt} involving developers. The other metrics, correctness and comparative analysis, were pretty similar in range and results to what was observed in Rename Method and Extract Method refactoring, with most studies not reporting explicit correctness metrics and only one performing tool-to-tool comparisons. Overall, while the available studies suggest that LLMs can assist in simplifying methods, the limited number of evaluations and infrequent reporting of correctness metrics showcase a need for more large-scale studies to validate reliability and performance of LLMs when it comes to Simplifying Method refactoring. \\
\textbf{Code Formatting.} Code Formatting refactoring involves changes to code style, such as indentation or ordering functions. The small sample size of this task \cite{ishizue2024improved, depalma2024exploring, guo2024exploring} meant that any solid patterns in the results of the evaluation metrics were difficult to identify. In fact, the metrics that the two studies’ report are relatively inconsistent from each other. Depalma \etal \cite{depalma2024exploring} conducted a quantitative evaluation on 40 files and reported a correctness Accuracy of 96.8\%. Their study also included qualitative feedback from 15 students, but did not perform a comparison against other refactoring tools. In contrast, Ishizue \etal \cite{ishizue2024improved} evaluated 1,800 programs in a large-scale quantitative analysis, with results from their three phase approach compared directly to ChatGPT, though they did not report any correctness metric nor qualitative evaluation. While both studies demonstrate that LLMs can have the capability of outputting stylistically consistent formatting, the lack of reported correctness data in \cite{ishizue2024improved} along with the absence of tool-to-tool comparisons in \cite{depalma2024exploring} make it difficult to assess the reliability of these results. However, Guo \etal \cite{guo2024exploring} was able to provide qualitative and quantitative data on various setting changes of ChatGPT and in comparison with the other state-of-the-art tools that exist, along with correctness checking with specific refactoring types, offering a conclusive usage of ChatGPT in refactoring. Similar to Split File refactoring, the scarcity of papers specifically discussing Code Formatting refactoring highlights the need for more widespread evaluations before coming to further conclusions about the effectiveness of LLM-based approaches for this task. \\
\textbf{Split File.} Split File refactoring involves separating functions, methods, or classes into different files for modularity. Only one study examined this type \cite{chavan2024analyzing}, mentioning that a few of the 54 refactoring-related conversations that developers had with ChatGPT had the goal of splitting the code to enhance readability and maintainability. The focus of this paper was to understand the nature of Developer-ChatGPT conversations and the model’s role in the evolution of software development. No qualitative developer feedback, or correctness metric related to this specific refactoring type was reported, and comparative analysis with other refactoring tools was absent. Due to this lack of focus on Split File refactoring, further comparative analysis of LLM-based refactoring tools for the Split File task is necessary for a more concrete understanding of their true effectiveness in supporting modularity and larger-scale software refactoring. \\
\textbf{Inline Method.} Inline Method refactoring involves replacing method calls with the method body itself, removing the need for separate calls. Multiple studies have examined this task as part of their study in passing \cite{wang2025testing, liu2024empirical, gautam2024refactorbench, guo2024exploring, wang2025towards, pucho2025refactoring, caumartin2025exploring, guo2025intention, tkachuk2025determining}. Gautam \etal \cite{gautam2024refactorbench} performed a quantitative evaluation with a dataset of nine Python projects, comparing LLM-produced refactorings with human developer implementations, but did not report any correctness metrics or qualitative developer feedback. Wang \etal \cite{wang2025testing} utilized 458 bug reports in their quantitative analysis, though no comparisons to other refactoring approaches or correctness measures were included. In contrast, Liu \etal \cite{liu2024empirical} evaluated 20 Java projects, integrating qualitative feedback from three human developers and comparing the results of prompting different LLMs with manual refactoring by human developers. However, this study did not report explicit correctness metrics. Multiple studies \cite{guo2024exploring, wang2025towards, guo2025intention, tkachuk2025determining, pucho2025refactoring, caumartin2025exploring}, have the authors reported quantitative, qualitative, correctness, and comparative data point on various tools refactoring. Overall, while these works collectively suggest that Inline Method refactoring with LLMs is being investigated in real-world contexts, the absence of correctness measures in some studies, makes it difficult to evaluate the reliability and effectiveness of this refactoring type.  \\
\textbf{JUnit Migration.} JUnit Migration refactoring involves converting existing test code to use the JUnit testing framework. Only one study examined this task \cite{gehring2023deterministic}, evaluating a single program and comparing the results of recipe-driven refactorings with outputs from ChatGPT-3.5 and StarChat--$\beta$. The study reported an Accuracy of 100\%; however, this statistic was calculated for refactorings related to incorrect switch statements (such as Java switch-case structures), not specifically for JUnit Migration. Additionally, no qualitative feedback from developers was collected. Similar to Split File refactoring, the limited empirical evidence on JUnit Migration, means that no solid patterns or conclusions can be drawn regarding the effectiveness of LLMs for this task. Thus, more extensive investigations and analyses are required before claims about the reliability and utility of LLM-assisted JUnit Migration can be made. \\
\textbf{Move Method.} Move Method refactoring involves creating a new method in the class that uses the method the most, and then moving code from the old method to the new location. The code of the original method is then replaced with a reference to the new method or removed entirely. Evaluations of this task \cite{alomar2024refactor, zhang2025move, cui2024three, wu2024ismell, menolli2024teaching, dong2025chatgpt, guo2024exploring, batole2025leveraging, wang2025towards, alomar2025chatgpt, caumartin2025exploring, guo2025intention, guenoune2025llm} spanned a wide range of datasets, from smaller sets of 4 Java projects in Menolli \etal \cite{menolli2024teaching} to large-scale analyses of over 11,000 Java projects in Cui \etal \cite{cui2024three} Metrics of correctness were reported in a subset of studies, with generally high performance where available, including Precision at 92.9\% in Dong \etal \cite{dong2025chatgpt}, and average Precision and Recall of up to 93.3\% and 85.4\%, respectively, in Cui \etal \cite{cui2024three}. However, in both studies, these correctness metrics were reported across the total set of refactoring cases involving multiple refactoring types, rather than being specified exclusively for the Move Method refactoring type. Several papers performed qualitative evaluations with human developers or students \cite{cui2024three, menolli2024teaching, dong2025chatgpt, guenoune2025llm, guo2025intention, caumartin2025exploring, alomar2025chatgpt}, with 68\% of those surveyed in Cui \etal \cite{cui2024three}. Finding the proposed LLM-driven tool, HMove, to be the most useful compared to two other refactoring tools. These qualitative evaluations suggest that involving human developers or students helps gauge not only the correctness but also the readability, maintainability, and practical relevance of LLM-generated Move Method refactorings. This feedback is valuable, as it indicates that LLMs could generate refactorings that are both technically accurate and aligned with developer expectations and coding standards. Comparative analyses were conducted in multiple papers, mainly against existing refactoring tools such as JDeodorant, RMove, PathMove, or manual refactoring \cite{zhang2025move, cui2024three, wu2024ismell, dong2025chatgpt, batole2025leveraging, wang2025towards, guenoune2025llm, guo2025intention, caumartin2025exploring}, demonstrating that LLM-based approaches can match or sometimes exceed the effectiveness of traditional tools in identifying suitable methods to move. Overall, while the studies collectively indicate that LLMs show strong potential for Move Method refactoring, demonstrated by promising Precision and Recall metrics as well as supportive feedback from human evaluators, the variability in reported metrics and the limited number of qualitative assessments underscore the need for more systematic and comprehensive analyses to fully understand LLM-assisted performance in this task. \\
\textbf{Loop Optimization.} Loop Optimization refactoring involves improving the efficiency or structure of loops in code, such as minimizing iterations, unrolling loops, or replacing loops with more efficient constructs. There were limited assessments of this task, with only two studies \cite{depalma2024exploring, alomar2024refactor} examining LLM-assisted approaches for Loop Optimization. Depalma \etal \cite{depalma2024exploring} conducted a quantitative evaluation on 40 files and included qualitative feedback from 15 college students (studying computer science or related field), and reported an Accuracy of 96.8\% when it came to preserving the behavior of original code segments. In contrast, AlOmar \etal \cite{alomar2024refactor} analyzed 470 commits, 69 issues, and 176 files in a large-scale quantitative study, but no explicit correctness metrics were reported. Neither study conducted comparative analyses against other refactoring tools, limiting our ability to assess the relative performance of LLM-based Loop Optimization approaches. Overall, while these studies suggest that LLMs may have potential for automating Loop Optimization, the small number of these studies suggests that before stronger conclusions about LLM effectiveness in this refactoring type can be drawn, more thorough and structured investigations are needed.
\input{Tables/Refactoring}
\input{Tables/Methods}

\begin{boxK}
\textit{\textbf{Summary for RQ$_5$.}  By addressing refactoring characteristics in LLM-driven refactoring in the frame of the various data points, we found that out of all of the code smells, Large Class and Long Method is the most often spoken code smells and that Extract Method is the most often seen Refactoring type applied. Extracting refactoring method is the most widely seen and studied refactoring method with high Accuracy rates reported when using LLMs to refactor showing the most promising results.} 
\end{boxK}

\color{black}

\subsection{RQ$_6$: What are the key challenges in automating refactoring using LLMs?} 

\textbf{Motivation.} As the use of LLMs in software engineering expands, their potential to assist in automating refactoring has become a point of focus in research. Refactoring is a necessary process for maintenance and improving code quality, but it requires a deep understanding of context, intent, and overall codebase structure. Researchers have begun evaluating the effectiveness of LLM-based refactoring tools and approaches by measuring key metrics such as lines of code (LOC), cyclomatic complexity (CC), and Precision and Recall rates among others.
While LLMs show promise in generating refactoring suggestions, their reliability and effectiveness are hindered by several common key challenges. 
Understanding these challenges is essential to determine what aspects of LLM-based refactoring can be improved to advance the automation of refactoring within software engineering. 

\textbf{Approach.} To examine this, we consolidated the results of all primary studies to identify the most prevalent challenges that researchers encountered when leveraging LLMs for refactoring tasks. \textcolor{black}{To prevent bias, two authors independently analyzed the primary studies to identify and extract challenges. Each researcher performed an initial, independent categorization of these challenges. The authors then convened to reconcile their categorizations through a structured discussion, resolving differences by consensus to establish a final, agreed-upon taxonomy. For example, the author initially disagreed on whether to include Pandini \etal \cite{pandini2025exploratory} in the category concerning LLMs struggling with complex refactoring tasks, but after discussion, they unanimously agreed to include it.}

\textbf{Results.} This section discusses the key challenges encountered when automating refactoring using LLMs. 



\noindent\textbf{LLMs generate erroneous refactored code and are unreliable.} 
    Despite their impressive natural language processing abilities, LLMs can often produce refactored code with errors, making them unreliable for automated refactoring if not properly validated. This behavior is displayed in many of the primary studies, such as  \cite{pomian2024assist}, which explored the effectiveness of an LLM-based Extract Method refactoring tool and found that 76.3\% of LLM suggestions were hallucinations–57.4\% were syntactically incorrect while 18.9\% were illogical (e.g.,  suggesting to extract an entire method body). Similarly, when tasked with creating Python files, GitHub Copilot produced code smells in about 15\% of its generated code \cite{zhang2024copilot}. \textcolor{black}{More concerning, one study evaluated ChatGPT's ability to refactor for-loops, finding that only 28.8\% of its refactored outputs were compilable} \textcolor{black}{noting ChatGPT’s difficulty handling complex control flows and implicit dependencies }\cite{midolo2025refactoring}. These findings call into question its usefulness for refactoring production-level code.
    
    Additionally, ChatGPT was shown to struggle with context awareness, leading to it misinterpreting the developer's original intent and, in some cases producing bugs \cite{alomar2024refactor}. For instance, DePalma \etal\cite{depalma2024exploring} found that ChatGPT frequently struggles to identify quality attributes like coupling and complexity and is inconsistent with performance as its effectiveness varies depending on the complexity of the refactoring task. Furthermore, Liu \etal \cite{liu2024empirical} observed that Gemini and ChatGPT produced buggy refactored code 5\% and 7.2\% of the time, respectively, when tasked with refactoring Java files. Although the refactoring tool iSMELL \cite{wu2024ismell} did increase the detection of code smells, LLMs sometimes fail to detect complex code smells such as feature envy, even though they excel in handling simple refactoring tasks. 
    \textcolor{black}{This inconsistency is reflected in De Sousa \etal \cite{de2025effectiveness}, where F1-scores for detecting unused variable or parameter smells dropped as low as 33\% (GPT-4o mini), perhaps due to ambiguity in the model's understanding of the code smell, and in Batole \etal \cite{batole2025leveraging}, where up to 80\% of LLM recommendations for Move Method refactoring were hallucinations \textcolor{black}{exhibiting significant difficulty in suggesting appropriate target classes} and demonstrating serious unreliability. \textcolor{black}{Kim \etal \cite{kim2025comparative} reveals that LLMs excel at implementing simple design patterns like Strategy but results are less consistent with more complex patterns like Factory Method and Observer.} More concerningly, when performing Code Refinement Before Review (CRB) tasks, which involve evaluating and refining code changes prior to manual review, LLMs exhibited an extremely low Real Change Accuracy (RCA), averaging only 17.5\% and 12.5\% across datasets, indicating that most modifications introduced were unnecessary \cite{wang2025exploration}.} Broadly, while LLMs can be effective tools for refactoring, their ability to produce reliable code is limited by their tendency to hallucinate, causing unintended modifications or errors. 
    
     \noindent\textbf{LLMs struggle with complex refactoring tasks involving multiple files or larger codebases.} While LLMs can effectively refactor code at a smaller scale, they struggle with more complex refactoring tasks that involve multiple files or large codebases, leading to incomplete or incorrect modifications if not carefully guided. This is evident in several of the primary studies, which found that LLM performance varied significantly based on task complexity \cite{shirafuji2023refactoring}. Similarly, it was observed that LLMs had difficulty handling larger codebases, failing to apply changes when working with multiple files \cite{dilhara2024unprecedented}. \textcolor{black}{In Batole \etal \cite{batole2025leveraging}, it is highlighted that Move Method refactoring is particularly difficult to implement because it requires project-level reasoning, but currently LLM’s prompt and input size limitation hinder global project reasoning. Furthermore, increasing the input size and passing the complete project as input distracts the LLM by introducing less relevant information requiring additional pre-processing in the form of semantic-based analysis and Retrieval Augmented Generation (RAG) before applying LLMs for Move Method refactoring eliminating hallucinations and increasing precision.}
     
     Structured prompting and Chain-of-Thought approaches have been shown to improve LLM performance compared to general approaches, indicating the need for continuous context reminders for more complex multi-file refactoring. For example, UTRefactor, which employs a structured, Chain-of-Thought prompt technique, achieved a significantly higher test smell elimination rate (89-91\%) compared to general LLM approaches (55-57\%) \cite{gao2024context}. White \etal \cite{white2024chatgpt} emphasized that while prompt techniques such as Few-Shot can aid LLMs' ability to operate on code, they can still produce inaccurate code examples if they encounter scenarios that were not represented in the given examples, which is likely when working with larger codebases. Choi \etal \cite{choi2024iterative} demonstrated that using LLMs for iterative refactoring helped decrease the overall cyclomatic complexity of the tested projects. However, 45.5\% of the generated refactored methods were initially non-plausible, highlighting the limitations of LLMs in producing reliable refactoring suggestions without additional validation. 
     
     Similarly, Gehring \cite{gehring2023deterministic} found that both StarChat-$\beta$ and ChatGPT produced incorrect refactoring instances when evaluated against the OpenRewrite tool. Additionally, these models struggled to output complete refactored code due to token limits, indicating that they are not ideal for refactoring tasks involving extremely large codebases.
 In addition, the effectiveness of LLMs in refactoring greatly depends on the prompting technique that is used. In one study, a One-Shot prompt proved to be the most effective approach for refactoring Java code, achieving a 34.51\% unit test pass rate and a 42.97\% code smell reduction rate, outperforming the 32.22\% pass and 42.34\% reduction rates for Chain-of-Thought and the 28.36\% pass and 39.45\% reduction rates of Zero-Shot \cite{cordeiro2024empirical}. Ultimately, the current inability of LLMs to grasp and retain the context of larger, multi-file codebases limits their effectiveness in refactoring. However, applying prompt techniques such as One-Shot and Chain-of-Thought can help reduce these limitations, though their reliability must be carefully reviewed. \\
  \noindent \textbf{LLMs misunderstand developers’ true intent and fail to detect qualified refactoring instances.} Another key challenge in LLM-driven refactoring is their tendency to misunderstand a developer’s true intent, leading to incorrect identification of refactoring opportunities or failure to detect qualified instances. As previously mentioned, one study found that LLMs often lack the necessary context awareness to make informed refactoring decisions \cite{alomar2024refactor}. \textcolor{black}{Their misunderstanding is further evidenced by AlOmar \etal\cite{alomar2025chatgpt}, where unstructured developer prompts required an average of 13.58 conversational turns to reach a satisfactory result, indicating a lack of understanding of the developer’s intentions in initial prompts. This combined with their limited refactoring abilities when working with larger codebases means that they struggle with grasping complex instructions, such as identifying possible refactoring instances. For example, ChatGPT was found to frequently fail in detecting code smells and therefore missed qualified refactoring opportunities, rendering it an unreliable educational tool on its own \cite{menolli2024teaching}. Similarly, Tkachuk \etal \cite{tkachuk2025determining} observed that general-purpose models struggled to identify problematic code, with both Gemini 2.0 Flash and Claude 3.7 Sonnet achieving only a 50\% success rate in detecting “bad” code or code smells.} Additionally, LLMs were observed to focus on individual methods rather than considering the broader project structure, \cite{pomian2024next}. This narrowed-down focus can restrict their ability to identify refactoring opportunities across multiple files. Cui \etal \cite{cui2024three} highlighted the importance of using short, concise prompts, timely feedback, and Few-Shot prompting to improve the accuracy of LLMs in detecting refactoring candidates, showing that LLMs still require very careful guidance to avoid misinterpretation.
Zhang \etal \cite{zhang2024refactoring} found that when their approach failed, it was due to LLMs either producing poor results when refactoring was too complex, or they failed to detect valid refactoring opportunities because they misinterpreted the provided code. 

Similarly, Alturayeif and Hassine \cite{alturayeif2025refactoring} also noted that in a few cases, LLMs misunderstood their task of refactoring linguistic elements such as complex sentences, opting to explain a concept that was mentioned in the prompt instead. Chavan \etal \cite{chavan2024analyzing} analyzed conversations between developers and ChatGPT and found that the model often makes its own assumptions about the output format and the quality attributes to enhance, even when not explicitly instructed to do so, which might not match the developer’s original intent. 
Overall, these findings indicate that developers should make use of very structured and detailed prompts when interacting with LLMs, especially when it comes to recognizing refactoring opportunities.

\textcolor{black}{\textbf{Observations.} The analysis reveals the percentages of challenges across
the primary studies. The most prevalent issue, observed in 25 out of 50 papers
(50\%), was LLMs generating erroneous or unreliable code, particularly in studies focused on transformations of many code components. This unreliable output often manifests as hallucinations, where suggestions are syntactically incorrect or illogical. For example, Midolo et al. \cite{midolo2025refactoring} used ChatGPT to refactor Java for-loops; however, only 29\% of the automatically refactored loops compiled successfully, while the remaining 71\% failed due to contextual or syntactic issues. Struggles with complex refactoring tasks involving multiple files were reported in 22 out of 50 studies (44\%), a challenge that was especially prevalent for studies where refactoring required more global project context. This is evident in the refactoring of Cyclic Dependencies (CDs), where performance declined as complexity increased (e.g., CDs of size 3 or 4) \cite{pandini2025exploratory}. Meanwhile, LLMs misunderstanding developers’ true intent and failing to detect qualified refactoring instances impacting 17 out of the 50 papers (34\%), a shortcoming most common in evaluations where refinement tasks came with vague instructions. In one study on automated code refinement, the main root causes for ChatGPT's underperformance were identified as a lack of domain knowledge, unclear location, and ambiguous changes in the review comments \cite{guo2024exploring}. Overall, the prevalence of these challenges
highlights that the reliability of LLMs for refactoring is highly dependent on the
complexity of the task and the specificity of the guidance provided.}

\begin{boxK}
\textit{\textbf{Summary for RQ$_6$.}  LLMs sometimes produce erroneous code, struggle with complex refactoring tasks and larger codebases, and misunderstand developers’ intent. To mitigate these limitations, developers must be very clear and intentional when crafting their prompts, check refactored code for errors specifically for more complex refactoring tasks, and implement techniques to decrease the size of codebases provided to LLMs.} 
\end{boxK}

\section{Discussion and Open Issues}
\label{Section:Discussion}
\noindent{\textcolor{black}{\textbf{ Lack of Standardization and Definitions on Refactoring Method Types.}}} 
 \textcolor{black}{One of the key challenges in the field of refactoring research, we noticed in RQ$_1$, is the lack of standardized terminology and definitions of refactoring methods. Across different studies and refactoring tools, the same refactoring techniques may be 
 referenced under different names, described without being identified, or grouped in different ways based on individual interpretations. For example, instead of describing code changes by ``reduce the software complexity" \cite{shirafuji2023refactoring} researchers should have directly mentioned the usage of ``Simplifying Method" or ``Extract Method", or instead of mentioning the bad smell of ``god class," or ``god class refactoring" \cite{wu2024ismell} researchers should have mentioned ``Extract Class" to address god classes. This inconsistency creates difficulties in comparing and reviewing research findings, as well as evaluating the impact of specific refactoring techniques on software quality. This is especially true with the usage of LLM-driven refactoring, as LLMs trained on different datasets may struggle to correctly interpret refactoring requests due to inconsistency in naming conventions across sources. LLMs need standardized and more concrete definitions for specifications and efficiencies. \\
}
    
\noindent\textbf{  What are the Best Method Combinations for Prompt Techniques.} From our observations in RQ$_3$, we know that the current existing prompt techniques are Zero-Shot, One-Shot, Few-Shot, Chain-of-Thought, Context-Specific, Output Constraints, and Ranking. We know about the efficiency of each of these prompting techniques, and we know the efficiency of their combinations, such as Few-Shot and Context-Specific \cite{pomian2024next}, Few-Shot and Chain-of-Thought \cite{pomian2024assist}, or Chain-of-Thought and Output Constraints \cite{white2024chatgpt}. So, the question lies in what is the best combination of prompting techniques to deploy to ensure the best output? We know from studies that Zero-Shot has a unit test pass rate of 28. 36\%, One-Shot has 34. 51\%, and Chain-of-Thought has 32. 22\% \cite{cordeiro2024empirical}. In another study, Shirafuji \etal discussed Few-Shots, showing that between Zero-Shot, One-Shot, and Few-Shots, Few-Shot performs the best \cite{shirafuji2023refactoring}, and in a study about built system refactoring, Ghammam \etal found that One-Shot outperformed Zero-Shot \cite{ghammam2025build}. This calls for extensive research to be done to more effectively decide which prompting techniques in combination will yield the best results. In these separate studies, due to the differences in datasets (how they are collected and the quality of the files) and how success is determined (differences in unit tests in testing for quality or programmers determining what is \textcolor{black}{``clean''} code) it is challenging to make a complete comparative empirical analysis to determine the best combination. The most effective way to successfully come to a conclusion on this would require expensive research done with a large dataset and the same tests and checks, but different technique combinations with an iterative study.  

\noindent{\textbf{ Future Considerations for Approaches for LLM-driven Refactoring.}} It has been shown in RQ$_1$ that when interacting with an LLM for refactoring there are various number of LLMs to choose from. Then, it also matters which LLM is used, for example, GPT-4 Turbo, GPT-3.5 Turbo, and Gemini, which have all shown different results in code refactoring in terms of efficiency and correctness \cite{gautam2024refactorbench}. This could mean various things, such as developers having different needs for each LLM. Are certain LLMs better at specific refactoring types than the other refactoring types? These topics have not been sufficiently explored yet. 

\noindent\textbf{ Effects of Developing LLMs on Refactoring Usage.} Another key point to consider when developing an efficient way to work with LLMs is the LLM's own development. In RQ$_6$, we asked about the key challenges in using LLMs, we can also consider it possible to be solved or worsened by developing LLMs. As an example, just between two versions of ChatGPT, such a ChatGPT-4 and ChatGPT-3.5, a decline in performance in various topics, such as math, answering sensitive questions, and prompt instabilities, has been observed. Within our topic, it has been found that from 3.5 to 4, ChatGPT has gotten worse at adhering to formatting instructions \cite{chen2024chatgpt}. 
\color{black} This is a critical problem with using LLMs for refactoring, as failure to adhere to formatting can lead to compilation issues. 

\noindent{\textbf{  Possible Language and Dataset Biases in LLMs. }} As LLMs become more internationally well-spread, different LLMs will be trained on different datasets, especially datasets that contain other languages. As we know, LLMs such as ChatGPT are capable of conversing in other languages, but it has been reported that replies fluctuate in quality across different languages. ChatGPT is known to hold certain biases in conversation, even in English, biases within gender, politics, and race\footnote{\url{https://hdsr.mitpress.mit.edu/pub/qh3dbdm9/release/2}} \color{black} are present. While those social problems held by ChatGPT are out of the scope of our paper, it is important to note if LLMs might also have biases in writing code and refactoring code due to their training sets. In RQ$_2$, we examined the datasets and benchmarks used by different research papers, and one of the important analyses that have been brought forth by Shirafuji \etal \cite{shirafuji2023refactoring} is the fact that code comments that are not written in English have noticeably been reduced, while English comments have been increased (95.11\% to 97.16\%). This change in proportion is due to the incorrect deletion of comments and translation of comments, both of which could lead to unnecessary and incorrect code refactoring. 

\noindent{\textbf{  Further Studies on Less Well-Known Techniques \color{black}  and Possible Tools. \color{black}}} While there are existing prompting techniques that are promising, it is important to explore further for new and more efficient techniques in an ever-growing and evolving LLM field. One prompting technique that we have noticed during RQ$_3$ that is less well-known and should be explored more is the Ranking technique. Where the developer would ask the LLM to provide multiple different options, or refactoring code along with a ranking of the code refactoring done \cite{pomian2024assist,pomian2024next}. \color{black} In addition to techniques, one other tool that can be seen as useful in the world of refactoring is the application of agentic AI. As agentic tools being further refined in recent years along with the creation of memory tools and other tools made available via Model Context Protocols such as search functions, access to cloud tools, access to databases, and more, refactoring can be taken to the next step, where LLMs can be made to being capable of self-fixing, compiling, checking, and running code. Enabling multi-step thinking and processing with tool access can help the LLM from making less mistakes and avoid hallucinations. Studies have been done on possibilities of conducting multi-agent systems, where there can be multiple LLM based agents can be communicating, checking and comparing work with each other based off of each of their own responses for software engineering related work. Including taking on roles in the software development life-cycle such as requirements engineering, code generation, Quality Assurance (QA) and maintenance. Agents can simulate different types of software models as well, such as waterfall, and agile methods \cite{he2025llm}. We hope to see that this can be studied further and have more instances to be referenced upon especially in the field of refactoring. \color{black}

\noindent{\textbf{  Evaluate Refactoring Tools on More Programming Languages.}} As identified in RQ$_2$, there is a broad diversity of LLM-based refactoring tools that have been developed, yet the majority of them have only been evaluated with datasets composed primarily of either Java or Python, with limited representation of other languages. While these are both widely used in industry and software engineering research, this dominance is disproportionate to the popularity and use within industry, open-source projects, and personal development of languages such as JavaScript, Kotlin, and C++, among others. This imbalance raises questions about the general effectiveness of LLM-based refactoring tools, since evaluation has been largely one-dimensional. Future studies should aim to widen the scope of evaluation of these tools to enhance their effectiveness as well as the amount of developers they could benefit.

\noindent{\textbf{  Lack of Standardized Performance Metrics for LLM-Based Refactoring.}} As shown in RQ$_4$, there are a few studies that evaluate LLMs based on established performance metrics such as Precision, Recall, and F1-score \cite{wu2024ismell, cui2024three,cui2024one, ghammam2025build}, there is a noticeable lack of standardization in how the effectiveness of these tools is measured across primary studies. The evaluation metrics used in the primary studies can vary a lot, making it hard to compare different LLM-based refactoring approaches. Some use metrics such as cyclomatic complexity (CC), lines of code (LOC) to measure effectiveness \cite{choi2024iterative,shirafuji2023refactoring}, while others rely on Accuracy rates that have very specific measurement criteria \cite{gehring2023deterministic}. Further, some studies use manual inspection to check the quality of the refactored code \cite{pomian2024next,depalma2024exploring}. \textcolor{black}{This inconsistency shows the need for more standardized evaluation metrics to
allow more robust comparisons in LLM-based refactoring research.}

\noindent{\textbf{  Maintainability of LLM-Refactored Code. }} As evidenced by RQ$_6$, integrating LLM-based refactoring tools into existing development workflows presents challenges due to their struggle with large codebases and occasional generation of erroneous code. Additionally, their inclination to misinterpret developer intent affects not only the execution but also the maintainability of the generated code. Liu \etal  \cite{liu2024refining} found that nearly 47\% of ChatGPT-generated code snippets suffer from code style and maintainability issues--such as having variables that should be private declared as public or using implementation types instead of interface--despite a majority compiling and running without errors. This highlights the importance of developers checking the style and maintainability of LLM-generated code to ensure it meets their standards before integrating it into an existing codebase, avoiding issues down the line. Future studies should evaluate the ability of LLMs to consistently comply with stylistic choices provided in their prompts.

\noindent{\textbf{  Integrating Human Feedback in LLM-Based Refactoring.}} While the refactoring tools introduced by the primary studies show the promising abilities of LLMs when it comes to refactoring, there is still concern about their reliability. As identified in RQ$_6$, LLMs have been observed to produce incorrect code suggestions  \cite{gehring2023deterministic} and hallucinations \cite{pomian2024assist}, there is a general need for manual verification of their output. \textcolor{black}{To address this, adding human experts who provide feedback to create a semi-automated refactoring process can increase the reliability and consequently the effectiveness of these LLM-based tools.}

\noindent{\textbf{  Security Concerns in Using LLM-Based Refactoring Tools in Industry.}} With RQ$_4$ showing promising results with LLMs achieving strong Accuracy in refactoring tasks (with Precision scores above 74\% and F1-measures reaching 87.8\%), it makes sense that companies would prioritize their integration. However, more importantly, the black-box nature and proprietary risks of LLMs hold back further integration into industry. Models such as ChatGPT are closed-source, with unclear internal logic \cite{bommasani2021opportunities}, therefore, it is not possible for outside businesses and users to understand how exactly refactorings are generated. Moreover, ChatGPT’s input is stored and used for training, which is a significant security risk as it may contain private business information \cite{bommasani2021opportunities, kibriya2024privacy}. This is especially concerning for companies whose systems handle sensitive personal user data such as health or financial information. \textcolor{black}{As a result, future work must focus on addressing these security risks, or aiding outside entities to develop their own LLMs to keep their data protected and in-house. This could involve leveraging open-source LLMs such as CodeGen or StarCoder, which can be deployed on local, private infrastructure, eliminating data exposure risks. Researchers can also explore on-premise fine-tuning to allow models to be trained on sensitive company data without sharing it externally. Prioritizing these approaches moving forward is the most effective way for organizations to mitigate significant safety concerns when integrating LLM-based refactoring tools.}

\textcolor{black}{\noindent{\textbf{ Lack of Data Leakage Prevention.}} As observed in our results from RQ$_2$, a majority of the studies concerning LLM-based refactoring, 59.61\% of the primary papers, make no mention of implementing any strategies to protect against data leakage of the models they are evaluating. However, this absence of authors outlining any direct techniques for combating this data leakage may be related to the novelty of research in this field, but also the nature of these language models. In the primary papers, when refactoring tools integrate the use of LLMs, they are simply using a pre-trained model such as ChatGPT, sometimes with fine-tuning or more minimal settings manipulation (e.g.,temperature). Due to this, researchers may assume that data leakage precautions were already taken by the model provider, which is partially true, yet, depending on the evaluation data in a study, data leakage can still occur. In addition, for some LLMs, information about the full extent of training data can be unavailable due to copyright issues or just a lack of transparency \cite{buick2025copyright}, and can therefore make it difficult to evaluate the model on completely new data. Sallou \etal \cite{sallou2024breaking} also highlight that the model may have been previously exposed to benchmark code through pre-training, potentially inflating reported performance claims. This suggests that implicit data leakage is a significant threat to validity in LLM-based software engineering research. It does remain that a large absence of consistent reporting on data leakage prevention can hinder the reproducibility of these studies, and therefore diminishes confidence in the performance of such models. }

\color{black}

\subsection{Actionable Insights}

\noindent{

\textbf{Include Comparative Analysis Between Various LLMs Within Studies.}
Adding to the discussion regarding RQ$_1$, we believe it would be more helpful for future works to consider comparing multiple LLMs in their studies and pursuits to not only find what is the best LLM for their specific case of usage but also to help find which LLMs excel in what field and why specific LLMs are better than others for that task. This can help LLMs grow in the field of refactoring.

\textbf{Further Studies on Less Well-Known Techniques and Possible Tools.}
For future development, for the field of code improvement with LLMs, it can benefit from more studies into using agentic AIs to further secure and better refactoring. There can be studies done on what part of software development are agents best at and how can we improve agents in other roles to improve refactoring and development, or what structure of development does agents excel in. 

\textbf{Standardize Performance Metrics for LLM-Based Refactoring.} Extending the insights from RQ$_4$, although some studies evaluate LLMs using established performance metrics like Precision, Recall, and F1-score, there remains a lack of standardization across primary studies, with many relying on varied or manual measures that make direct comparison more complicated. Future work could extensively adopt standardized reporting and evaluation statistics, including Accuracy, Precision, Recall, and F1-score, to systematically assess LLM outputs across multiple projects. Furthermore, establishing go-to community repositories of standardized LLM refactoring test cases would allow consistent, reproducible benchmarking across studies.

\textbf{Establish Standard Naming Conventions for Code Smells, Code Refactoring, and Prompt Engineering; Future Works Should State Clearly Approaches.}
Adding on to the topic of standardization, as mentioned in regards to RQ$_3$ and RQ$_5$, there should be a more uniform way of discussing prompting techniques, code smells and code refactoring. Future work should mention by specific names what their study works on specifically, such as what prompting techniques they used, what code smells they address, and what refactoring methods they are conducting. This helps create a more uniform research environment and can aid in analysis helping furthering the field.

\textbf{Address Security Concerns in Using LLM-Based Refactoring Tools in Industry.} Further, while LLMs have demonstrated strong potential in automated refactoring, achieving high Precision and F1 scores that suggest feasible industry value, the integration of such tools comes with concerns around data privacy. The lack of transparency of Large Language Models like ChatGPT makes the adoption of these technologies challenging for businesses that handle regulated and proprietary information. As a result, future work must focus on addressing these security risks, or aiding outside entities to develop their own LLMs to keep their data protected and in-house. This could involve leveraging open-source LLMs such as CodeGen or StarCoder, which can be deployed on local, private infrastructure, eliminating data exposure risks. Researchers can also explore on-premise fine-tuning to allow models to be trained on sensitive company data without sharing it externally. Prioritizing these approaches moving forward is the most effective way for organizations to mitigate significant safety concerns when integrating LLM-based refactoring tools.

\textbf{Integrating Human Feedback in LLM-Based Refactoring.} Building on the potential of LLMs in automated refactoring, it is important to note that their reliability remains a significant concern. As reported in RQ$_6$, these models can generate incorrect code suggestions and hallucinations, making manual verification necessary to ensure code quality. To address this, adding human experts who provide feedback to create a semi-automated refactoring process can increase the reliability, and consequently, the effectiveness of these LLM-based tools. Future studies should explore the implementation of interactive reinforcement learning loops, in which developers validate or fix LLM suggestions, then feed corrections back into the model using frameworks such as Hugging Face’s TRL library. Additionally, code review workflows on platforms like GitHub could be improved with LLM-assisted suggestion systems that allow human verification.

\textbf{Prevent Data Leakage.} Another point stemming from our analysis of RQ$_2$ is that most studies on LLM-based refactoring do not report implementing strategies to prevent data leakage, perhaps because they rely on pre-trained models whose internal safeguards are assumed but not independently verified. A big takeaway for any future work is that authors must explicitly report the precautions they used (e.g., training/testing splits, held-out datasets, unseen repositories), even if partial, so that results across studies are more easily comparable. Going further, framing results with the context of the level of potential leakage would aid other researchers in drawing clear and accurate interpretations of the findings.

\textcolor{black}{\textbf{Incorporate Guidance for Targeted Refactoring.} Based on our analysis in RQ$_6$, LLMs struggle with refactoring large projects and code files, and incorporating semantic analysis into LLM-based refactoring enhances efficiency and precision by detecting and differentiating between related and irrelevant methods when refactoring. Incorporating analysis to guide LLMs to identify refactoring opportunities allows for more LLMs to focus on refactoring relevant instances rather than hallucinating refactorings. Future work should discuss other techniques to decrease irrelevant code inputs, provide LLM’s with guidance identifying refactoring opportunities, and explore how semantic similarity is measured and applied using embedding or other representations to ensure consistency across studies.}

\textcolor{black}{\textbf{Need for Standardization in Model Classification and Identification.} It is important to note that various studies differed on the categorization of different language models as LLMs, pretrained models, or Small Language Models, creating confusion for the reader. Additionally, while some papers specified the exact model or version of the LLM they were using, others referred broadly to "ChatGPT", "DeepSeek" and other models without clarification. Establishing clearer standards for model classification and identification would help improve consistency across future work.}
\color{black}

\section{Threats To Validity} 
\label{Section:Threats}
\textbf{Internal and Construct threats to validity:}
The search process for retrieving a fully representative set of literature publications for SLR may be a threat to validity. To mitigate this threat, we followed the guidelines for SLR outlined in the following: \cite{kitchenham2007guidelines, wohlin2014guidelines, kitchenham2004procedures, brereton2007lessons, kitchenham2009systematic}. When considering possible limitations and issues of our research, it is important to examine the tools we and our citations have used. Since our main focus is on examining the interactions between LLMs and developers, and specifically ChatGPT and developers, it is crucial to ask whether our results have been affected by the continuous development of ChatGPT.

Over time, even within the duration of this research, OpenAI has introduced GPT-4.5, and there is news from Sam Altman that ChatGPT-5 is already in development\footnote{\url{[https://x.com/sama/status/1889755723078443244]}}. Based on the official documentation from OpenAI, ChatGPT is also updated roughly around every 2 weeks\footnote{\url{[https://help.openai.com/en/articles/6825453-chatgpt-release-notes]}}. So, how exactly do we decide a research piece done on ChatGPT has been rendered completely obsolete? \\
Additionally, developers and researchers have noticed that while ChatGPT has become better at certain things, writing code is something it has become worse at \cite{chen2024chatgpt}. We are not exactly sure of the implications that this has on code refactoring yet, but we have noted the difficulties that ChatGPT has on certain refactoring tasks within our own testing, specifically with markdown files and formatting. However, because of the nature of LLMs and how fast they have been developing, it is hard to decide when a research piece is truly obsolete. Within the scope of our research, we do have to consider if the increase in efficiency we saw in developer and ChatGPT interaction has come from our decisions in creating an efficient prompting format or simply from the development that OpenAI has done on ChatGPT. 
Another threat to validity to consider is ChatGPT-3.5's tendency to delete and translate comments when producing refactored code candidates. Shirafuji \etal \cite{shirafuji2023refactoring} found that the comment ratio of the original code samples decreased from 2.39\% to 0.89\% in the generated code samples. Also, the proportion of English comments in the code went from 95.11\% to 97.16\%, subsequently decreasing the proportion of non-English comments. This reduction and translation can negatively impact the code’s maintainability and reliability, as it removes helpful context for understanding the code while possibly creating errors through ChatGPT's unreviewed translation.

Concerning the subjectivity of the data collection and assessments of the primary studies, the primary studies were collected and then two authors independently reviewed then cross-reviewed each other's notes. \color{black}To ensure complete and correct collections of data, we also asked ChatGPT-4 to check over our written evaluations and methodologies to see if there are important data points missed and added to some of our notes. This process included sending ChatGPT-4 all of the written notes of summaries with the original papers, asking if there are any more notes or topics that should be included in the summary. This usually resulted in improved grammar, added with some sentences for extra details. In total there were 16 out of 50 papers (32\%) that were used in this way. \color{black} At the end of each iteration, the authors met and discussed any disagreements that had arisen in interpretations and understanding to reach a consensus.

\textbf{External threats to validity:} The collected papers mostly consist of academic studies, providing a solid foundation for drawing conclusions relevant to academic research. Although we cannot claim that the same methods for LLM-driven refactoring detection and execution are employed in industry. Moreover, our conclusions are within the realm of software refactoring, therefore, we cannot generalize our findings beyond this field.

\section{Conclusion} \label{Section:Conclusion} In this paper, we map and review the body of knowledge on LLM-based refactoring tools and approaches within software engineering. We systematically reviewed \textcolor{black}{50} papers and classified them. Our findings show that (i) \color{black}LLM-driven refactoring studies utilized various different external frameworks and tools to aid in studying and enhancing LLM abilities in refactoring; \color{black}
(ii) the datasets and benchmarks used to evaluate LLMs for refactoring tasks vary greatly in both their composition and scale, making their standardization for more consistent comparison difficult;  
(iii) numerous prompt techniques were used for refactoring tasks, with Few-Shot and Context-Specific being the most common; 
(iv) Accuracy rates of the tools were consistently high (above 87.8\%), with strong F1-measure (above 74\%) and Recall (mainly above 84.3\%), yet Precision varied depending on the context of each study; 
\color{black}(v) Extract Method is the most popular and promising refactoring type employed by LLMs with high Accuracy rates and qualifying, quantifying, comparative, and correctness data points; \color{black}
and (vi) LLMs sometimes produce erroneous code, struggle with complex refactoring tasks, and misunderstand developers’ intent. This research provides the community with information about the emerging LLM-based refactoring approaches to guide future LLM-based refactoring tools. Future work includes standardizing refactoring method terminology, assessing optimal prompt technique combinations, and expanding the evaluation of LLM-based refactoring to less prevalent programming languages.
{\footnotesize\bibliography{references.bib}}
\end{document}

%% file: Tables/Venue.tex
\begin{table}[h]
\begin{center}
\caption{\textcolor{black}{Publication venues.}}
\label{Table:Publication_Sources}
\begin{adjustbox}{width=0.9\textwidth,center}
\rowcolors{2}{white}{gray!25}
\begin{tabular}{llll}\hline
\toprule
\bfseries Venue & \bfseries Publication Venue & \bfseries PSs  \\
\midrule
Journal & Expert Systems with Application & 
\cite{depalma2024exploring}\\
Journal & Automated Software Engineering   & \cite{liu2024empirical} \\
Journal & IEEE Transactions on Software Engineering   & \cite{zhang2024refactoring}  \\
Journal & Electronics & \cite{baumgartner2024ai} \\
Journal & Information Sciences & \cite{zhang2025move} \\
Journal & Software and Systems Modeling & \cite{alturayeif2025refactoring} \\
\textcolor{black}{Journal} & \textcolor{black}{Journal of Systems and Software} & \textcolor{black}{\cite{kim2025comparative}} \\
\textcolor{black}{Journal} & \textcolor{black}{ACM Transactions on Software Engineering and Methodology} & \textcolor{black}{\cite{wang2025towards}}\\
\textcolor{black}{Journal} & \textcolor{black}{International Journal of Software \& Informatics} & \textcolor{black}{\cite{wang2025exploration}}\\ 
\textcolor{black}{Journal} & \textcolor{black}{Future Internet} & \textcolor{black}{\cite{midolo2025refactoring}}\\ 
\textcolor{black}{Journal} & \textcolor{black}{ACM Transactions on Design Automation of Electronic Systems} & \textcolor{black}{\cite{xu2025hlsrewriter}}\\ 
\textcolor{black}{Journal} & \textcolor{black}{Technology audit and production reserves} & \textcolor{black}{\cite{tkachuk2025determining}}\\ 
Conference & International Conference on Mining Software Repositories  & \cite{alomar2024refactor,chavan2024analyzing,ghammam2025build,ksontini2025refactoring}  \\
Conference & International Conference on Automated Software Engineering  & \cite{cui2024three, cui2024one, gao2024preference, wu2024ismell, zhang2024copilot}  \\
Conference & International Conference on the Foundations of Software Engineering   & \cite{dilhara2024unprecedented,gao2024context, pomian2024assist}  \\
Conference & International Conference on Software Maintenance and Evolution   & \cite{gehring2023deterministic, pomian2024next}\textcolor{black}{\cite{batole2025leveraging}} \\
Conference & Asia-Pacific Software Engineering Conference   & \cite{shirafuji2023refactoring} \\
Conference & International Symposium on Search-Based Software Engineering & \cite{choi2024iterative} \\
Conference & Generative AI for Effective Software Development & \cite{white2024chatgpt} \\
Conference & Technical Symposium on Computer Science Education  & \cite{ishizue2024improved} \\
Conference & International Conference on Service-Oriented Computing & \cite{ben2024assessing} \\
Conference & International conference on AI Foundation Models and Software Engineering  & \cite{wang2025testing} \\
Conference & International Conference on Program Comprehension & \cite{midolo2025automated} \\
Conference & Brazilian Symposium on Software Quality  & \cite{menolli2024teaching} \\
Conference & International Symposium on Software Testing and Analysis & \cite{wang2024hecs} \\
Conference & International Conference on Computing and Machine Intelligence  & \cite{icmi2025evaluating} \\
Conference & International Conference on Software Engineering & \cite{dong2025chatgpt}\textcolor{black}{\cite{guo2024exploring,guo2025intention}} \\
\textcolor{black}{Conference} & \textcolor{black}{International Conference on Evaluation and Assessment in Software Engineering} & \textcolor{black}{\cite{palit2025reinforcement}} \\
\textcolor{black}{Conference} & \textcolor{black}{SoutheastCon} & \textcolor{black}{\cite{rajendran2025multi}} \\
\textcolor{black}{Conference} & \textcolor{black}{International Conference on Collaborative Advances in Software and
Computing} & \textcolor{black}{\cite{alomar2025chatgpt}} \\
\textcolor{black}{Conference} & \textcolor{black}{International Conference on Software Analysis, Evolution and Reengineering} & \textcolor{black}{\cite{caumartin2025exploring}} \\
\textcolor{black}{Conference} & \textcolor{black}{International Conference on Software Architecture Companion} & \textcolor{black}{\cite{pandini2025exploratory}} \\
\textcolor{black}{Conference} & \textcolor{black}{International Joint Conference on Conceptual Knowledge Structures} & \textcolor{black}{\cite{guenoune2025llm}} \\
\textcolor{black}{Conference} & \textcolor{black}{Simp{\'o}sio Brasileiro de Engenharia de Software} & \textcolor{black}{\cite{amaral2025improving,pucho2025refactoring}} \\
\textcolor{black}{Conference} & \textcolor{black}{Simp{\'o}sio Brasileiro de Banco de Dados} & \textcolor{black}{\cite{de2025effectiveness}} \\
Workshop & NeurIPS 2024 Workshop on Open-World Agents  & \cite{gautam2024refactorbench} \\
\bottomrule
\end{tabular}
\end{adjustbox}
\end{center}
\end{table}

%% file: Tables/Approach.tex
\begin{table*}[htbp]
  \centering
	 \caption{\textcolor{black}{Related work in LLM-driven refactoring (RQ$_1$).}}
	 \label{Table:Related_Work_in_llms_Refatoring}
\begin{adjustbox}{width=1.5\textwidth,center}
\rowcolors{2}{white}{gray!25}
\begin{tabular}{lcllllllll}\hline
\toprule
\bfseries Study & \bfseries Year &  \bfseries Focus & \bfseries Tools/Models & \bfseries LLMs & \bfseries Language      \\
\midrule
Gehring \cite{gehring2023deterministic} & 2023 & Introduces OpenRewrite, a tool for large-scale code refactoring. & Moderne, OpenRewrite, SonarQube & StarChat, ChatGPT-3.5 & Java, Python, JavaScript, COBOL\\ 
White \etal \cite{white2024chatgpt} & 2023 & Prompt engineering for code refactoring & IntelliJ IDEA & ChatGPT-3.5 Turbo, ChatGPT-4 & Python \\
Shirafuji \etal \cite{shirafuji2023refactoring} & 2023 & Effectiveness of few shot prompting & Aizu online judge, YAPF, Radon, Python tokenizer & ChatGPT-3.5 & Python \\
 Depalma \etal \cite{depalma2024exploring} & 2024 & Correctness/Behavior Preservation/Documentation & PMD & ChatGPT-3.5 & Java  \\
 AlOmar \etal \cite{alomar2024refactor} & 2024 & Refactoring on documentations, quality, and software repositories & Hacker News, SQLite, FastText, Google Translator, NLTK and Spacy, Dask & ChatGPT & Python, Java, JavaScript, C++, C\#, TypeScript, Go, Ruby \\
 Chavan \etal \cite{chavan2024analyzing} & 2024 & Examines how developers interact with ChatGPT, how they prompt & Hacker News, Google Translator & ChatGPT & Python, JavaScript \\ & & it to refactor code, and how many prompts they make on average & & & \\ & & before reaching a conclusion & & & \\
Pomian \etal \cite{pomian2024assist} & 2024 & The effectiveness of EM-Assist in applying refactoring, especially & EM-Assist (developed plug-in tool), IntelliJ IDEA APIs, JExtract& GPT-3.5 Turbo& Java, Kotlin \\ & & the extract method & & & \\
Pomian \etal \cite{pomian2024next} & 2024 & The creation and need for EM-Assist, the inspection of how we can & EM-Assist (developed plug-in tool), IntelliJ IDEA APIs, JExtract, Kolin and Java codebases & ChatGPT-3.5, ChatGPT-4, PaLM & Java, Kotlin \\ & & create a new refactoring tool by harnessing the power of LLMs & & & \\
Dilhara \etal \cite{dilhara2024unprecedented} & 2024 & How effective LLM is towards different ways of prompting for refactoring & PyCraft, R-CPATMiner, PyEvolve, Mutmut& ChatGPT-3.5, ChatGPT-4, PaLM & Python \\
Gao \etal \cite{gao2024context} & 2024 & How can adding context using the approach of UTRefactor improve & GitHub, Maven, tsDetect, UTRefactor, TESTAXE, JUnit 5, Llama & ChatGPT-4o & Java \\ & &  the refactoring outputs from LLMs & & & \\
Zhang \etal \cite{zhang2024refactoring} & 2024 & Refactoring of non-idiomatic python code to pythonic idioms using & ARIs, RIdiom & ChatGPT-3.5 Turbo & Python \\ & & hybrid approach that combines rule-based systems and LLMs; addresses & & & \\ & & code detection, correctness, idiomatic transformation, evaluates & & & \\ & & the proposed method against benchmarks and explores its scalabilities & & & \\
Choi \etal \cite{choi2024iterative} & 2024 & Testing how much a LLM-based refactoring pipeline effects the Cyclomatic & EvoSuite ,Defects4J, Lizard & ChatGPT-3.5 Turbo & Java\\ & & Complexity of various Java projects, measured with the Lizard library & & & \\
Wu \etal \cite{wu2024ismell} & 2024 & Testing if combining the output of code smell detection tool sets & PMD, JDeodorant, Organic, DesigniteJava, JSpIRIT, FeTruth, JMove, \textcolor{black}{CodeBERT, CKmetrics} & ChatGPT-3.5 turbo, ChatGPT-4.0 turbo, LLAMA3-70B, CodeLlama-34B & Java\\ & & recommended by Mixture of Experts(machine learning model) with LLMs & & & \\ & &  will improve their code smell detection capabilities & & & \\
Ishizue \etal \cite{ishizue2024improved} & 2024 & Code repair/refactoring, Automated feedback systems & Refractory, Google Cloud Platform’s Google Compute Engine & ChatGPT-3.5 Turbo, ChatGPT-4 & Python \\
Gautam \etal \cite{gautam2024refactorbench} & 2024 & Testing a benchmark for multi-file refactoring tasks, State-ful reasoning & SWE-bench &  ChatGPT-4 Turbo, Claude 3.5 Sonnet & Python \\ & & State-aware behavior of LMMs. LLMs(ChatGPT) are used to generate tasks & & & \\ & &  to test the LMMs on & & &\\
Baumgartner \etal \cite{baumgartner2024ai} & 2024 &Testing different techniques of implementing LLMs(ChatGPT) into a & ArgoUml & ChatGPT-4 Turbo, ChatGPT-3.5 & Java\\ & & refactoring workflow/pipeline using prompt engineering. These include & & & \\ & & data clump detection, and refactoring tasks & & & \\
Cui \etal \cite{cui2024three} & 2024 & Move Method refactoring & Graph Neural Network models: GraphSAGE, GAT, HGCN, HGNN, GCN, & Code Llama, DeepSeek, ChatGPT-4, ChatGPT-3.5 & Java \\ 
 & & & Refactoring tools: LLM Refactor, FeTruth, FeDeep, PathMove, JDeodorant, JMove, RMove, FeGNN, and FePM & & \\
Zhang \etal \cite{zhang2025move} & 2024 & Move Method refactoring & Deep Learning Models: CNN, LSTM, GRU, Natural language processing models: RoBERTa, ALBERT,  XLNet, & ChatGPT-3.5 Turbo & Java \\ & & & Refactoring tools: PathMove, JDeodorant, JMove, RMove,Pydriller, CodeSplitJava, RefactoringMiner, iPlasma,& & \\ & & & PMD, Cayenne, Pinpoint & & \\
Ben Mrad \etal \cite{ben2024assessing} & 2024 & Programming/Method Nomenclature Prediction and Automation & Nvidia T4 GPU & ChatGPT-3.5, Llama2-6.74B, Mistral-instruct-v0.2-7.24B & Java \\
Wang \etal \cite{wang2024hecs} & 2024 & Introduces the design and uses of HECS, a tool that gives extract class & Refactoring Tools: JDeodorant, SSECS, LLMRefactor, Graph Neural Networks: HGNN, SAGE, GCN, GAT,& ChatGPT-3.5&Java\\ & & refactoring suggestions & Communities: GitHub, F-Droid, and Apach, Code Models: CodeBERT, CodeT, CodeGPT, GraphCodeBERT, & & \\ & & & PLBART, CoTexT & &\\
Gao \etal \cite{gao2024preference} & 2024 & The effectiveness of RRG (Retrieve, Refactor, Generate) & Javalang, Python AST & ChatGPT-2, ChatGPT, CodeGEN, PolyCoder, CodeT5 & Java, Python \\
Cui \etal \cite{cui2024one}& 2024 & Explores the usefulness of a new tool for extract class refactoring suggestions. & Refactoring Tools: JDeodorant , SSECS, LLMRefactor, Graph Neural Networks: HGNN, SAGE, GCN, GAT, & GPT-3.5 & Java\\ & & It combines hypergraph learning with an LLM& Communities: GitHub, F-Droid, and Apache, Code Models: CodeBERT, CodeT ,CodeGPT, GraphCodeBERT, & & \\ & & & PLBART, CoTexT & & \\
Zhang \etal \cite{zhang2024copilot} & 2024 & The code smells/quality of the code generated by copilot, effectiveness of & The code smells/quality of the code generated by copilot, effectiveness of copilot in fixing code & GitHub Copilot (derived from OpenAI Codex) & Python\\ & & copilot in fixing code & & & \\
Menolli \etal \cite{menolli2024teaching} & 2024 & Evaluating the effectiveness of a refactoring teaching method & Refractory& ChatGPT-3.5 & Java \\
Wang \etal \cite{wang2025testing} & 2025 & The reliability and testing of refactoring engines in IDEs such as IntelliJ & JVM Compiler, Eclipse's Bugzilla, GitHub issue trackers, and IntelliJ IDEA's YouTrack, Eclipse, IntelliJ IDEA, & ChatGPT-4 & Java \\ & & and Eclipse. uses RETESTER that use historical bug reports and LLMs to & ASTGen, SAFEREFACTOR & & \\ & & generate test inputs. automated detection of bugs in refactoring engines, & & & \\ & & evaluates the effectiveness of different input program characteristics & & & \\
Midolo \etal \cite{midolo2025automated} & 2025 & The effectiveness of ChatGPT-4 in recommending and suggesting idiomatic & RIdiom, GitHub & ChatGPT-4 & Python\\ & & refactoring actions. automated refactoring & & & \\
Ksontini \etal \cite{ksontini2025refactoring} & 2025 & The dockerfile quality checking, refactoring capabilities, automated vs & DRMiner, PARFUM, BM-25, Google BigQuery Github, automated build system & ChatGPT-4o & Docker, Python, bash, SQL\\ & & manual, CI/CD pipelines  & & & \\
Alturayeif \& Hassine \cite{alturayeif2025refactoring} & 2025 & Decreasing bad smells when using GORE, types of linguistic bad smells & Stanza, BERT, jUCMNav & ChatGPT-3.5 turbo & Python (but the focus is on english) \\ & & in goal models & & & \\
Divyansh \etal \cite{icmi2025evaluating} & 2025 & Effectiveness of ChatGPT on the cyclomatic complexity, cognitive complexity, & SonarQube, Node, Stepik and Hyperskill & ChatGPT-3.5 & Python, Javascript(for dataset extraction) \\ & & code smells, and time debt after refactoring & & & \\
Liu \etal \cite{liu2024empirical} & 2024 & The potentials of LLMs in automated software refactoring & ReExtractor, RefactoringMiner, IntelliJ IDEA, Dice Coefficient, Point-Biserial Correlation Coefficient & ChatGPT-4, Gemini & Java \\
\textcolor{black}{Guo \etal \cite{guo2024exploring}} & \textcolor{black}{2024} & \textcolor{black}{Investigate ChatGPT's ability to perform automated code refinement based} & \textcolor{black}{CodeReviewer} & \textcolor{black}{GPT-3.5 Turbo, GPT-4}& \textcolor{black}{C, C++, C\#, Go, Java, JavaScript, PHP, Python, Swift,}\\& & \textcolor{black}{on code review comments} & & & \textcolor{black}{Objective-C, Kotlin, SQL, Perl, Scala, R}\\
Dong \etal \cite{dong2025chatgpt} & 2025 & The use of ChatGPT to test refactoring engines & IntelliJ, Eclipse, NetBeans, VSCode-Java & ChatGPT & Java \\
Ghammam \etal \cite{ghammam2025build} & 2025 & Current refactoring practices and technical debt in build systems & Maven, Ant, Gradle & ChatGPT-4o & 30 languages but mainly: Java, Kotlin, PHP \\
\textcolor{black}{Batole \etal \cite{batole2025leveraging}} & \textcolor{black}{2025} & \textcolor{black}{Implementing move method refactoring using LLMs} & \textcolor{black}{JMove, FeTruth, RefactoringMiner, VoyageAI} & \textcolor{black}{GPT-4o}& \textcolor{black}{Java}\\
\textcolor{black}{Palit \& Sharma \cite{palit2025reinforcement}} & \textcolor{black}{2025} & \textcolor{black}{Combining supervised fine-tuning with reinforcement learning for automated} & \textcolor{black}{SEARTtool, RefactoringMiner, tree-sitter} & \textcolor{black}{Code-T5,  PLBART, CodeGPT-adapt, CodeGen}& \textcolor{black}{Java}\\ & & \textcolor{black}{extract method refactoring} & & & \\
\textcolor{black}{Rajendran \etal \cite{rajendran2025multi}} & \textcolor{black}{2025} & \textcolor{black}{Specializing LLM-based multiple agents based on different needs of refactoring} & \textcolor{black}{Not Specified} & \textcolor{black}{ChatGPT-4}& \textcolor{black}{Java}\\
\textcolor{black}{Kim \cite{kim2025comparative}} & \textcolor{black}{2025} & \textcolor{black}{Comparing the validity of results from LLMs identifying relevant design patterns} & \textcolor{black}{Not Specified} & \textcolor{black}{ChatGPT-4o, Claude 3.5 Sonnet, Microsoft 365 Copilot,}& \textcolor{black}{Java}\\ & & \textcolor{black}{and then applying them to a location in a program} & & \textcolor{black}{Gemini 1.5, Llama 3.1-405B} & \\
\textcolor{black}{Wang \etal \cite{wang2025towards}} & \textcolor{black}{2025} & \textcolor{black}{Refactoring engine bugs} & \textcolor{black}{Eclipse, IntelliJ IDEA, NetBeans, SAFEREFACTOR-based tools, ASTGen} & \textcolor{black}{ChatGPT-4o mini}& \textcolor{black}{Java}\\
\textcolor{black}{Wang \etal \cite{wang2025exploration}} & \textcolor{black}{2025} & \textcolor{black}{Evaluate the performance of LLMs vs pre-trained models in code refinement tasks} & \textcolor{black}{LitGPT} & \textcolor{black}{CodeLLaMA, StableCode3B, LLaMA3, and ChatGPT-4o,}& \textcolor{black}{C, C++, C\#, Go, Java, JavaScript, PHP, Python, Ruby}\\ & & \textcolor{black}{and proposes a technique to mitigate the disadvantages of LLMs} & & \textcolor{black}{T5-Review, CodeT5, CodeReviewer, and CodeBERT} & \\
\textcolor{black}{AlOmar \etal \cite{alomar2025chatgpt}} & \textcolor{black}{2025} & \textcolor{black}{Investigate refactoring conversations between developers and ChatGPT} & \textcolor{black}{SQLite, FastText, Google Translator, NLTK, Spacy, Dask} & \textcolor{black}{ChatGPT-4, ChatGPT-3.5}& \textcolor{black}{Python, JavaScript, Bash}\\
\textcolor{black}{Midolo \etal \cite{midolo2025refactoring}} & \textcolor{black}{2025} & \textcolor{black}{ChatGPT's abilities in refactoring for-loops into streams} & \textcolor{black}{JavaParser, Apache Maven (seen in code)} & \textcolor{black}{ChatGPT-3.5 Turbo}& \textcolor{black}{Java}\\
\textcolor{black}{Pucho \etal \cite{pucho2025refactoring}} & \textcolor{black}{2025} & \textcolor{black}{Automated Python code refactoring using LLM-based multi-agent systems in} & \textcolor{black}{MetaGPT, PyRef, PyDriller, Radon, Lizard} & \textcolor{black}{Mixtral-8x22b}& \textcolor{black}{Python}\\ & & \textcolor{black}{machine learning projects} & & & \\
\textcolor{black}{Caumartin \etal \cite{caumartin2025exploring}} & \textcolor{black}{2025} & \textcolor{black}{Comparing small open-source models to ChatGPT in code refinement tasks} & \textcolor{black}{CodeReviewer} & \textcolor{black}{Llama 2 7b, CodeLlama, GPT-3.5 Turbo}& \textcolor{black}{C, C++, C\#, Go, Java, JavaScript, PHP, Python, Swift,}\\& & & & & \textcolor{black}{Objective-C, Kotlin, SQL, Perl, Scala, R}\\
\textcolor{black}{Pandini \etal \cite{pandini2025exploratory}} & \textcolor{black}{2025} & \textcolor{black}{The use of LLMs to suggest refactoring for Cyclic Dependency architectural smells} & \textcolor{black}{Arcan, LangChain, Ollama, Amazon Bedrock} & \textcolor{black}{Claude 3.5 Sonnet, Qwen}& \textcolor{black}{Java}\\
\textcolor{black}{Guo \etal \cite{guo2025intention}} & \textcolor{black}{2025} & \textcolor{black}{Automated code refinement to generate revised code based on identified intentions} & \textcolor{black}{CodeReviewer, T5CR, GitHub REST API, BM25} & \textcolor{black}{GPT-3.5, GPT-4o, DeepSeekV2, DeepSeek7B, CodeQwen7B}& \textcolor{black}{C, C++, C\#, Go, Java, JavaScript, PHP, Python}\\
\textcolor{black}{Xu \etal \cite{xu2025hlsrewriter}} & \textcolor{black}{2025} & \textcolor{black}{
Automation of refactoring and optimization of C/C++ code for High-Level Synthesis} & \textcolor{black}{Catapult HLS Tool, Synopsys Design Compiler, Sentence Transformer} & \textcolor{black}{GPT-4 Turbo}& \textcolor{black}{C/C++}\\
\textcolor{black}{Tkackuk \cite{tkachuk2025determining}} & \textcolor{black}{2025} & \textcolor{black}{Investigate the capability of LLMs to refactor code that improves maintainability and} & \textcolor{black}{Not Specified} & \textcolor{black}{ChatGPT, Copilot, Gemini, Claude}& \textcolor{black}{Java, Swift}\\ & & \textcolor{black}{efficiency} & & & \\
\textcolor{black}{Guenoune \etal \cite{guenoune2025llm}} & \textcolor{black}{2025} & \textcolor{black}{Combining Formal Concept Analysis (FCA), Relational Concept Analysis (RCA),} & \textcolor{black}{RCAtool, PlantText modeling tool} & \textcolor{black}{Gemini pro 2.5, Grok 3 thinking, ChatGPT o1, ChatGPT o3,}& \textcolor{black}{UML model that can be used to generate Python, C++, Java}\\ & & \textcolor{black}{and LLMs to help access the relevancy of derived concepts and name software} & & \textcolor{black}{DeepSeek V3, Claude 3.7 thinking 2, Claude 3.7 Sonnet no think,} & \\ & & \textcolor{black}{entities appropriately} & & \textcolor{black}{Phind 2.0 70B} & \\
\textcolor{black}{Amaral \cite{amaral2025improving}} & \textcolor{black}{2025} & \textcolor{black}{Investigate the potential of LLMs to automatically refactor JavaScript test smells} & \textcolor{black}{GitHub Copilot Chat, Amazon CodeWhisperer, SNUTS.JS, Steel, TestSmellDetectorJs, VSCode} & \textcolor{black}{GPT-4o (powers Github Copilot Chat), Amazon CodeWhisperer}& \textcolor{black}{JavaScript}\\
\textcolor{black}{De Sousa \etal \cite{de2025effectiveness}} & \textcolor{black}{2025} & \textcolor{black}{Evaluate LLMs and small language models ability in detection issues and bad smells} & \textcolor{black}{Not Specified} & \textcolor{black}{GPT-4o, Gemini 2.0 Flash, GPT-4o mini, Phi-4}& \textcolor{black}{PL/SQL}\\ & & \textcolor{black}{in PL/SQL} & & & \\
\bottomrule
\end{tabular}
\end{adjustbox}
\vspace{-.3cm}
\end{table*}

%% file: Tables/Dataset.tex
\begin{table*}[ht!]
  \centering
	 \caption{\textcolor{black}{Dataset in LLM-driven refactoring (RQ$_2$).}}
	 \label{Table:Dataset_in_llms_Refatoring}

\begin{adjustbox}{width=1.4\textwidth,center}
\rowcolors{2}{white}{gray!25}
\begin{tabular}{llllllllll}\hline
\toprule
\bfseries Study & \bfseries Projects &  \bfseries Dataset &   \bfseries Language  & \bfseries Manual / Automated & \bfseries Year & \bfseries Data leakage precautions       \\
\midrule
Gehring \cite{gehring2023deterministic} & openrewrite/rewrite, 
Spring Cloud, Netflix & 1 recipe, 4 files & Java & \textcolor{black}{Yes/No} & \textcolor{black}{2023} & \textcolor{black}{No (not mentioned)}\\ 
White \etal \cite{white2024chatgpt} & Previous work\cite{white2023prompt} & 14 prompt patterns & Python & \textcolor{black}{Yes/No} & \textcolor{black}{2023} & \textcolor{black}{No(not mentioned)}\\
Shirafuji \etal \cite{shirafuji2023refactoring} & Aizu Online Judge & 8,000,000 submissions: 880 selected (avg. LOC: 14.82)
 & Python & \textcolor{black}{Yes/Yes} & \textcolor{black}{2023} & \textcolor{black}{No(not mentioned)}\\
 Depalma \etal \cite{depalma2024exploring} &  Previous work \cite{ma2023scope} &   40 files: 1 class (LOC: 15 - 35)    & Java  & \textcolor{black}{Yes/Yes} & \textcolor{black}{2023} & \textcolor{black}{No(not mentioned)}\\
 Alomar \etal \cite{alomar2024refactor} &  DevGPT \cite{xiao2024devgpt} &   470 commits, 69 issues, 176 files    & Python, Java, Javascript, C++, C\#, TypeScript, Go, Ruby  & \textcolor{black}{Yes/Yes} & \textcolor{black}{2023} & \textcolor{black}{No(not mentioned)}\\
 Chavan \etal \cite{chavan2024analyzing} & DevGPT \cite{xiao2024devgpt} & 447 conversations: 1,246 records & Python, Javascript & \textcolor{black}{Yes/Yes} & \textcolor{black}{2023} & \textcolor{black}{No(not mentioned)}\\
Pomian \etal \cite{pomian2024assist} & Various open-source & 1,756 refactorings: (avg. LOC: 30) & Java, Kotlin & \textcolor{black}{No/Yes} & \textcolor{black}{2024} & \textcolor{black}{No(not mentioned)}\\
Pomian \etal \cite{pomian2024next} & MyWebMart, SelfPlanner, WikiDev, JHotDraw, JUnit & 122 methods & Java, Kotlin & \textcolor{black}{Yes/Yes} & \textcolor{black}{2024} & \textcolor{black}{No(not mentioned)}\\
Dilhara \etal \cite{dilhara2024unprecedented} & Various open-source & 9,325 code change patterns & Python & \textcolor{black}{Yes/Yes} & \textcolor{black}{2023} & \textcolor{black}{No(not mentioned)}\\
Gao \etal \cite{gao2024context} & google/gson, jhy/jsoup, apache/commons, jfree/jfreechart & 6 projects: 9,149 tests, 2,375 code smells (total LOC: 174,510) & Java & \textcolor{black}{No/Yes} & \textcolor{black}{2024} & \textcolor{black}{No (not mentioned)}\\
 Zhang \etal \cite{zhang2024refactoring} & LLM-generated ARIs & 3,311 methods, 4678 code pairs & Python & \textcolor{black}{No/Yes} & \textcolor{black}{2024} & \textcolor{black}{Yes: unseen projects}\\
Choi \etal \cite{choi2024iterative} &  Defects4J & 17 projects & Java & \textcolor{black}{No/Yes} &  \textcolor{black}{2023} & \textcolor{black}{Not clear}\\
Wu \etal \cite{wu2024ismell} & Previous work \cite{palomba2018diffuseness}\cite{arcelli2016comparing}\cite{khomh2012exploratory} & 3 widely-used datasets & Java & \textcolor{black}{No/Yes} &  \textcolor{black}{2024} & \textcolor{black}{Yes: cross-validation}\\
Ishizue \etal \cite{ishizue2024improved} & Large public university & 361 students: 5 assignments: 1,800 submissions & Python & \textcolor{black}{No/No} & \textcolor{black}{2023} & \textcolor{black}{No (not mentioned)}\\
Gautam \etal \cite{gautam2024refactorbench} & Various open-source & 9 projects: 100 multi-file refactoring tasks & Python & \textcolor{black}{Yes/Yes} & \textcolor{black}{2024} & \textcolor{black}{Yes: unseen projects}\\
Baumgartner \etal \cite{baumgartner2024ai} & ArgoUml & LOC: 180,000 & Java & \textcolor{black}{Yes/No} & \textcolor{black}{2024} & \textcolor{black}{No(not mentioned)}\\
Cui \etal \cite{cui2024three} & GitHub, F-Droid, Apache & 11,169 projects & Java & \textcolor{black}{Yes/Yes} & \textcolor{black}{2024} & \textcolor{black}{Yes: train/test split}\\
Zhang \etal \cite{zhang2025move} & Various open-source & 58 projects: 639,010 total methods & Java & \textcolor{black}{No/Yes} & \textcolor{black}{2024} & \textcolor{black}{Yes: train/test split}\\
Ben Mrad \etal \cite{ben2024assessing} & Previous work\cite{8816971} & 116 methods & Java & \textcolor{black}{No/Yes} & \textcolor{black}{2024} & \textcolor{black}{No (not mentioned)}\\
Wang \etal \cite{wang2024hecs} & GitHub, Apache, F-Droid, GanttProject, Xerces & 11,336 projects: 41,311 refactoring cases & Java & \textcolor{black}{Yes/Yes} & \textcolor{black}{2024} & \textcolor{black}{No (not mentioned)}\\
Gao \etal \cite{gao2024preference} & CodeSearchNet, Concode & 966,629 code samples: 30,000 selected & Java, Python & \textcolor{black}{Yes/Yes} & \textcolor{black}{2024} & \textcolor{black}{No (not mentioned)}\\
Cui \etal \cite{cui2024one}& GitHub, Apache, F-Droid, GanttProject, Xerces & 11,336 projects: 41,311 refactoring cases & Java & \textcolor{black}{No/Yes} & \textcolor{black}{2024} & \textcolor{black}{Yes: held-out projects}\\
Zhang \etal \cite{zhang2024copilot} & GitHub Copilot & 1,204 files & Python & \textcolor{black}{Yes/Yes} & \textcolor{black}{2023} & \textcolor{black}{No (not mentioned)}\\
Menolli \etal \cite{menolli2024teaching} &  Unknown GitHub Projects & 23 students selected 4 projects & Java & \textcolor{black}{Yes/No} & \textcolor{black}{2024}  & \textcolor{black}{No (not mentioned)}\\
Wang \etal \cite{wang2025testing} & Eclipse, GitHub, IntelliJ IDEA & 458 bug reports & Java & \textcolor{black}{No/Yes} & \textcolor{black}{2025} & \textcolor{black}{No (not mentioned)}\\
Midolo \etal \cite{midolo2025automated} &  idiomaticrefactoring/pythonidiomsrefactor & 736 projects: 1,061 methods total & Python & \textcolor{black}{Yes/No} & \textcolor{black}{2024} & \textcolor{black}{Not clear}\\
Ksontini \etal \cite{ksontini2025refactoring} & Google BigQuery GitHub Public Dataset & 600 dockerfiles & Python, Bash, SQL & \textcolor{black}{Yes/Yes} & \textcolor{black}{2024} & \textcolor{black}{Yes: train/test split}\\
Alturayeif \& Hassine \cite{alturayeif2025refactoring} & Goal-oriented Requirements Language & 4 categories: 17 linguistic smells total & Python & \textcolor{black}{Yes/No} & \textcolor{black}{2024} & \textcolor{black}{Yes: held-out datasets}\\
Divyansh \etal \cite{icmi2025evaluating} & Previous work \cite{birillo2022hyperstyle} & 1,032 files & Python & \textcolor{black}{Yes/Yes} & \textcolor{black}{2024} & \textcolor{black}{No (not mentioned)}\\
Liu \etal \cite{liu2024empirical} & Various open-source & 20 projects: 180 refactorings & Java & \textcolor{black}{Yes/Yes} & \textcolor{black}{2024} & \textcolor{black}{Yes: unseen repositories}\\
\textcolor{black}{Guo \etal \cite{guo2024exploring}} & \textcolor{black}{GitHub/CodeReview} & \textcolor{black}{467 projects, 14,568 samples} & \textcolor{black}{16 languages including: C, C++, Swift, Kotlin} & \textcolor{black}{Yes/Yes} & \textcolor{black}{2023}& \textcolor{black}{Yes: unseen repositories}\\
Dong \etal \cite{dong2025chatgpt} & Eclipse, IntelliJ IDEA, and VSCode & 700 bug reports and 185 test cases & Java & \textcolor{black}{Yes/Yes} & \textcolor{black}{2024} & \textcolor{black}{No (not mentioned)}\\
Ghammam \etal \cite{ghammam2025build} & GitHub & 725 commits from  609 projects & 30 languages but mainly: Java, Kotin, PHP & \textcolor{black}{Yes/Yes} & \textcolor{black}{2024} & \textcolor{black}{Yes: train/test split}\\
\textcolor{black}{Batole \etal \cite{batole2025leveraging}} & \textcolor{black}{Various open-source} & \textcolor{black}{10 projects, 210 refactorings instances} & \textcolor{black}{Java} & \textcolor{black}{Yes/Yes} & \textcolor{black}{2024}& \textcolor{black}{No (not mentioned)}\\
\textcolor{black}{Palit \& Sharma \cite{palit2025reinforcement}} & \textcolor{black}{GitHub} & \textcolor{black}{1,618 projects, 33,477 samples} & \textcolor{black}{Java} & \textcolor{black}{Yes/Yes} & \textcolor{black}{2024}& \textcolor{black}{Yes: train/test split}\\
\textcolor{black}{Rajendran \etal \cite{rajendran2025multi}} & \textcolor{black}{Various open-source} & \textcolor{black}{4 agents: dataset composition not specified} & \textcolor{black}{Java} & \textcolor{black}{Yes/No} & \textcolor{black}{2024}& \textcolor{black}{No (not mentioned)}\\
\textcolor{black}{Kim \cite{kim2025comparative}} & \textcolor{black}{Various open-source} & \textcolor{black}{2 programs: 11 and 6 classes} & \textcolor{black}{Java} & \textcolor{black}{Yes/No} & \textcolor{black}{2024}& \textcolor{black}{Not clear}\\
\textcolor{black}{Wang \etal \cite{wang2025towards}} & \textcolor{black}{Eclipse, GitHub, IntelliJ IDEA} & \textcolor{black}{1,651 bug reports} & \textcolor{black}{Java} & \textcolor{black}{Yes/Yes} & \textcolor{black}{2024}& \textcolor{black}{No (not mentioned)}\\
\textcolor{black}{Wang \etal \cite{wang2025exploration}} & \textcolor{black}{GitHub and Gerrit} & \textcolor{black}{800 samples} & \textcolor{black}{9 languages including: C, C++, C\#, Go, Java, JavaScript} & \textcolor{black}{Yes/No} & \textcolor{black}{2024}& \textcolor{black}{Not clear}\\
\textcolor{black}{AlOmar \etal \cite{alomar2025chatgpt}} & \textcolor{black}{DevGPT \cite{xiao2024devgpt}} & \textcolor{black}{470 commits, 69 issues, and 176 files} & \textcolor{black}{Python, JavaScript, Bash} & \textcolor{black}{Yes/Yes} & \textcolor{black}{2024}& \textcolor{black}{No (not mentioned)}\\
\textcolor{black}{Midolo \etal \cite{midolo2025refactoring}} & \textcolor{black}{GitHub} & \textcolor{black}{4 projects, 2,132 loops} & \textcolor{black}{Java} & \textcolor{black}{Yes/Yes} & \textcolor{black}{2025}& \textcolor{black}{Not clear}\\
\textcolor{black}{Pucho \etal \cite{pucho2025refactoring}} & \textcolor{black}{Various open-source} & \textcolor{black}{2,000 files} & \textcolor{black}{Python} & \textcolor{black}{No/Yes} & \textcolor{black}{2024}& \textcolor{black}{No (not mentioned)}\\
\textcolor{black}{Caumartin \etal \cite{caumartin2025exploring}} & \textcolor{black}{GitHub/CodeReview-New} & \textcolor{black}{467 projects, 14,568 samples} & \textcolor{black}{16 languages including: C, C++, Swift, Kotlin} & \textcolor{black}{No/Yes} & \textcolor{black}{2024}& \textcolor{black}{Yes: train/test split}\\
\textcolor{black}{Pandini \etal \cite{pandini2025exploratory}} & \textcolor{black}{Keyword, Various open-source} & \textcolor{black}{5 projects} & \textcolor{black}{Java} & \textcolor{black}{No/Yes} & \textcolor{black}{2024}& \textcolor{black}{No (not mentioned)}\\
\textcolor{black}{Guo \etal \cite{guo2025intention}} & \textcolor{black}{GitHub/CodeReview} & \textcolor{black}{15 projects, 1,100 code reviews} & \textcolor{black}{C, C++, C\#, Go, Java, JavaScript, PHP, and Python} & \textcolor{black}{Yes/Yes} & \textcolor{black}{2024}& \textcolor{black}{Yes: held-out datasets}\\
\textcolor{black}{Xu \etal \cite{xu2025hlsrewriter}} & \textcolor{black}{Previous work\cite{thomas2016synthesis}} & \textcolor{black}{74 files, 24 examples} & \textcolor{black}{C/C++} & \textcolor{black}{Yes/No} & \textcolor{black}{2024}& \textcolor{black}{No (not mentioned)}\\
\textcolor{black}{Tkackuk \cite{tkachuk2025determining}} & \textcolor{black}{Unknown} & \textcolor{black}{4 code snippets} & \textcolor{black}{Swift, Java} & \textcolor{black}{Yes/No} & \textcolor{black}{2024}& \textcolor{black}{No (not mentioned)}\\
\textcolor{black}{Guenoune \etal \cite{guenoune2025llm}} & \textcolor{black}{Previous work\cite{vernier2013eis}} & \textcolor{black}{2 projects} & \textcolor{black}{Python, C++, and Java are mentioned} & \textcolor{black}{Yes/No} & \textcolor{black}{2023}& \textcolor{black}{No (not mentioned)}\\
\textcolor{black}{Amaral \cite{amaral2025improving}} & \textcolor{black}{GitHub} & \textcolor{black}{10 projects, 148 test smells} & \textcolor{black}{JavaScript} & \textcolor{black}{Yes/Yes} & \textcolor{black}{2024}& \textcolor{black}{No (not mentioned)
}\\
\textcolor{black}{De Sousa \etal \cite{de2025effectiveness}} & \textcolor{black}{Various open-source} & \textcolor{black}{44 code snippets} & \textcolor{black}{PL/SQL} & \textcolor{black}{Yes/No} & \textcolor{black}{2025}& \textcolor{black}{No (not mentioned)
}\\
\bottomrule
\end{tabular}
\end{adjustbox}

\vspace{-.3cm}
\end{table*}


\pgfplotsset{compat=1.18}
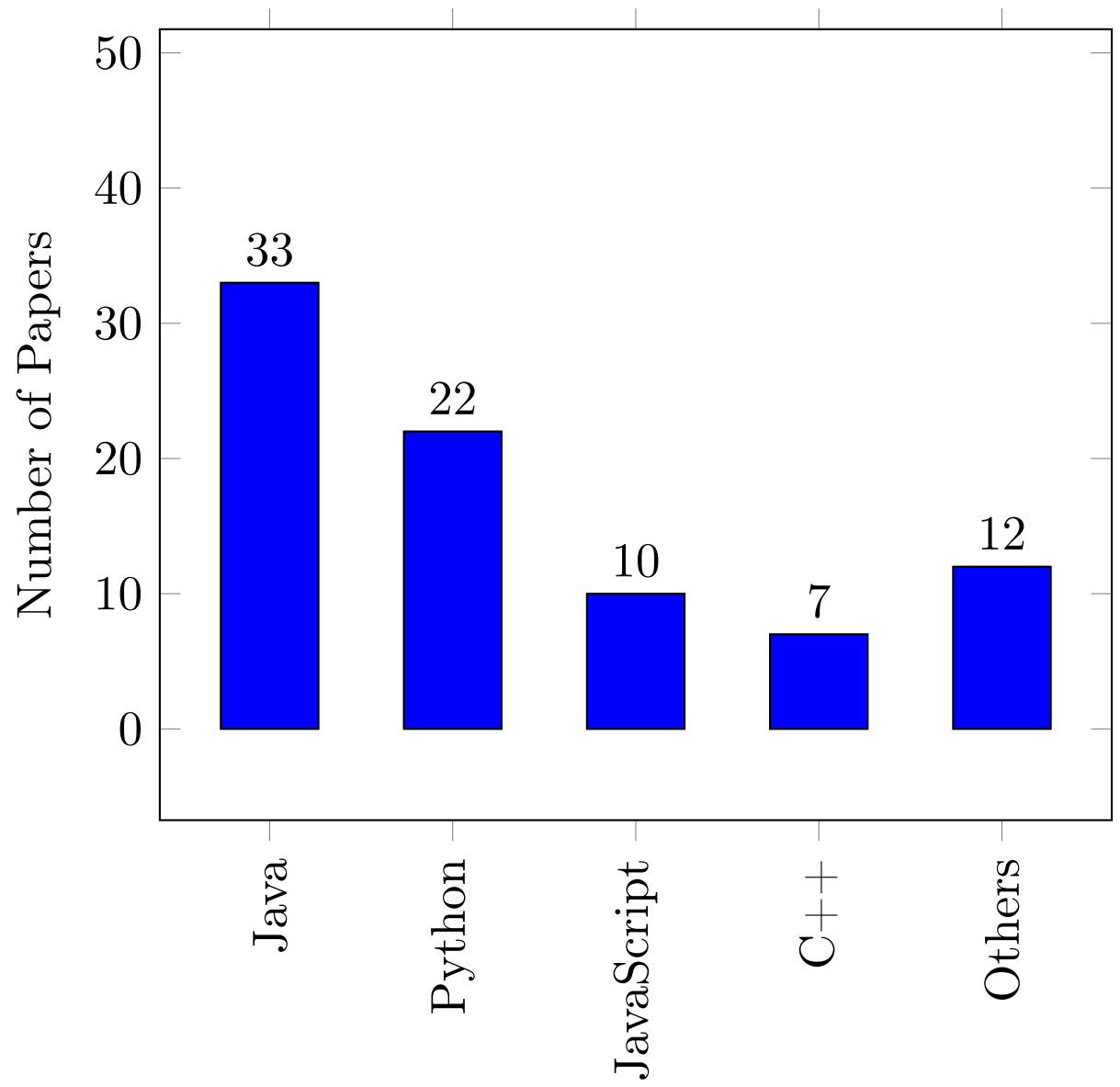
\begin{figure}
\centering
\begin{tikzpicture}
\begin{axis}[
    ybar,
    bar width=20pt,
    enlargelimits=0.15,
    ylabel={Number of Papers},
    symbolic x coords={Java, Python, JavaScript, C++, Others},
    xtick=data,
    nodes near coords,
    x tick label style={rotate=90, anchor=east},
    ymin=0, ymax=45,
    legend style={at={(0.5,-0.15)}, anchor=north, legend columns=-1}
]
\addplot[fill=blue] coordinates {
    (Java,33) 
    (Python, 22) 
    (JavaScript,10) 
    (C++,7)
    (Others,12)
};

\end{axis}
\end{tikzpicture} 
   \caption{\textcolor{black}{Programming languages across primary studies.}}
    \label{fig:language}
\end{figure}

%% file: Tables/Prompt.tex
\begin{table*}[ht!]
  \centering
	 \caption{\textcolor{black}{Prompt technique used in LLM-driven refactoring (RQ$_3$).}}
	 \label{Table:prompt_in_llms_Refatoring}

\begin{adjustbox}{width=1.45\textwidth,center}
\rowcolors{2}{white}{gray!25}
\begin{tabular}{llLLlllll}\hline
\toprule
\bfseries Study & \bfseries Prompt Technique &  \bfseries \color{black} Best Approach \color{black} &  \bfseries Final Prompt     \\
\midrule
Gehring \cite{gehring2023deterministic} & Zero-Shot, Chain-of-Thought &  \color{black} Using ChatGPT-3.5 with Chain-of-Thought, resulting 16 out of 20 repository results being correctly refactored \color{black} & Repeat the code making sure that any default case is in the last case of a switch statement: [code] \\ 
White \etal \cite{white2024chatgpt} & Context-Specific, Chain-of-Thought, Output Constraints, Few-Shot, One-Shot & \color{black}Using Context-specific prompting with Chain-of-Thought can best help improve LLM-generated refactoring\color{black} & Multiple prompts are featured (see paper for individual prompts) \\
Shirafuji \etal \cite{shirafuji2023refactoring} & Zero-Shot, One-Shot, Few-Shot & \color{black}Using Few-Shot prompting leads to a 95.68\% of successful refactoring\color{black} & "Refactor the given Python program to a more readable, efficient, and maintainable one. You can assume that the given program is semantically correct. Do not change the external behavior of the program, and keep the syntactic and semantic correctness. Python programs should be in a code block. Do not explain anything in natural language."" provide snippets of code and ask for variety. \\
 Depalma \etal \cite{depalma2024exploring} & Zero-Shot & \color{black}Using Few-Shot prompting yielded best results from ChatGPT\color{black} &  \say{With no explanation refactor the Java code to improve its quality and [quality attribute]:}  \\
 & & \color{black}\color{black}  & \say{Does this Java code preserve the behavior of the above code segment? Explain in detail why or why not:} \\
 & & \color{black}\color{black}  & \say{Provide concise commit messages that describe the overall goal, refactoring changes, and impact of quality of the following code: [include refactored code here].} \\
 AlOmar \etal \cite{alomar2024refactor} & Zero-Shot, One-Shot, Few-Shot & \color{black}Using Few-Shot prompting along with context-specific prompting yielded best results \color{black} & No Final Prompt Provided \\
 Chavan \etal \cite{chavan2024analyzing} & Context-Specific, Zero-Shot, Output Constraints, One-Shot & \color{black}No best approach was mentioned. The paper broke down how and what developers used LLMs for; there was two primary ways developers used ChatGPT for refactoring and it was via Chain-of-Thought prompting or One-Shot prompting\color{black} & No Final Prompt Provided \\
Pomian \etal \cite{pomian2024assist} & Context-Specific, Few-Shot, Ranking & \color{black}The tool EM-Assist uses Few-Shot prompting provided with Context Specifcs with and Rankings to recommend for best refactoring options\color{black} & No Final Prompt Provided \\
Pomian \etal \cite{pomian2024next} & Context-Specific, Few-Shot, Ranking & \color{black}The tool EM-Assist was able to successfully perform refactoring in 53.4\% of the cases by using Few-Shot, Context-Specific, and Rankings\color{black} & https://github.com/llm-refactoring/llm-refactoring-plugin/tree/main\#prompt-example
summary of the repo:
You are an expert software developer skilled in Java and Kotlin refactoring.
Your task is to identify and extract a method from the given long method below....
[Actual Code Here] \\
Dilhara \etal \cite{dilhara2024unprecedented} & Few-Shot, Chain-of-Thought & \color{black}Using Chain-of-Thought reasoning and few-shot prompting, the tool PyCraft was able provide input example and test cases that can identify help identify code change patterns successfully up to an overall success rate of 96.6\%\color{black} & Provide an snippet of the CPAT transformation example to set the context, then generate unit tests to verify correctness, set the temperature and iteration settings(T = 1.2, I = 5 for random; T = 0.5, 0.7, I = 3, I\_f = 5 for structured generation), check for structural intent, and the filter out the incorrect variants\\
Gao \etal \cite{gao2024context} & Chain-of-Thought & \color{black}Using Chain-of-Thought, the tool UTRefactor is capable of achieving an 89\% reduction of code smells, from 2,375 of the testing data to 265\color{black} & "Role Assignment: ""You are a Java testing expert, and now you need to refactor the test code which contains bad test code smells.""

Step-by-Step Reasoning Using CoT:

Understand the intent of the test.
Recognize the code smells present in the test code.
Comprehend the refactoring rules and measures.
Refactor the test code accordingly.
Provided Information:

Source code of the unit test.
Context of the unit test (focal class, method, related functions).
Detected smells and descriptions.
Refactoring steps in DSL."
 \\
Zhang \etal \cite{zhang2024refactoring} & Few-Shot, Context-Specific & \color{black}Using Few-Shot and Context-Specific, the authors were able to achieve over 90\% in improving python code idomization in accuracy and precision\color{black} & No Final Prompt Provided \\
Choi \etal \cite{choi2024iterative} & One-Shot, Few-Shot, Chain-of-Thought, Output Constraints, Context-Specific & \color{black}Using Chain-of-Thought, LLMs was able to break complex methods to multiple simple methods, which allows the reduction of Cyclomatic Complexity to be reduced up to 10.4\%.\color{black} & "```java
Target Method
```
Refactor the provided Java method to enhance its readability and maintainability. You can assume that the given method is functionally correct. Ensure that you do not alter the external behavior of the method, maintaining both syntactic and semantic correctness. Provide the Java method within a code block. Avoid using natural language explanations."
 \\
Wu \etal \cite{wu2024ismell} & Context-Specific, One-Shot & \color{black}The tool iSMELL uses both One-Shot and Context-Specific methods and apporaches to identify code smells, reaching an accuracy of 75.17\%.\color{black} & "<Instruction>: In computer programming, a code smell is any characteristic in the source code of a program that possibly indicates a deeper problem. I will now tell you the definition about <code smell>. Please read the definition and refactor the code according to the target to eliminate this smell. The definition of <code smell> is: <Definition> 
Here is an example for refactoring code which has <code smell> smell, 
Please read it carefully and learn to show only the most modified critical code after refactoring. You can replace function code with comments. 
[code]
Now based on the example, refactor the following code to eliminate the <code smell>. \\
Ishizue \etal \cite{ishizue2024improved} & Output Constraints, Context-Specific, One-Shot & \color{black}Using Output Constraints, Context-Specific, One-Shot methods combined with Refactory, the authors are able to achieve near 88.8\% - 99.5\% of correct refactoring.\color{black} & "This is a final prompt that will be provided if the previous prompt is unsuccessful(see paper pg. 571):..."
\\
Gautam \etal \cite{gautam2024refactorbench} & Few-Shot, Output Constraints, Context-Specific & \color{black}Using Few-Shot, Output Constraints, Context-Specific methods combined, the authors are able to reach 43.9\% of the testcases resolved.color{black} & The prompt used on the SWE-agent is very long(see paper pg.17-18), there are also additional prompts used to create the instruction set(see pg. 16) \\
Baumgartner \etal \cite{baumgartner2024ai} & Few-Shot, One-Shot, Zero-Shot & \color{black}The study showed that the Few-Shot method is best used to find areas of refactoring.\color{black} & No Final Prompt Provided
 \\
Cui \etal \cite{cui2024three} & Few-Shot, Output Constraints, Context-Specific & \color{black}Using Few-Shot, Output Constraints, and Context-Specific the authors were able to achieve an increase of 27.8\% in precision, 2.5\% in recall and 18.5\% in f1 compared to other leading tools.\color{black} & This is an example prompt: 
1. There are now seven Pre-conditions for move method refactoring(P1-P7):
2. Consider the following test move method candidates(if the candidate move method is related to the Superclass, the corresponding Superclass is given below, otherwise it is empty): [Code]
3. Please verify whether the candidate meet pre-conditions P1-P7 by extracting the precondition relevant entries from the candidate and presenting the verification results in tabular form.
(see paper for defnintions of P1-P7)"
 \\
Zhang \etal \cite{zhang2025move} & Zero-Shot, Context-Specific, Output Constraints & \color{black}By using Zero-Shot, Context-Specific, Output Constraints the tool MoveRec was able to improve the F1 from 9.4\% to 53.4\%.\color{black} & You are a software engineer who is proficient in refactoring. \\
& & \color{black}\color{black} & You will do the work of code summary. Please make a concise
summary based on the following code features. + [CODE]
\\
Ben Mrad \etal \cite{ben2024assessing} & Zero-Shot, Context-Specific, Output Constraints & \color{black}By using Zero-Shot, Context-Specific, Output Constraints the tool MoveRec was able to improve the F1 from 9.4\% to 53.4\%.\color{black} & No Final Prompt Provided
\\
Wang \etal \cite{wang2024hecs}  & Few-Shot, Context-Specific & \color{black}By using Few-Shot, Context-Specific, the tool HECS was able to produce a precision of 76.8\%, recall of 84.3\%, and F-1 of 80.4\%.\color{black} & No Final Prompt Provided\\
Gao \etal \cite{gao2024preference} & Context-specific, Few-Shot, Output Constraint & \color{black}By using Context-specific, Few-Shot, and Output Constraints, the tool RRG was able to increase similarity of code generated and referenced by up to 28\%.\color{black} & No Final Prompt Provided \\
Cui \etal \cite{cui2024one} & Few-Shot, Context-Specific & \color{black}Using Few-Shot, Context-Specific, HECS was able to achieve 76.8\% in precision, 84.3\% in recall, and 80.4\% in F1-Measure. \color{black} & The P1-P8 preconditions are defined in detail in the paper.         1. Pre-condition rules (P1-P8) as follows:
…
2. Consider the following test extract class candidate:
Source class:...
Extract class candidate: - field: A.f2 - method: A.m2()
Please verify whether the candidate meet pre-conditions P1-P8 by
extracting the pre-condition relevant entities from the candidate
and presenting the verification results in tabular form.
 \\
Zhang \etal \cite{zhang2024copilot}  & Few-Shot, Chain-of-Thought & \color{black}Using Few-Shot with Chain-of-Thought, Copilot Chat was able to achieve a successful refactoring rate of 87.1\% in a code base of 102 code smells.\color{black} & General Fix Prompt: Fix the problem in the selected code
Code Smell Fix Prompt: Fix the code smell in the selected code Specific Code Smell Fix Prompt: Fix the [code smell name] (e.g., Long Method) code smell in the selected code
 \\
 Menolli \etal \cite{menolli2024teaching} & Context-Specific Prompting, Zero-Shot Learning & \color{black}No best prompt was mentioned, but by using Context-Specific prompting, the participants were able to use ChatGPT to learn programming and successfully refactored 86\% of the code smells.\color{black} &  No Final Prompt Provided
 \\
Wang \etal \cite{wang2025testing} & Chain-of-Thought, Few-Shot, Context-Specific, Output Constraints & \color{black}Using Chain-of-Thought, Few-Shot, Context-Specific and Output Constraints, RETester was able to identify new bugs in codebases that was never found before.\color{black} & Now, I will give the definition of the current refactoring, you need to understand it. You need to make sure the original refactoring could still be applied on the variant. 1. {Refactoring Type}: {Definition} 2. To expose more bugs in the refactoring engines, please generate edge case variant considering the {Characteristic} in current refactoring scenario. You need to generate the variant according to the Input Program Structure Template, it is {Template}. 3. You should give me the variant, the program elements to be refactored, and the procedures to refactoring. 4. The generated variant should not contain any syntax errors. The Java program you generated should conformance with the JDK {Version} standard. Please generate one edge case variant considering different edge usage scenarios of {Characteristic} based on the template. The variant format should be {Format}. \\
Midolo \etal \cite{midolo2025automated} & Context-Specific, Output Constraints, Zero-Shot & \color{black}Using Context-Specific, Output Constraints, and Zero-Shot, the authors was able to produce 90.7\% accurate refactoring suggestions with GPT-4.\color{black} & System: ”You are a software developer, skilled in writing
Python source code and refactoring Python code using
[pythonic idiom]”
User: ”Refactor the code using the [pythonic idiom] idiom, and provide the refactored Python code along with the
number of [pythonic idiom] refactoring you have made:
[code]”
 \\
Ksontini \etal \cite{ksontini2025refactoring} & Zero-Shot, One-Shot, Few-Shot & \color{black}Using Few-Shot (specifically 50-shot) the authors were able to achieve a success build rate of 63\% with refactoring dockerfiles.\color{black} & 1. In context with Few-Shot 2. Prompt components mentioned in paper 3. Demonstration selection \\
Alturayeif \& Hassine \cite{alturayeif2025refactoring} & Context-Specific, Few-Shot, Output Constraints, Chain-of-Thought & \color{black}Using Context-Specific, Few-Shot, Output Constraints and Chain-of-Thought, the authors were able to detect and refactor linguistic bad smells.\color{black} & No Final Prompt Provided \\
Divyansh \etal \cite{icmi2025evaluating} & Few-Shot & \color{black}By using Few-Shot, authors was able to reduce code debt significantly.\color{black} & No Final Prompt Provided\\
\textcolor{black}{Guo \etal \cite{guo2024exploring}} & \textcolor{black}{Chain-of-Thought, Context-Specific, Zero-Shot} & \textcolor{black}{By using Context-Specific prompting in Prompt 2, the authors were able to achieve a 83.58\% similarity with human refactored code.} & \textcolor{black}{Prompt 1 (P1): the simplest prompt.}  \\
& & & \textcolor{black}{Prompt 2 (P2): P1 + Scenario Description.} \\
& & & \textcolor{black}{Prompt 3 (P3): P1 + Detailed Requirements.} \\
& & & \textcolor{black}{Prompt 4 (P4): P1 + Concise Requirements.} \\
& & & \textcolor{black}{Prompt 5 (P5): P4 + Scenario Description.} \\
& & & \textcolor{black}{Examples:
Prompt 1 (P1):
[INST]
code snippet:
```original code```
code review:
review comment
Please generate the revised code according to the review. 
In your response, put the revised code between triple backticks and avoid mentioning the programming language between the backticks.
[/INST]
Prompt 2 (P2): P1 + Scenario Description
[INST]
As a developer, imagine you've submitted a pull request 
and your team leader requests you to make a change to a 
piece of code. The original code being referred to in the 
hunk of code changes is: 
```original code```
There is the code review for this code:
review comment
Please generate the revised code according to the review. In your response, put the revised code between triple backticks and avoid mentioning the programming language between the backticks.
[/INST]}\\
Dong \etal \cite{dong2025chatgpt} & Context-Specific & \color{black}Using Context-Specific prompting with ChatGPT, the authors was able to identify previously unknown refactoring bugs.\color{black} & For Testing: 
Description:
"Please generate ten java test programs that test the move method refactoring and meet the following constraints"

Conditions:
"The original class of the method to be moved contains a field of the target class type."
"The name of the method to be moved is ‘methodToBeMoved’."
"The target class name for the method to be moved is ‘TargetClass’."

Features:
"The method to be moved calls a private field within the original class."
"The modifier of the method to be moved is ‘protected’."
"The target class of the method to be moved is an anonymous class."

For feature extraction using ChatGPT: 
You are an expert in testing refactoring engines, and your task is to summarize the kind of features a test Java program should have so that it could trigger the same bug in [refactoring type] as the following test program triggers.
Notably, 'features' refer to code entities with specific characters as well as relationships among such entities.
The given test program that triggers a bug in [refactoring type]: [source code]
The error messages of [refactoring type]: [error message]
To make it clear, we present the following three examples to illustrate how to extract features that cause bugs during [refactoring type] process.
Example 1
Test Program: "[source code]"
Error Message: "[error message]"
Summarized Features: "[features]"
Example 2
Test Program: "[source code]"
Error Message: "[error message]"
Summarized Features: "[features]"
Example 3
Test Program: "[source code]"
Error Message: "[error message]"
\\
Ghammam \etal \cite{ghammam2025build} & Zero-Shot, One-Shot & \color{black}Using One-Shot, BuildRefMiner was able to obtain a F1 score of 76\% across refactoring build systems.\color{black} & Task: Given the commit changes below that are applied on {build system} build scripts which will be delimited with {delimiter} characters, identify any occurrences of the listed refactoring types. Provide the results strictly in JSON format with the following keys: RefactoringType and Details. If there are multiple refactorings, return them as a list of JSON objects, where each object contains the following: - "RefactoringType": The type of refactoring detected. - "Details": A Description with further information about the change. Here are two examples of the format I expect: .... If no refactorings are detected, return the message: "No refactorings found." This is a list of 24 refactoring types in build files: ... {commit} \\
\bottomrule
\end{tabular}
\end{adjustbox}

\vspace{-.3cm}
\end{table*}

\begin{table*}[ht!]
  \centering
	 \caption{\textcolor{black}{Prompt technique used in LLM-driven refactoring (RQ$_3$) - Cont'd.}}
	 \label{Table:prompt_in_llms_Refatoring-continue}

\begin{adjustbox}{width=1.5\textwidth,center}
\rowcolors{2}{white}{gray!25}
\begin{tabular}{llLlllllll}\hline
\toprule
\bfseries Study & \bfseries Prompt Technique &  \bfseries Best Approach &  \bfseries Final Prompt     \\
\midrule 
\textcolor{black}{Batole \etal \cite{batole2025leveraging}} & \textcolor{black}{Chain-of-Thought} & \textcolor{black}{Using Chain-of-Thought, MM-Assist was able to achieve a Recall@3 of 80\%, a 2.4x improvement compare to the standard state of the art tools.} & \textcolor{black}{No Final Prompt Provided} \\
\textcolor{black}{Palit \& Sharma \cite{palit2025reinforcement}} & \textcolor{black}{N/A} & \textcolor{black}{The paper didn’t talk about best prompting methods but by using Supervised Fine-Tuning and Reinforcement Learning the Code-T5 model was able to increase unit tests passes from 41 to 66.} & \textcolor{black}{No Final Prompt Provided} \\
\textcolor{black}{Rajendran \etal \cite{rajendran2025multi}} & \textcolor{black}{N/A} & \textcolor{black}{The paper does not mention a best prompting approach, rather, it mentions different usages of Agent to best refactor, breaking down tasks for Specialized agents.} &\textcolor{black}{No Final Prompt Provided} \\
\textcolor{black}{Kim \cite{kim2025comparative}} & \textcolor{black}{Zero-Shot} & \textcolor{black}{The paper identified Claude combined with Context-Specific Prompts as the most effective LLM with the highest PIQS of 89.51 followed by Meta’s 88.98.} & \textcolor{black}{Prompt 1: Identify applicable design patterns for the given program. Prompt 2: Apply [design pattern] to} \\ & & & \textcolor{black}{[location] in the given program. Prompt 3: Can [design pattern] be applicable to [location]?}  \\
\textcolor{black}{Wang \etal \cite{wang2025towards}} & \textcolor{black}{Few-Shot} & \textcolor{black}{Using Few-shot and Context-Specific prompting with GOT-2o-mini, the authors was able to reach 98\% precision for refactoring identification, and 80\% for refactoring.} & \textcolor{black}{You are a software testing expert. I will give you some historical bug reports for the refactoring engines.}\\ & & & \textcolor{black}{You need to extract the following information from the bug reports: 1. Refactoring type; 2. Bug symptom;} \\ & & & \textcolor{black}{3. Input program characteristic. The following are examples: {Example}. The extracted information format}\\ & & & \textcolor{black}{should be: {Format}.}\\
\textcolor{black}{Wang \etal \cite{wang2025exploration}} & \textcolor{black}{Few-Shot} & \textcolor{black}{The paper did not talk about best prompting methods, but by using Ensemble Learning the paper was able achieve higher EM.} & \textcolor{black}{Multiple prompts are featured (see paper for individual prompts)}\\
\textcolor{black}{AlOmar \etal \cite{alomar2025chatgpt}} & \textcolor{black}{Few-Shot, One-Shot, Zero-Shot, Context-Specific, Output Constraints, Chain-of-Thought} & \textcolor{black}{By using context-specific, few-shot, and output-constrained prompting the authors was able to reduce prompting turns from 13.58 average to 1.45.} & \textcolor{black}{You are an expert in programming, refactoring, and providing advice in software quality. Given the following} \\ & & & \textcolor{black}{ prompt/code snippet, modify/write the code so that it best matches what the user wants, and provide} \\ & & & \textcolor{black}{comments explaining your work. If needed, ask clarifying questions to check both your and the user’s} \\ & & & \textcolor{black}{understanding of the prompt. Working Set: [input code]
Context: ...
Refactoring Task: ...
Steps to Follow:} \\ & & & \textcolor{black}{1) Analyze the Code... 2) Apply Refactoring Strategies ... 3) Provide Multiple Solutions... 4) Validate} \\ & & & \textcolor{black}{Functionality... 5) Comment and Document... Output Format: ...
Example start: 
[example]
Example end}\\ & & & \textcolor{black}{If there are any unclear instructions, please ask clarifying questions.} \\
\textcolor{black}{Midolo \etal \cite{midolo2025refactoring}} & \textcolor{black}{One-Shot} & \textcolor{black}{Using Context-Specific Prompting with GPT-3.5-Turbo, the authors was able to refactor 2101 out of 2132 java for-loops into stream pipelines.} & \textcolor{black}{System: ‘You are a software developer, skilled in writing Java source code and refactoring for loops into stream-based code’} \\ & & & \textcolor{black}{User: ‘Please refactor this code to use streams, and provide only the Java code without any markdown, explanation, or comments: *code*’} \\
\textcolor{black}{Pucho \etal \cite{pucho2025refactoring}} & \textcolor{black}{Zero-Shot} & \textcolor{black}{Using Context-Specific and multi-llm agents with MetaGPT the authors was able to reduce cyclomatic complexity by 26\%.} & \textcolor{black}{No Final Prompt Provided} \\
\textcolor{black}{Caumartin \etal \cite{caumartin2025exploring}} & \textcolor{black}{Chain-of-Thought, Context-Specific, Zero-Shot} & \textcolor{black}{By using Context-Specific and Zero-Shot prompting the authors are able to improve CodeLlama to achieve comparable performance to ChatGPT (77.13 vs 76.44 BLEU-T)} & \textcolor{black}{Prompt 1 (P1): the simplest prompt. 
Prompt 2 (P2): P1 + Scenario Description. 
Prompt 3 (P3): P1 + Detailed Requirements.}\\ & & & \textcolor{black}{Prompt 4 (P4): P1 + Concise Requirements. 
Prompt 5 (P5): P4 + Scenario Description.} \\ & & & \textcolor{black}{Examples:
Prompt 1 (P1):
[INST]
code snippet:
```original code```
code review:
review comment
Please generate the revised code according to the review.} \\ & & & \textcolor{black}{In your response, put the revised code between triple backticks and avoid mentioning the programming language between the backticks.
[/INST]} \\ & & & \textcolor{black}{Prompt 2 (P2): P1 + Scenario Description
[INST]
As a developer, imagine you've submitted a pull request 
and your team leader requests you to make a change to a 
piece of code.} \\ & & & \textcolor{black}{The original code being referred to in the hunk of code changes is: 
```original code```} \\ & & & \textcolor{black}{There is the code review for this code:
review comment
Please generate the revised code according to the review.} \\ & & & \textcolor{black}{In your response, put the revised code between triple backticks and avoid mentioning the programming language between the backticks.
[/INST]} \\
\textcolor{black}{Pandini \etal \cite{pandini2025exploratory}} & \textcolor{black}{Chain-of-Thought, Context-Specific} & \textcolor{black}{Using Context-Specific prompting with Claude 3.5 Sonnet the authors was able to reach up to 75\% result preference from study praticipents.}  & \textcolor{black}{Multiple prompts are featured (see paper for individual prompts)} \\
\textcolor{black}{Guo \etal \cite{guo2025intention}} & \textcolor{black}{Chain-of-Thought, Context-Specific, Few-Shot, Zero-Shot} & \textcolor{black}{Using Few-Shot prompting the authors was able to achieve 79\% accuracy in intention extraction and 66\% in code refactoring.} & \textcolor{black}{Multiple prompts are featured (see paper for individual prompts)} \\
\textcolor{black}{Xu \etal \cite{xu2025hlsrewriter}} & \textcolor{black}{Chain-of-Thought, Context-Specific, Few-Shot, Output Constraints} & \textcolor{black}{Using Chain-of-Thought prompting, HLSRewriter was able to achieve an 86.67\% refactoring pass rate.} & \textcolor{black}{Multiple prompts are featured (see paper for individual prompts)} \\
\textcolor{black}{Tkackuk \cite{tkachuk2025determining}} & \textcolor{black}{Zero-Shot} & \textcolor{black}{Using Context-Specific prompting with Claude, the authors was able to reach the highest success rate of refactoring by 78.8\% which ChatGPT followed with 76.6\%.} & \textcolor{black}{No Final Prompt Provided} \\
\textcolor{black}{Guenoune \etal \cite{guenoune2025llm}} & \textcolor{black}{Chain-of-Thought, Context-Specific, Few-Shot} & \textcolor{black}{Using Few-Shot and Chain-of-Thought with Gemini the authors was able to reach a highest score of 98.75\% in matching with human naming.} & \textcolor{black}{No Final Prompt, but the steps are described: 
Step 1: Instructs the LLM and check its understanding of the RCS + dot syntax
Step 2: Ask LLM its criteria about class relevance.} \\ & & & \textcolor{black}{Step 3: Ask LLM evaluate each concept (explanation + score) and name the candidate class
Step 4: Asks the LLM to format the result} \\
\textcolor{black}{Amaral \cite{amaral2025improving}} & \textcolor{black}{Zero-Shot} & \textcolor{black}{Using Zero-Shot and Constext-Specific prompting, Copilot (ChatGPT-4o) was able to achieve up to 58.78\% successful code smell removals.} & \textcolor{black}{Context:I'm refactoring test smells from a test file to improve code quality.} \\ & & & \textcolor{black}{Issue Details:
Smell Category: {Test smell category}
Smell Location:- Line Range: startLine: {startLine}, endLine: {endLine} 
or- Line: {line}, Column: {column}, Index: {index}} \\ & & & \textcolor{black}{Description:
{Detailed test smell description}
Request:
Refactor the affected code to eliminate the {Test smell category}.
Ensure the test remains correct, readable, and maintainable.} \\
\textcolor{black}{De Sousa \etal \cite{de2025effectiveness}} & \textcolor{black}{Chain-of-Thought, Output Constraints} & \textcolor{black}{Using Context-Specific prompting with GPT-4o, the authors was able to achieve a F1-scores of 79\% for detecting PL/SQL bad smell.} & \textcolor{black}{You are an AI assistant specialized in PL/SQL code analysis and linting. 
Consider the PL/SQL code smells listed below:} \\ & & & \textcolor{black}{Unused local variable or unused parameter.
Call to procedure from package DBMS\_OUTPUT.
Use of an IF statement instead of EXIT WHEN to exit from a loop.} \\ & & & \textcolor{black}{Empty code block (BEGIN NULL; END;).
Parameter mode omission (parameter declaration without explicit
IN, OUT, or IN OUT).} \\ & & & \textcolor{black}{Use of SELECT * (projection of all columns).
COMMIT or ROLLBACK in a non-autonomous transaction.} \\ & & & \textcolor{black}{SELECT INTO statement without an associated exception handler covering potential query exceptions (e.g. NO\_DATA\_FOUND, TOO\_MANY\_ROWS).} \\ & & & \textcolor{black}{Analyze the provided PL/SQL code snippet and check for those code smells. If any smells are detected, present them as a list where each item matches this format:} \\ & & & \textcolor{black}{- Location: line(s) where the smell occurs
- Issue: smell number from the list above} \\ & & & \textcolor{black}{If none of the specified smells are identified, respond with this sentence: 
“No smell detected.” 
Here is the PL/SQL code snippet: [code]} \\
\bottomrule
\end{tabular}
\end{adjustbox}

\vspace{-.3cm}
\end{table*}

\begin{figure}
\centering 

\centering

\begin{tikzpicture}
\begin{axis}[
    ybar,
    bar width=20pt,
    enlargelimits=0.15,
    ylabel={Number of Papers},
    symbolic x coords={Zero-Shot, One-Shot, Few-Shot, Chain-of-Thought, Context Specific, Output Constraints, Ranking, Not Specified},
    xtick=data,
    nodes near coords,
    x tick label style={rotate=90, anchor=east},
    ymin=0, ymax=45,
    legend style={at={(0.5,-0.15)}, anchor=north, legend columns=-1}
]
\addplot[fill=blue] coordinates {
    (Zero-Shot, 19) 
    (One-Shot, 12) 
    (Few-Shot, 25) 
    (Chain-of-Thought, 18) 
    (Context Specific, 27) 
    (Output Constraints, 15) 
    (Ranking, 2) 
    (Not Specified, 2)
};

\end{axis}
 \end{tikzpicture}
 \caption{\textcolor{black}{Prompt engineering techniques across primary studies.}} \label{prompt-techniques}
\end{figure}
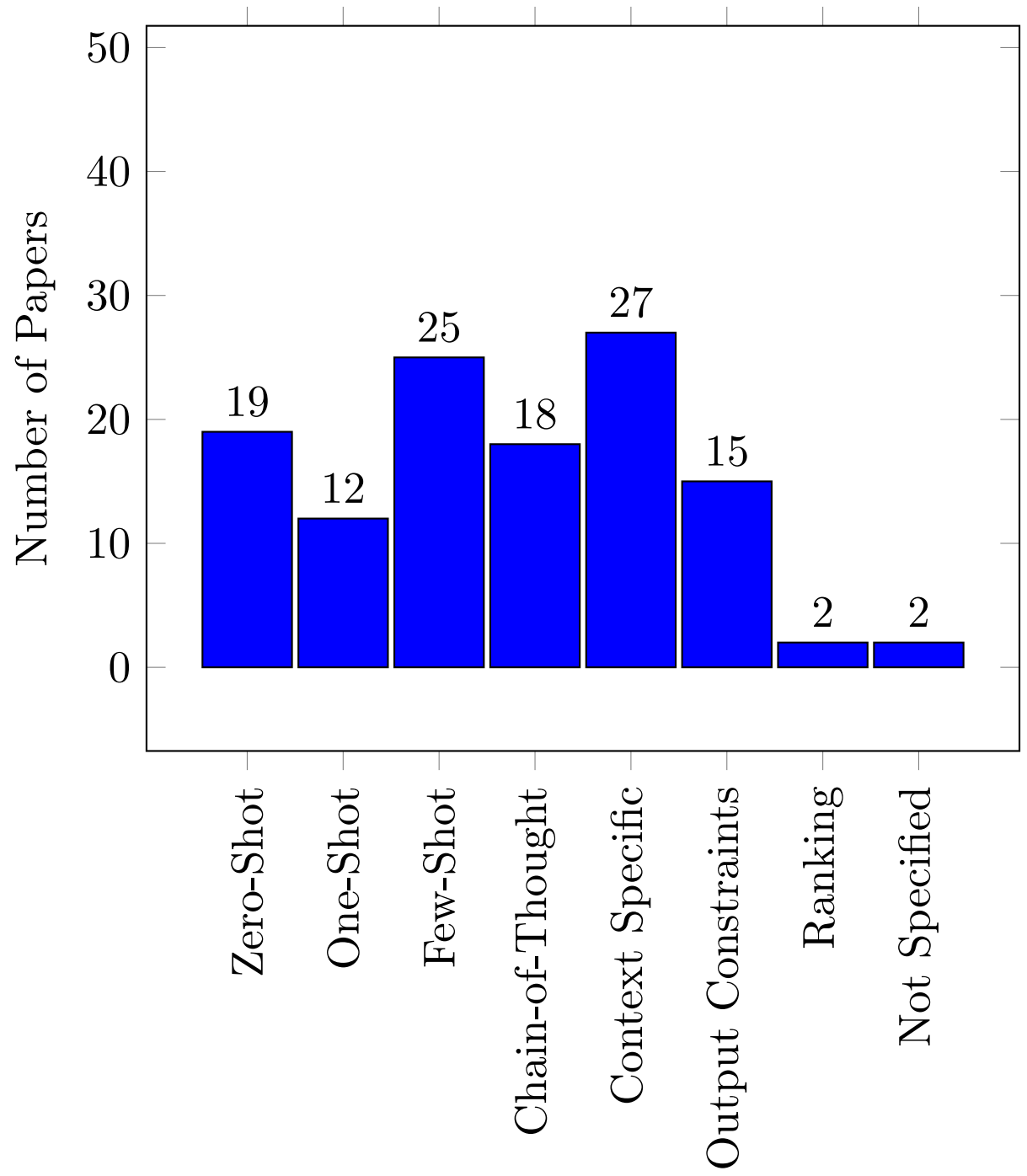%

%% file: Tables/Analysis.tex
\begin{table*}[ht!]
  \centering
	 \caption{\textcolor{black}{Quantitative, qualitative, and comparative analysis of LLM-driven refactoring (RQ$_4$).}}
	 \label{Table:Analysis_in_llms_Refatoring}
\begin{adjustbox}{width=1.0\textwidth,center}
\rowcolors{2}{white}{gray!25}
\begin{tabular}{lcllllllll}\hline
\toprule
\bfseries Study & \bfseries Quantitative &  \bfseries Qualitative &   \bfseries Comparative &  \bfseries Correctness     \\
\midrule
Gehring \cite{gehring2023deterministic} & 1 project & No & w/ ChatGPT3.5 and  StarChat-$\beta$ & Accuracy: 100\%\\ 
White \etal \cite{white2024chatgpt} & 14 prompt patterns  & No & No & Unknown \\
Shirafuji \etal \cite{shirafuji2023refactoring} & 880 programs & No & No & Unknown\\
 Depalma \etal \cite{depalma2024exploring} & 40 files & w/ 15 students & No & Accuracy: 96.8\% \\
  AlOmar \etal \cite{alomar2024refactor} & 470 commits, 69 issues, 176 files & No & No & Unknown \\
 Chavan \etal \cite{chavan2024analyzing} & 447 conversations & No & No & Unknown\\
Pomian \etal \cite{pomian2024assist} & 1756 refactorings & w/ 18 developers & w/ JExtract & Unknown\\
Pomian \etal \cite{pomian2024next} & 122 methods & w/ 16 developers & w/ JDeodorant, JExtract, REMS, & Unknown\\
& & & GEMS, SEMI, and LiveRef\\
Dilhara \etal \cite{dilhara2024unprecedented} & 9325 code change patterns & No & w/ PyEvolve & F1-measure: 96.6\% \\
Gao \etal \cite{gao2024context} & 6 projects & No & w/ TESTAXE, Llama3-70B, & Unknown\\
& & & and GPT-4o & \\
 Zhang \etal \cite{zhang2024refactoring} & 3,311 methods & No & w/ RIdiom and Prompt-LLM & Precision: increase to 100\%,\\
 & & & & decrease to 93.8\%\\
 & & & & Accuracy: increase to 87.8\%-100\% \\
 & & & & Recall: increase to 87.8\%-100\% \\
 & & & & F1-measure: increase to 92.5\%-100\% \\
Choi \etal \cite{choi2024iterative} & 17 projects & No & No & Unknown \\
Wu \etal \cite{wu2024ismell} & 3 datasets & No & w/ PMD, JDeodorant, Organic, & Precision: decrease to 82.13\%\\
& & & DesigniteJava, JSPIRIT, & Accuracy: increase to 95.14\% \\
& & & FeTruth, and JMove & Recall: increase to 92.41\% \\
& & & & F1-measure: increase to 95.52\% \\  
Ishizue \etal \cite{ishizue2024improved} & 1,800 methods & No & w/ ChatGPT & Unknown\\
Gautam \etal \cite{gautam2024refactorbench} & 9 projects & No & w/ manual refactoring & Unknown \\
Baumgartner \etal \cite{baumgartner2024ai} & 1 project & No & No & Recall(Median): 48\% \\
Cui \etal \cite{cui2024three} & 11169 projects & w/ 50 software engineers & w/ LLMRefactor, FeTruth, FeDeep, & Precision(AVG): increase to 93.3\%\\
& & & PathMove, JDeodorant,
JMove, & Recall(AVG): 85.4\%\\
& & & RMove, FeGNN, and FePM & F1-measure(AVG): 80.7\%\\
Zhang \etal \cite{zhang2025move} & 58 projects & No & w/ PathMove, JDeodorant, JMove, & Precision: increase to 60.6\%-91.4\%\\
& & & and RMove & Recall: decrease to 58.8\%-91\%\\
& & & & F1-measure: 74\%-91\%\\
Ben Mrad \etal \cite{ben2024assessing} & 116 methods & No & w/ Mistral-7B and Llama 2-7B & Unknown \\
Wang \etal \cite{wang2024hecs} & 11,336 projects & w/ 50 software engineers & w/ JDeodorant, SSECS, & Precision(AVG): increase to 76.8\%\\
& & & and LLMRefactor & Recall(AVG): increase to 84.3\%\\
& & & & F1-measure(AVG): increase to 80.4\%\\
Gao \etal \cite{gao2024preference} & 30,000 code samples & No & w/ GPT-2, CodeGPT, CodeGEN, & Unknown \\
& & &  and PolyCoder & \\
Cui \etal \cite{cui2024one}& 11,336 projects & w/ 50 software engineers & w/ JDeodorant, SSECS, & Precision(AVG): increase to 76.8\%\\
& & & and LLMRefactor & Recall(AVG): increase to 84.3\%\\
& & & & F1-measure(AVG): increase to 80.4\%\\
Zhang \etal \cite{zhang2024copilot} & 1,204 files & No & No & Unknown\\
Menolli \etal \cite{menolli2024teaching} & 4 projects & w/ 23 students & No & Unknown\\
Wang \etal \cite{wang2025testing} & 458 bug reports & No & Unknown & Unknown\\
Midolo \etal \cite{midolo2025automated} & 736 projects & No & w/ SOTA benchmark & Unknown \\
Ksontini \etal \cite{ksontini2025refactoring} & 600 dockerfiles & No & w/ PARFUM and manual refactoring & Unknown\\
Alturayeif \& Hassine \cite{alturayeif2025refactoring} & 17 linguistic smells & w/ 2 professors & w/ manual refactoring & Precision(AVG): 57.5\%\\
& & and 11 graduate students & & Recall(AVG): 100\% \\
& & & & F1-measure(AVG): 74\% \\
Divyansh \etal \cite{icmi2025evaluating} & 1,032 files & No & No & Unknown \\
Liu \etal \cite{liu2024empirical} & 20 projects & w/ 3 human developers & w/ manual refactoring & Unknown\\
\textcolor{black}{Guo \etal \cite{guo2024exploring}} & \textcolor{black}{467 projects} & \textcolor{black}{No} & \textcolor{black}{w/CodeReviewer} & \textcolor{black}{Unknown} \\
Dong \etal \cite{dong2025chatgpt} & 700 bug reports and 185 test cases & w/ original developers & w/ manual refactoring & Precision: 92.9\%\\
Ghammam \etal \cite{ghammam2025build} & 609 projects & w/ 60 developers & w/ prompting technique & Precision: 79\% \\ & & & & Recall: 78\% \\ & & & & F1-measure: 76\% \\
\textcolor{black}{Batole \etal \cite{batole2025leveraging}} & \textcolor{black}{10 projects} & \textcolor{black}{w/30 grad students} & \textcolor{black}{w/JMove, FeTruth, HMove,} & \textcolor{black}{Recall: 80\%} \\
& & & \textcolor{black}{and GPT-4o} & &\\
\textcolor{black}{Palit \& Sharma \cite{palit2025reinforcement}} & \textcolor{black}{1,618 projects} & \textcolor{black}{No} & \textcolor{black}{w/Code-T5, PLBART, CodeGPT-adapt} & \textcolor{black}{Unknown} \\
& & & \textcolor{black}{and CodeGen} & &\\
\textcolor{black}{Rajendran \etal \cite{rajendran2025multi}} & \textcolor{black}{4 agents} & \textcolor{black}{No} & \textcolor{black}{w/GPT-4, SonarQube, ESLint} & \textcolor{black}{Unknown} \\
\textcolor{black}{Kim \cite{kim2025comparative}} & \textcolor{black}{2 projects} & \textcolor{black}{No} & \textcolor{black}{w/ChatGPT-4o, Claude 3.5 Sonnet, Microsoft 365 Copilot,} & \textcolor{black}{PIQS(AVG): 89.51\%} \\
& & & \textcolor{black}{Gemini 1.5, Llama 3.1} & &\\
\textcolor{black}{Wang \etal \cite{wang2025towards}} & \textcolor{black}{1,651 bug reports} & \textcolor{black}{No} & \textcolor{black}{w/SAFEREFACTOR and ASTGen} & \textcolor{black}{Unknown} \\
& & & \textcolor{black}{CodeT5, and Unixcoder} & \textcolor{black}{Precision(AVG): 96.40\%}\\
& & & & \textcolor{black}{Recall(AVG): 93.38\%}\\
& & & & \textcolor{black}{F1-measure: 94.83\%}\\
\textcolor{black}{Wang \etal \cite{wang2025exploration}} & \textcolor{black}{800 samples} & \textcolor{black}{No} & \textcolor{black}{w/CodeT5, CodeReviewer, T5-Review,} & \textcolor{black}{Unknown} \\
& & & \textcolor{black}{and CodeBERT} &\\
\textcolor{black}{AlOmar \etal \cite{alomar2025chatgpt}} & \textcolor{black}{470 commits, 69 issues, and 176 files} & \textcolor{black}{No} & \textcolor{black}{No} & \textcolor{black}{Unknown} \\
\textcolor{black}{Midolo \etal \cite{midolo2025refactoring}} & \textcolor{black}{4 projects} & \textcolor{black}{No} & \textcolor{black}{w/previous SOTA approaches} & \textcolor{black}{Unknown} \\
\textcolor{black}{Pucho \etal \cite{pucho2025refactoring}} & \textcolor{black}{2,000 files} & \textcolor{black}{No} & \textcolor{black}{w/manual refactoring} & \textcolor{black}{Unknown} \\
\textcolor{black}{Caumartin \etal \cite{caumartin2025exploring}} & \textcolor{black}{467 projects} & \textcolor{black}{No} & \textcolor{black}{w/CodeLlama, Llama 2, Llama 3.1} & \textcolor{black}{BLEU-T: 77.13\%} \\
\textcolor{black}{Pandini \etal \cite{pandini2025exploratory}} & \textcolor{black}{5 projects} & \textcolor{black}{w/10 students} & \textcolor{black}{w/Qwen2.5-Coder, Claude 3.5 Sonnet} & \textcolor{black}{Unknown} \\
\textcolor{black}{Guo \etal \cite{guo2025intention}} & \textcolor{black}{15 projects} & \textcolor{black}{No} & \textcolor{black}{w/CodeReviewer and T5CR} & \textcolor{black}{Accuracy: 78.61\%} \\
\textcolor{black}{Xu \etal \cite{xu2025hlsrewriter}} & \textcolor{black}{74 files} & \textcolor{black}{No} & \textcolor{black}{w/GPT-4 Turbo} & \textcolor{black}{Accuracy:  86.67\%}\\
\textcolor{black}{Tkackuk \cite{tkachuk2025determining}} & \textcolor{black}{4 code snippets} & \textcolor{black}{No} & \textcolor{black}{No} & \textcolor{black}{Accuracy: 78.8\%} \\
\textcolor{black}{Guenoune \etal \cite{guenoune2025llm}} & \textcolor{black}{2 projects} & \textcolor{black}{No} & \textcolor{black}{w/Gemini, ChatGPT, Claude,} & \textcolor{black}{Accuracy: 98.75\%}\\
& & &  \textcolor{black}{DeepSeek and manual refactoring} & &\\
\textcolor{black}{Amaral \cite{amaral2025improving}} & \textcolor{black}{10 projects} & \textcolor{black}{No} & \textcolor{black}{w/SNUTS.JS and Steel} & \textcolor{black}{Accuracy: 58.78\%} \\
\textcolor{black}{De Sousa \etal \cite{de2025effectiveness}} & \textcolor{black}{44 code snippets} & \textcolor{black}{No} & \textcolor{black}{w/GPT-4o, Gemini 2.0, GPT-4o mini,} & \textcolor{black}{Precision: 70\%} \\
& & & \textcolor{black}{and Phi-4} & \textcolor{black}{Recall: 93\%} &\\
& & & & \textcolor{black}{F1-measure: 79\%} &\\
\bottomrule
\end{tabular}
\end{adjustbox}
\vspace{-.3cm}
\end{table*}

\pgfplotsset{compat=1.18}
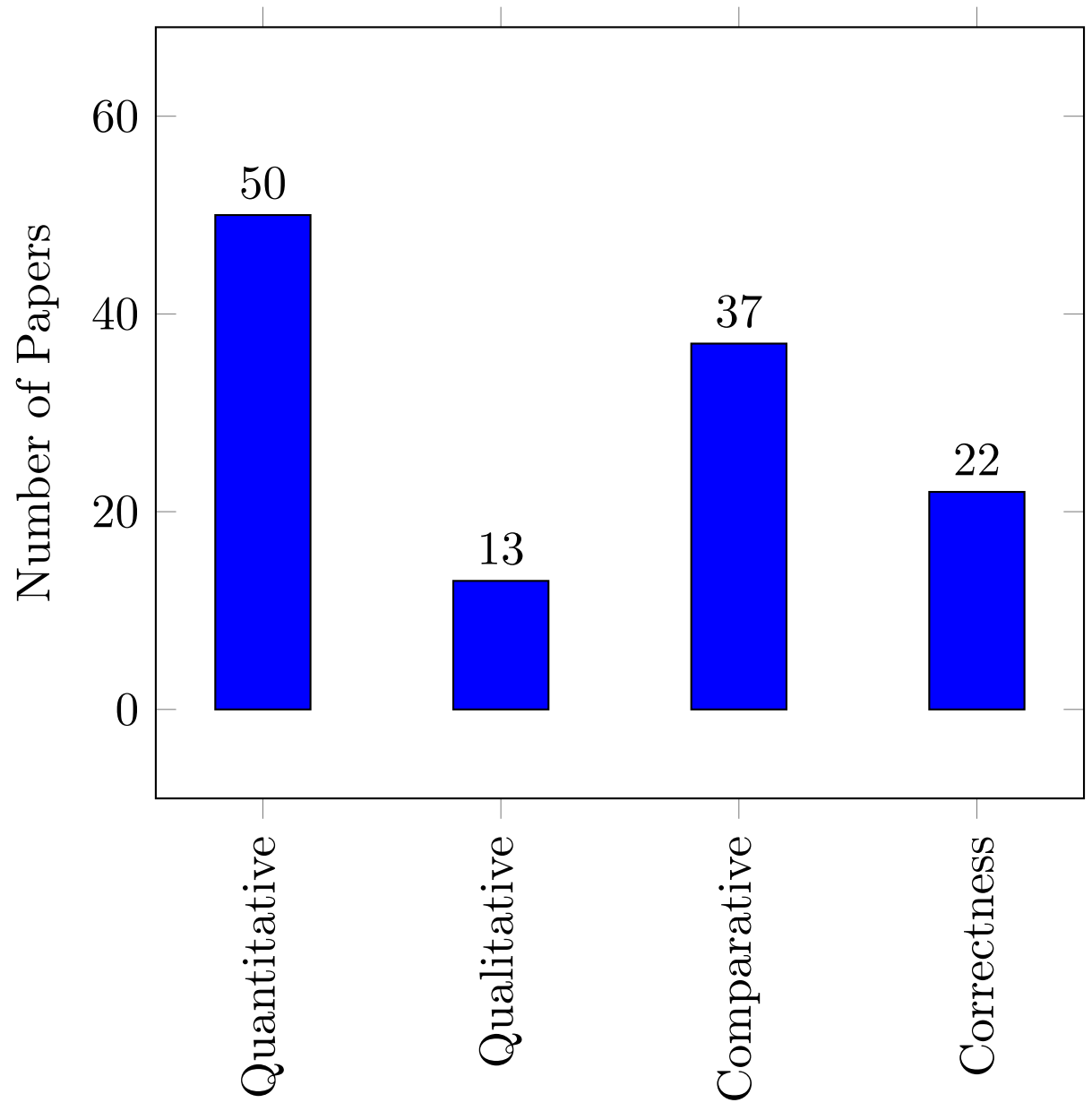
\begin{figure}
\centering
\begin{tikzpicture}
\begin{axis}[
    ybar,
    bar width=20pt,
    enlargelimits=0.15,
    ylabel={Number of Papers},
    symbolic x coords={Quantitative, Qualitative, Comparative, Correctness},
    xtick=data,
    nodes near coords,
    x tick label style={rotate=90, anchor=east},
    ymin=0, ymax=60,
    legend style={at={(0.5,-0.15)}, anchor=north, legend columns=-1}
]
\addplot[fill=blue] coordinates {
    (Quantitative,50) 
    (Qualitative,13) 
    (Comparative,37) 
    (Correctness,22)
};

\end{axis}
\end{tikzpicture} 
   \caption{\textcolor{black}{Quantitative, qualitative, and comparative analysis across primary studies.}}
    \label{fig:metrics}
\end{figure}

%% file: Tables/Code_Smells.tex
\color{black}
\begin{table*}[ht!]
  \centering
	 \caption{\textcolor{black}{Code Smell Definitions (RQ$_5$).}}
	 \label{Table:Code_Smells_Defintion}
\begin{adjustbox}{width=1.5\textwidth,center}
\rowcolors{2}{white}{gray!25}
\color{black}
\begin{tabular}{llllllllll}\hline
\toprule
\bfseries Code Smell & Related Papers & \bfseries Definition \\
\midrule
Code Duplication & \cite{shirafuji2023refactoring, depalma2024exploring, alomar2024refactor, icmi2025evaluating, liu2024empirical, ghammam2025build, ghammam2025build, batole2025leveraging, rajendran2025multi, caumartin2025exploring, tkachuk2025determining} & Identical code that has the same functionalities in different places of the code base. \\ 
Switch Statements & \cite{shirafuji2023refactoring, depalma2024exploring, zhang2024copilot, midolo2025automated, dong2025chatgpt} & Long sequence of if statements and switch operators. \\
Feature Envy & \cite{depalma2024exploring, wu2024ismell, cui2024three, zhang2025move, gao2024preference, midolo2025automated, batole2025leveraging} & When one class/method accesses another’s fields for computation too often. \\
Large Class & \cite{depalma2024exploring, wu2024ismell, zhang2025move, wang2024hecs, gao2024preference, cui2024one, zhang2024copilot, menolli2024teaching, icmi2025evaluating, liu2024empirical, dong2025chatgpt, ghammam2025build, batole2025leveraging, rajendran2025multi, pucho2025refactoring} & A class that contains too many lines of code, taking in too many fields or has too many methods. \\
Long Method & \cite{depalma2024exploring, alomar2024refactor, chavan2024analyzing, pomian2024assist, pomian2024next, choi2024iterative, zhang2025move, menolli2024teaching, icmi2025evaluating, liu2024empirical, ghammam2025build, dong2025chatgpt, rajendran2025multi, pucho2025refactoring} & A method that is way too long and confusing to understand. \\
Refused Bequest & \cite{wu2024ismell, gao2024preference} & When the child class inherits the parent class’s methods and properties but does not use it or is largely redefined for other purposes. \\
Data Clumps & \cite{baumgartner2024ai} & Repeated code with the same groups of variables. \\
Poor Naming & \cite{ben2024assessing, icmi2025evaluating, dong2025chatgpt, guo2024exploring, dong2025chatgpt, pucho2025refactoring, caumartin2025exploring} & When a method/class/object does not have an intuitive name. \\
Long Parameter List & \cite{zhang2024copilot, tkachuk2025determining} & When a method holds too many parameters. \\
Lazy Class & \cite{wang2025testing} & When a class is too small or does not do enough to be a class of its own. \\
Comments & \cite{icmi2025evaluating, pucho2025refactoring, caumartin2025exploring} & When a method/class has too many unclear comments. \\
Message Chain & \cite{batole2025leveraging} & When there are a long list of methods calling each other. \\
Lingustic smells & \cite{alturayeif2025refactoring} & Specialized code smell specificlly for language; such as prompts and comments. This was not mentioned in the refactoring guru. \\
Test code smells & \cite{gao2024context, amaral2025improving} & Specialized code smell specifically for test cases, not mentioned in refactoring guru. Defined by the paper: Eager Test, Duplicate Assert, Assertion Roulette. \\
Docker file code smells & \cite{ksontini2025refactoring} & Specialized code smell specifically for Dockerfiles, not mentioned in refactoring guru. Defined by the paper: Temporary File Smells, Shell/Script Smells. \\
Architectural code smells & \cite{pandini2025exploratory} & Specialized code smell specifically for code on an architectural level, not mentioned in refactoring guru. Defined by the paper: Cyclic Dependency, Unstable Dependency, God Component, Hublike Dependency \\
Hardware code smells & \cite{xu2025hlsrewriter} & Specialized code smell specifically for hardware, not mentioned in refactoring guru. Defined by the paper: Dynamic Pointers, Recursion, and Dynamic Memory. \\
SQL code smells & \cite{de2025effectiveness} & Specialized code smell specifically for hardware, not mentioned in refactoring guru. Defined by the paper: UVP, CDO, IEW, ECB, PMO, USA, CAT, and QEH. \\
\bottomrule
\end{tabular}
\color{black}
\end{adjustbox}
\vspace{-.3cm}
\end{table*}

 \begin{figure}[t]
 	\centering
 	\includegraphics[width=1.0\columnwidth]{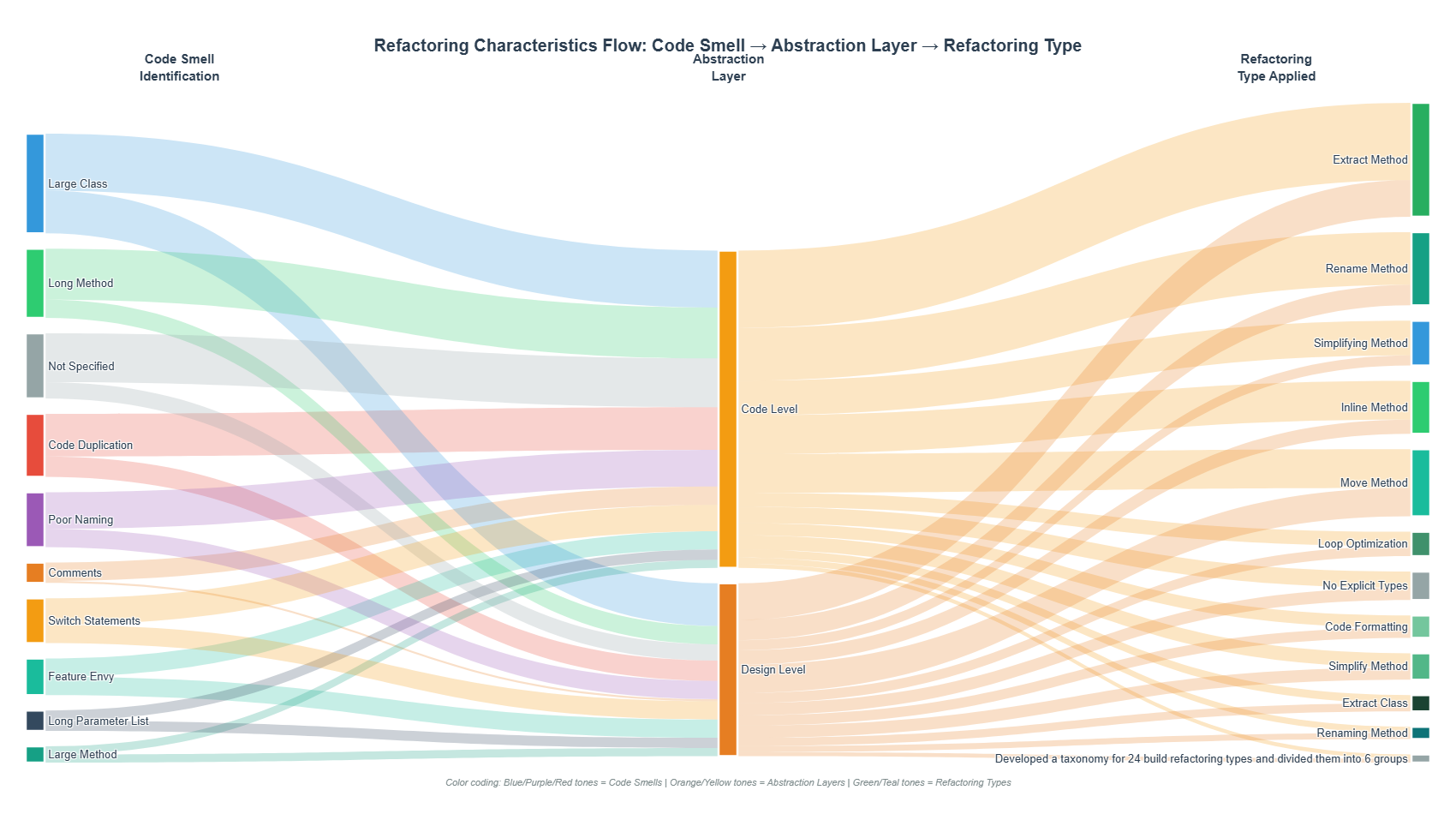}
 	\caption{\textcolor{black}{The relationship among refactoring type applied and code smell identification in LLM-driven refactoring studies.}}
 	\label{fig:mapping-refactoring}
\end{figure}

%% file: Tables/Refactoring.tex
\begin{table*}[ht!]
  \centering
	 \caption{\textcolor{black}{Refactoring characteristics in LLM-driven refactoring (RQ$_5$).}}
	 \label{Table:refactoring_in_llms_Refatoring}
\color{black}
\begin{adjustbox}{width=1.4\textwidth,center}
\rowcolors{2}{white}{gray!25}
\begin{tabular}{lLllllllll}\hline
\toprule
\bfseries Study & \bfseries Refactoring Identification  &  \bfseries Abstraction Layer &  \bfseries Refactoring Types &   \bfseries Behavior Preservation     \\
\midrule
Gehring \cite{gehring2023deterministic} &Not Specified & Design Level & Rename Method, Junit Migration & Manual Analysis \\
White \etal \cite{white2024chatgpt} & Not Specified & Design Level & No Explicit Refactoring Types & Manual Analysis \\
Shirafuji \etal \cite{shirafuji2023refactoring} & Code Duplication, Switch Statements & Code Level & Simplifying Method, Rename Method, Extract Method & Manual Analysis \\
Depalma \etal \cite{depalma2024exploring} & Code Duplication, Feature Envy, Large Class, Long Method, Switch Statements & Code Level, Design Level & Rename Method, Code Formatting, Loop Optimization, Simplifying Method & Static Code Analysis\\
AlOmar \etal \cite{alomar2024refactor} & Long Method, Code Duplication & Code Level & Rename Method, Extract Method, Move Method, Loop Optimization, Simplifying Method & Manual Analysis \\
Chavan \etal \cite{chavan2024analyzing} & Long Method & Code Level & Split file, Simplifying Method & Manual Analysis\\
Pomian \etal \cite{pomian2024assist} & Long Method & Code Level & Extract Method & Static Code Analysis \\
Pomian \etal \cite{pomian2024next} & Long Method & Code Level & Extract Method & Static Code Analysis\\
Dilhara \etal \cite{dilhara2024unprecedented} & Not Specified & Code Level & No Explicit Refactoring Types & Static and Dynamic Code Analysis\\
Gao \etal \cite{gao2024context} & Eager Test, Duplicate Assert, Assertion Roulette & Code Level & No Explicit Refactoring Types & Manual Analysis\\
Zhang \etal \cite{zhang2024refactoring} & Not Specified & Code Level & Extract Method, Simplify Method & Manual Analysis \\ 
Choi \etal \cite{choi2024iterative} & Long Method & Code Level & No Explicit Refactoring Types & Manual Analysis  \\
Wu \etal \cite{wu2024ismell} & Refused Bequest, Large Class, and Feature Envy & Code Level, Design Level & Extract Method, Move Method & Manual Analysis, Static Code Analysis Tool\\
Ishizue \etal \cite{ishizue2024improved} & Not Specified & Code Level & Rename Method, Extract Method, Simplifying Method, Code Formatting & Manual Analysis\\
Gautam \etal \cite{gautam2024refactorbench} & Not Specified & Code Level & Extract Method, Inline Method, Rename Method & Manual Analysis, Static Code Analysis \\
Baumgartner \etal \cite{baumgartner2024ai} & Data Clumps & Code Level, Design Level & Extract Class & Manual Analysis \\
Cui \etal \cite{cui2024three} & Feature Envy & Code Level, Design Level & Move Method & Manual Analysis \\
Zhang \etal \cite{zhang2025move} & Feature Envy, Long Method, Large Class & Code Level, Design Level & Move Method & Static Code Analysis \\
Ben Mrad \etal \cite{ben2024assessing} & Poor Naming & Code Level & Rename Method & Manual Analysis\\
Wang \etal \cite{wang2024hecs} & Large Class & Code Level, Design Level & Extract Class & Manual Analysis, Static Code Analysis\\
Gao \etal \cite{gao2024preference} & Large Class, Feature Envy, Refused Bequest & Code Level, Design Level & Extract Method, Simplify Method & Manual Analysis, Static Code Analysis \\
Cui \etal \cite{cui2024one} & Large Class & Code Level, Design Level & Extract Class & Manual Analysis, Static Code Analysis \\
Zhang \etal \cite{zhang2024copilot} & Long Parameter List, Long Method, Large Class, Switch Statements & Code Level, Design Level & No Explicit Refactoring Types & Manual Analysis, Static Code Analysis \\
Menolli \etal \cite{menolli2024teaching} & Duplicate code, Long Method, Large Class & Code Level, Design Level & Move Method, Extract Class, Extract Method & Manual Analysis, Static Code Analysis\\
Wang \etal \cite{wang2025testing} & Lazy Class & Code Level, Design Level & Extract Method, Inline Method, Simplifying Method & Manual Analysis, Static Code Analysis \\
Midolo \etal \cite{midolo2025automated} & Switch Statements, Feature Envy & Code Level & Simplifying Method & Manual Analysis \\
Ksontini \etal \cite{ksontini2025refactoring} & Temporary File Smells, Shell/Script Smells & Design Level & Extract Method, Rename Method & Manual Analysis, Dynamic Code Analysis\\
Alturayeif \& Hassine \cite{alturayeif2025refactoring} & Linguistic Smells & Design Level & No Explicit Refactoring Types & Manual Analysis \\
Divyansh \etal \cite{icmi2025evaluating} & Code Duplication, Long Methods, Poor Naming, Comments, Large Class & Code Level, Design Level & Extract Method & Manual Analysis, Static Code Analysis \\
Liu \etal \cite{liu2024empirical} & Code Duplication, Large Class, Long Method & Code Level & Extract Method, Rename Method, Inline Method & Manual Analysis, Dynamic and Static Code Analysis \\
Guo \etal \cite{guo2024exploring} & Poor naming & Code Level & Rename Method, Move Method, Extract Method, Code Formatting, Inline Method & Manual Analysis, Static Code Analysis\\
Dong \etal \cite{dong2025chatgpt} & Poor Naming, Switch Statements, Large Method, Large Class & Code Level, Design Level & Rename Method, Move Method, Extract Method, Inline Method & Manual Analysis, Dynamic and Static Code Analysis \\
Ghammam \etal \cite{ghammam2025build} & Code Duplication, Large Class, Long Methods & Code Level, Design Level & Developed a taxonomy for 24 build refactoring types and divided them into 6 groups & Manual Analysis \\
Batole \etal \cite{batole2025leveraging} & Feature Envy, God Class, 
Duplicated Code, Message Chain & Class Design & Move Method & Static Code Analysis \\
Palit \& Sharma \cite{palit2025reinforcement} & Not Specified & Code Level & Extract Method & Manual Analysis, Dynamic and Static Code Analysis \\
Rajendran \etal \cite{rajendran2025multi} & Long Methods, Large Classes, Duplicated Code & Code Level, Design Level, Architecture Level & No Explicit Refactoring Types & Manual Analysis, Static and Dynamic Code Analysis \\
Kim \cite{kim2025comparative} & Not Specified & Design Level & No Explicit Refactoring Types & Dynamic Code Analysis \\
Wang \etal \cite{wang2025towards} & Not Specified & Code Level & Inline Method, Move Method, Extract Method, Move Method, Rename Method & Manual Analysis, Static and Dynamic Code Analysis \\
Wang \etal \cite{wang2025exploration} & Not Specified & Code Level & No Explicit Refactoring Types & Manual Analysis \\
AlOmar \etal \cite{alomar2025chatgpt} & Not Specified & Code Level, Design Level & Extract Method, Move Method, Simplify Method & Manual Analysis \\
Midolo \etal \cite{midolo2025refactoring} & Not Specified & Code Level & No Explicit Refactoring Types & Manual Analysis, Static Code Analysis \\
Pucho \etal \cite{pucho2025refactoring} & Large Class, Long Method, Poor Naming Comments & Code Level & Rename Method, Extract Method, Inline Method, Simplifying Method & Static Code Analysis \\
Caumartin \etal \cite{caumartin2025exploring} & Poor Naming, Duplicated Code, Comments & Code Level & Rename Method, Extract Method, Inline Method, Move Method & Static Code Analysis \\
Pandini \etal \cite{pandini2025exploratory} & Architectural Smells: Cyclic Dependency, Unstable Dependency, God Component, Hublike Dependency & Architecture Level & No Explicit Refactoring Types & Manual Analysis \\
Guo \etal \cite{guo2025intention} & Not Specified & Code Level & Extract Method, Inline Method, Move Method & Manual Analysis \\
Xu \etal \cite{xu2025hlsrewriter} & Hardware level refactoring: dynamic pointers, recursion, and dynamic memory & Code Level, Design Level & Simplifying Method & Static Code Analysis \\
Tkackuk \cite{tkachuk2025determining} & Long Parameter List, Poor Naming, Code Duplication & Code Level, Design Level & Simplify Method, Renaming Method, Extract Method, Inline Method & Manual analysis \\
Guenoune \etal \cite{guenoune2025llm} & Not Specified & Design Level & Extract Method, Move Method & Manual Analysis \\
Amaral \cite{amaral2025improving} & Specific test code smells identified by the paper:  Conditional Test Logic, Overcommented Test, Suboptimal Assert, Test Without Description, Sensitive Equality, Assertion Roulette, Duplicate Assert, Magic Number, Lazy Test Redundant Print & Code Level & No Explicit Refactoring Types & Manual Analysis, Static Code Analysis \\
De Sousa \etal \cite{de2025effectiveness} & Specific SQL code smells identified by the paper: UVP, CDO, IEW, ECB, PMO, USA, CAT, QEH & Code Level & No Explicit Refactoring Types & N/A \\
\bottomrule
\end{tabular}
\end{adjustbox}

\vspace{-.3cm}
\end{table*}

%% file: Tables/Methods.tex
\begin{table*}[ht!]
  \centering
	 \caption{\textcolor{black}{Study Methods of Correction and Detection (RQ$_5$).}}
	 \label{Table:Methods_of_Correction_and_Detection}
\begin{adjustbox}{width=1.0\textwidth,center}
\rowcolors{2}{white}{gray!25}
\color{black}
\begin{tabular}{lLLlllllll}\hline
\toprule
\bfseries Study & \bfseries Method of Detection & \bfseries Method of Correction \\
\midrule
Gehring \cite{gehring2023deterministic} & Pattern-based static code analysis using OpenRewrite recipes & OpenRewrite recipes using lossless semantic tree using visitor pattern \\
White \etal \cite{white2024chatgpt} & Using ChatGPT and structured prompt patterns to detect code smells & Using ChatGPT and structured prompt patterns to refactor \\
Shirafuji \etal \cite{shirafuji2023refactoring} & Using radon for CC calculation, calculated Levenshtein, counted LOC, Chars, Tokens with tokenize, checked if the code compiled, and validated against hidden AOJ test cases & Collected correct programs as examples from Aizu, performed one-shot and few-shot prompting with examples \\
Depalma \etal \cite{depalma2024exploring} & N/A & Prompt Enginnering, with focuses on prompting structures and grammar \\
AlOmar \etal \cite{alomar2024refactor} & N/A & The authors examined DevGPT interactions; Prompt Enginnering approaches such as structures and grammar \\
Chavan \etal \cite{chavan2024analyzing} & N/A & The authors examined DevGPT interactions; Prompt Enginnering approaches such as structures and grammar \\
Pomian \etal \cite{pomian2024assist} & N/A & EM-Assist builds a few-shot prompt with the selected method and send it to the LLM several times to gather possible refactoring ways, using IntelliJ IDEA to check for safety, executes one of the options \\
Pomian \etal \cite{pomian2024next} & N/A & EM-Assist builds a few-shot prompt with the selected method and send it to the LLM several times to gather possible refactoring ways, using IntelliJ IDEA to check for safety, executes one of the options\\
Dilhara \etal \cite{dilhara2024unprecedented} & N/A & Mined for CPATs with R-CPATMiner, generate variants via LLMs, PyCraft validated the variants via static and dynamic checks, using PyEvolve turns valiants and original code to generalized rules, using these PyCraft automatically rewrote matching code fragments \\
Gao \etal \cite{gao2024context} & N/A & Build an test refactoring knowledge base, UTRefactor then used the code base to Chain-of-Thought prompting to refactor\\
Zhang \etal \cite{zhang2024refactoring} & Define smells via ASTscenario, ASTcomponent and conditions & Generate prompts using Prompt-LLM depending on the smells detected, reviewed by humans, then invoke ARIs to rewrite code\\
Choi \etal \cite{choi2024iterative} & Using Lizard library to measure CC, the highest CC is treated as refactoring target & Prompting LLM for refactoring, check with human-tests and generated-tests, check for compilability, recaluates CC, repeat 20 times  \\
Wu \etal \cite{wu2024ismell} & Using Mixture of Experts (MoE) such as PMD and JDeodorant to find the code smell & iSMELL uses LLMs for refactoring from the detection results by creating structured prompts \\
Ishizue \etal \cite{ishizue2024improved} & N/A & Refractory uses ChatGPT to improve automatic program repair via analyzing structure and syntax \\
Gautam \etal \cite{gautam2024refactorbench} & N/A & RefactorBench sets up LLMs, references, and AST testing with LM agents such as SWE-agent to refactor the code base \\
Baumgartner \etal \cite{baumgartner2024ai} & Using AST, deterministic rules, and LLMs to detect data clumps/code smells & After clumps are detected, LLM performs refactoring on the data clumps, and validates via CI/CD \\
Cui \etal \cite{cui2024three} & By creating an inter-class code entity dependency hypergraph using pre-train code models and LLMs, where the nodes are the methods and fields and the edges are dependencies, the paper detects feature envy specifically & The trained models output a set of Move Method candidates, the ChatGPT-3.5 completes code analysis and applied via building a VS code prototype \\
Zhang \etal \cite{zhang2025move} & N/A & MoveRec uses static analysis, LLMs, and semantic similarity to decide how to use move method refactoring to directly target feature envy \\
Ben Mrad \etal \cite{ben2024assessing} & N/A & Using LLMs to suggest new method names after code modification by referencing a dataset of 116 real world methods \\
Wang \etal \cite{wang2024hecs} & HECS detetcts extract class opportunities by building intra-class dependency hypergraphs using pre-trained code models, and training an enhanced version of the hypergraphs to identify clusters of methods/fields & N/A\\
Gao \etal \cite{gao2024preference} & N/A & RRG framework refactors by finetuning a CodeT5-style model to compressing relevant code snippets, then using PPO-based reinforcement learning to align with preferences to generate refactored code \\
Cui \etal \cite{cui2024one} & HECS detetcts extract class opportunities by building intra-class dependency hypergraphs using pre-trained code models, and training an enhanced version of the hypergraphs to identify clusters of methods/fields & N/A \\
Zhang \etal \cite{zhang2024copilot} & Pysmell automatically detected code smell via the Turning Machine Strategy & Created three types of prompts to refactor code by using Copilot, then reevaulated by Pysmell again \\
Menolli \etal \cite{menolli2024teaching} & Manual and using tools such as JDeodorant and SonarLint to identify code smells & Students used ChatGPT to refactor their code and used it to aid learning programming \\
Wang \etal \cite{wang2025testing} & N/A & RETESTER uses historical bug reports help train and guide LLMs using few-shot prompting \\
Midolo \etal \cite{midolo2025automated} & N/A & Using ChatGPT-4 with prompt engineering (system and user) to refactor non-idiomatic python code to idiomatic python code \\
Ksontini \etal \cite{ksontini2025refactoring} & N/A & With refactored dockerfiles examples collected, they used ChatGPT-4o with few-shot prompting to generate refactored dockerfiles \\
Alturayeif \& Hassine \cite{alturayeif2025refactoring} & By using LLMs and NLPs with prompt engineering the authors was able to to identify linguistic based smells & GPT was used with custome prompts (system, user) for each smell to be refactored \\
Divyansh \etal \cite{icmi2025evaluating} & Used SonarQube to automatically detect code smells & The code was automatically refactored via a pipeline with scripts and datasets to refactor by using the ChatGPT API \\
Liu \etal \cite{liu2024empirical} & With LLMs and prompt engineering the authors asked for refactoring opportunities & With LLMs and prompt engineering java files are automatically rewritten according to specific refactoring tasks\\
Guo \etal \cite{guo2024exploring} & Using existing tools the authors randomly selected 400 samples and manually annotated them for refactoring opportunities & Author utilized ChatGPT and prompt engineering by using the code and the review messages with Zero-Shot prompting \\
Dong \etal \cite{dong2025chatgpt} & N/A & By using Chat-GPT and prompt engineering to generate test programs and analyzing the bug reports, the authors created a feature library, then used GPT to produce new test programs from the results of feature library \\
Ghammam \etal \cite{ghammam2025build} & BuildRefMiner used 725 build-related commits and GPT-4o with prompt engineering to detect and classify refactorings that happened in version controls & N/A \\
Batole \etal \cite{batole2025leveraging} & MM-Assist uses LLMs with Chain-of-Thought reasoning to identify code smells & MM-Assist filters out valid methods then sends the most problematic methods using vectors then using LLMs to rank which methods to move via Chain-of-Thought\\
Palit \& Sharma \cite{palit2025reinforcement} & The authors used RefactoringMiner and SEART to detect code smells to create their dataset & The authors generated refactoring by fine-tuning code-t5 and plbart models based off of proximal policy optimization reinforcement learning training on good and bad behaviors \\
Rajendran \etal \cite{rajendran2025multi} & The authors mentions of using a maintainability agent who specicializes in static analysis, rule-based smell detection and LLM fine tuning specifically on maintainability patterns & The authors mentioned the use of multiple specialized LLM agents that can do different parts of refactoring such as agents who are specialized in their specific parts, they each suggest refactoring and collaborate and negotiate \\
Kim \etal \cite{kim2025comparative} & The authors used formal predicate logic patterns to detect code smells & The authors used prompt engineering to refactor based off of the patterns in designs that was detected \\
Wang \etal \cite{wang2025towards} & The authors manually analyzed 518 real bug reports & The authors refactored using LLMs and prompt template and manually tagging and labeling bugs, with ten-shot prompting \\
Wang \etal \cite{wang2025exploration} & The authors used a Unidiff-based quantitative metric, that compared target and real code changes to detect code smells & The authors used fine-tuning or few-shot methods with three open-source LLMs: CodeLLaMA, LLaMA3, and StableCode3B \\
AlOmar \etal \cite{alomar2025chatgpt} & N/A & The authors manually examined conversations between developers and ChatGPT and analyzed how refactoring was carried out by ChatGPT \\
Midolo \etal \cite{midolo2025refactoring} & The authors used Abstract Syntax Tree analysis with JavaParser to automatically extract loops to be refactored & Using ChatGPT-3.5 Turbo and prompt engineering the authors was able to refactor the code \\
Pucho \etal \cite{pucho2025refactoring} & The authors feed the code file to the Code Quality Agent which performs static code analysis and reports code smells & The Refactoring Strategy Agent takes in the report from the Code Quality Agent and planned refactoring editing and sent it to the Code Transformation Agent to apply it to the code \\
Caumartin \etal \cite{caumartin2025exploring} & The authors collected samples from the CodeReview with code smells identified in code reviews & The author adjusted temperature settings and tested five different prompt templates with the LLM models Llama 2 and CodeLlama to refactor the code \\
Pandini \etal \cite{pandini2025exploratory} & The authors used Arcan, that automatically detected Architectural Smells. Arcan uses denedency graph analysis and machine-learning based assessment & The authors experimented with prompts (detailed vs not), using retrieval-augmented generation for context, and different LLMS such as Qwen and Claude Sonnet \\
Guo \etal \cite{guo2025intention} & The authors used code review comments along with prompt engineering to extract code smells & Utillizing prompt engineering and code reviews, the authors tried different methods of prompting with retrieval-augmented generation \\
Xu \etal \cite{xu2025hlsrewriter} & C/C++ code is compiled using the Hardware Level Synthesis tool and errors are flagged & Using LLMs with Retrieval-Augmented Generation the authors were able to create prompts that helped them create loops instead of recrusions, pointers to arrays, static allocations instead of malloc() etc \\
Tkachuk \cite{tkachuk2025determining} & N/A & The authors compare Gemini, ChatGPT and Claude to compare how each did in refactoring, the authors also asked for reasons why they refactored the code \\
Guenoune \etal \cite{guenoune2025llm} & The authors find where is needed to refactor by using Formal Concept Analysis and Relational Concept Analysis, that together they can find redundant or overlapping code & THE RCA and FCA builds on each other iteratively and using LLMs the authors was able to refactor the code \\
Amaral \cite{amaral2025improving} & The authors used Steel and SNUTS.JS to detect and find code smells (rule-based static analysis) & The authors opened the code in VSCode and manually prompted Copilot and CodeWhisperer with zero-shot prompting to refactor the code \\
De Sousa \etal \cite{de2025effectiveness} & The authors used GPT-4o mini and Phi-4 compared to GPT-4o and Gemini 2.0 Flash by using set prompt templates with the identified SQL code smells; the best ones were the GPT-4o and  Phi-4 with F1 score of 79\%. & N/A \\
\bottomrule
\end{tabular}
\color{black}
\end{adjustbox}
\vspace{-.3cm}
\end{table*}